\documentclass{aa}  

\usepackage{graphicx}
\usepackage{txfonts}
\usepackage{comment}
\usepackage{multirow}
\usepackage[dvipsnames]{xcolor}
\usepackage{hyperref}
\usepackage{booktabs,multirow, threeparttable}
\usepackage{placeins}
\usepackage{lscape}
\newcommand{\ergs}{erg s$^{-1}$}

\begin{document}

   \title{ The Success of Optical Variability in Uncovering AGNs in Low-stellar Mass Galaxies  }
   \subtitle{}

   \author{ S. Bernal\inst{1,2}
          \and
          P. Sánchez-Sáez\inst{3}
          \and
          P. Arévalo\inst{1,2,5}
          \and
          F.E. Bauer\inst{4,5,6}
          \and
          P. Lira\inst{7,2}
          \and
          B. Sotomayor\inst{1}
          }

  \institute{Instituto de F\'isica y Astronom\'ia, Universidad de Valpara\'iso, Gran Breta\~na 1111, Valpara\'iso, Chile\\
\email{santiago.bernal@postgrado.uv.cl}
\and Millennium Nucleus on Transversal Research and Technology to Explore Supermassive Black Holes (TITANS)
\and European Southern Observatory, Karl-Schwarzschild-Str. 2, 85748, Garching, Germany
\and Instituto de Astrof{\'{\i}}sica and Centro de Astroingenier{\'{\i}}a, Facultad de F{\'{i}}sica, Pontificia Universidad Cat{\'{o}}lica de Chile, Campus San Joaquín, Av. Vicuña Mackenna 4860, Macul Santiago, Chile, 7820436
\and Millennium Institute of Astrophysics (MAS), Nuncio Monseñor Sótero Sanz 100, Providencia, Santiago, Chile
\and Space Science Institute, 4750 Walnut Street, Suite 205, Boulder, Colorado 80301
 \and Departamento de Astronom{\'{\i}}a, Universidad de Chile, Camino el Observatorio 1515, Santiago, Chile}
 
   \date{Received August 13, 2024 ; accepted December 10, 2024}

  \abstract
  % context heading (optional)
  % {} leave it empty if necessary  
   {The origins of supermassive black holes (SMBHs) at the centers of massive galaxies are a topic of intense investigation. One way to address this subject is to identify the seeds of SMBH as intermediate-mass black holes (IMBHs; $100M_\odot < M_{BH} < 10^6M_\odot$). IMBHs are expected to be found at the centers of low-stellar-mass galaxies (LSMGs).}
  % aims heading (mandatory)
   {Our goal is to complete the census of SMBHs in LSMGs. In this work, we aim to establish the purity of Active Galactic Nuclei (AGN) selection by algorithms based on optical variability and characterize the black hole population found through this method. }
  % methods heading (mandatory)
   {We used random forest algorithms to classify all objects in a large portion of the sky, using optical light curves obtained, or built from images provided, by the Zwicky Transient Facility (ZTF). We compare different selection sets based on alerts (flux changes with at least $5\sigma$ significance) or complete light curves derived from different photometric selection algorithms. The AGN candidates thus selected are cross-matched with objects in the NASA-Sloan Atlas (NSA) of local galaxies, with $M_*<2\times10^{10}M_\odot$. The AGN nature of these candidates is verified and characterized using archival optical spectra from SDSS. We further establish the fraction of candidates with counterparts in the eROSITA data release 1 catalog of X-ray sources. }
  % results heading (mandatory)
   {From an initial sample of 506 candidates, 415 have good-quality spectra. Among these 415 objects, we found significant broad Balmer lines in the spectra for 86\% (357) of the candidates. When considering BPT classifications, an additional 5 candidates were confirmed, resulting in 87\% (362) confirmed candidates. Specifically,  broad Balmer lines were detected in 94\%-98\% of the AGN candidates selected from complete light curves and in 80\% of those selected from the less frequent ZTF alerts. The black hole masses estimated from the spectra range from $2.2\times10^6M_\odot$ to $4.2\times10^7M_\odot$, reaching lower values for the candidates selected using the more sensitive light curves. The black hole masses obtained cluster around $0.1\%$ of the stellar mass of the host from the NSA catalog. Two-thirds of the AGN candidates are classified as Seyfert or Composite by their narrow emission line ratios (BPT diagnostics) while the rest are star-forming. Almost all the candidates classified as Seyfert and over $50\%$ of those classified as star-forming have significant BELs. We found X-ray counterparts for $67\%$ of the candidates that fall in the footprint of the eROSITA-DE DR1. Considering only the candidates with significant BELs the matches raise to $75\%$, regardless of where they appear in the BPT diagnostics diagrams.}
  % conclusions heading (optional), leave it empty if necessary 
   {}


   \maketitle
%
%-------------------------------------------------------------------

\section{Introduction}

The presence of supermassive black holes (SMBHs; $M_{BH}>10^6$$M_\odot$) at the center of massive galaxies is well established, yet the origins of these entities remain a topic of intense investigation ( \citealt{BHseeds2016PASA...33...51L}, \citealt{2020ARA&A..58...27IBHseeds}). One plausible pathway involves the formation of intermediate-mass black holes (IMBHs; $100M_\odot<M_BH<10^6M_\odot$), which serve as seeds that grow through accretion and mergers to become SMBHs (\citealt{2012NewAR..56...93A} ). Understanding the properties of IMBHs, particularly those formed at early epochs, will allow us to understand the evolution of SMBHs \citep{2010A&ARv..18..279VFsmbh}. Unfortunately, our technological resources are still incapable of detecting the expected seeds at high redshifts ($z>6-12$), but a census of the local low-mass black holes can provide alternative constraints. Given the relation observed between the mass of a central massive black hole and the mass of its host galaxy \citep{Reines2015} in the local universe searching for IMBHs in low-stellar mass galaxies (LSMGs) in the near Universe ($z<0.15$) is a plausible option to study the origins of SMBHs. An advantage in these LSMGs is that we can study the processes of black holes without the complex merge histories found in more massive galaxies (\citealt{2010A&ARv..18..279VFsmbh}, \citealt{2012NatCo...3.1304G}), this allows to observe the black hole evolution driven primarily by internal process and compare these findings with those in more massive, merger-rich galaxies to determine the effects of different growth mechanisms. Hence, nearby LSMGs are currently the best laboratories for elucidating the initial conditions and growth mechanisms of black holes in the early Universe. 

The challenge in detecting IMBHs in LSMGs comes from the faintness and weakness of their observational signatures (\citealt{Reines2015},\citealt{Baldassare18}). Despite this, the search for active galactic nuclei (AGNs) in LSMGs has produced various results using different approaches. For instance, the first searches were in optical spectroscopic surveys using narrow emission-line (NEL) diagnostic diagrams, known as BPT-diagrams \citep{baldwin1981classification}, to distinguish between stellar or AGN ionization sources, together with potential detections of broad Balmer emission lines used to estimate black hole masses (e.g. \citealt{reines2013}, \citealt{2014AJ....148..136Moran}). However, emission line diagnostics can fail to identify AGNs in these LSMGs, as shown by photoionization models \citep{2019ApJ...870L...2Cann}. For example, \citet{2020MNRAS.492.2268Birchall} find that among 61 dwarf galaxies ($M_*\leq 3\times 10^{9}$$M_\odot$) that exhibit AGN X-ray activity, 85 percent are not classified as AGNs by BPT-diagrams. Another approach is the identification of AGN activity using Integral Field Unit spectroscopy observations in dwarf galaxies (\citealt{Mezcua17}, \citealt{2024MNRAS.528.5252Mezcua}), which is significantly more successful than searches with single-fiber spectra because it allows for the separation of the galaxy emission from the light emitted by the gas ionized by the AGN. Nevertheless, this method is limited by the expensive data required.

Another explored method to search for AGNs in LSMGs is the detection of variability from these sources; stochastic changes in brightness over time can indicate the presence of an AGN due to the dynamic processes occurring near the black hole at the center of the galaxy. For example, \citet{2020ApJ...889..113MP_IMBH_photo_var} found 502 AGN candidates via nuclear optical variability among 12,300 galaxies with z<0.35 and Sloan Digital Sky Survey (SDSS) legacy spectra, of which 22 were confirmed as AGN using BPT diagnostic diagrams. In another study, \citet{Burke2022} looked for the expected short variability timescales, constraining the characteristic variability timescale as $log(\tau/day) \leq 1.5)$, in six-year light curves acquired with the Dark Energy Camera. From a parent sample of 63721 galaxies, they selected 706 AGN candidates including 26 LSMGs ($M_*< 10^{9.5}$$M_\odot$ ). However, spectroscopic confirmation was achieved for only one of the five candidates with available spectra. Other authors (e.g., \citealt{Baldassare2020}, \citealt{Kimura2020}, \citealt{Ward2022}) have selected candidates by characterizing the variability amplitude in light curves from different photometric surveys, such as the Palomar Transient Factory (PTF), Hyper Suprime-Cam Subaru Strategic Program (HSC-SSP), and Zwicky Transient Facility (ZTF). However, confirmation of these candidates through X-ray observations has yielded a low success rate. In some cases, this is attributed to the depth limits of the X-ray observations, while in others, it may be due to the AGN candidate selection method. These cases will be discussed later in Section \ref{Sec:Comparison}.

The increasing number of AGN candidates and their confirmation in large numbers is possible by the advantages offered by large surveys. Therefore, it is anticipated that upcoming observations from new large observatories and surveys will significantly advance our understanding of AGNs in faint, low-steallar mass galaxies. For instance, the Vera C. Rubin-Large Synoptic Survey Telescope (LSST; \citealt{Ivezic2019}) is expected to probe weaker variability despite host contamination. The Extremely Large Telescope (ELT; \citealt{2007Msngr.127...11GELT}) will offer improved spatial resolution and sensitivity, and the 4-meter Multi-Object Spectroscopic Telescope (4MOST; \citealt{2022SPIE12184E..14D}) will provide enhanced spectral resolution. Together, these instruments will serve as powerful tools in the search and characterization of AGNs in these galaxies.

Here, we present the successful identification of AGNs in LSMGs by spectroscopic confirmation of candidates selected through variability together with color and morphology. These candidates were selected by the application of a random forest classifier to ZTF light curves, later limited to objects in the NASA-Sloan Atlas v1.0.1 catalog of nearby galaxies \footnote{https://www.sdss4.org/dr17/manga/manga-target-selection/nsa/} (NSA), with stellar mass $M_*<2\times 10^{10}$$M_\odot$ and redshift $z<0.15$. For spectroscopic confirmation, we used archival spectra from SDSS-DR17. Furthermore, we present the success in finding X-ray counterparts using the recent eRASSv1.1 catalog \citep{Merloni2024}.

This work is organized as follows: Section \ref{sec: Selection of candidates} describes the selection of AGN-candidates; Section \ref{Sec:Spectro data and alysis} presents the methods used for the spectroscopic confirmation and characterization of candidates; Section \ref{sec: results} provides the results of the characterization; In section \ref{sec:xrays} we present a comparison with the X-ray catalog, and in Section \ref{sec: Discussion} the consistency between indicators of AGN activity, and the comparison with previous similar studies; Conclusions are summarized in section \ref{sec:conclusions}. When needed, we assume a $\Lambda$CDM cosmology with parameters  $\Omega_M=0.3$, $\Omega_{\lambda}=0.7$ and $H_0=100h\ kms^{-1} Mpc^{-1}$ with $h=70$.

%--------------------------------------------------------------------

\section{Selection of AGN candidates in low mass galaxies}\label{sec: Selection of candidates}
We selected AGN candidates by means of their optical variability. The variability features were measured from the multi-epoch photometry on a large portion of the sky provided by ZTF. The selections were made using hierarchical random forest (RF) algorithms.

ZTF is a Northern sky survey ($\text{Dec}> -30$) that has been in operation since 2018, with a typical ABmag depth limits of $20.8$ and $20.6$ in $g$ and $r$-band, respectively, and with a cadence of ${\sim}3\ \text{days}$ \citep{Bellm19}. ZTF provides different data products, including an alert stream, data release (DR) light curves, DR images, and DR catalogs, among others.
Using these, we made four different selections. The first selection was made using the ALeRCE (Automatic Learning for the Rapid Classification of Events) broker \citep{Forster21}, specifically utilizing the classifications provided by the ALeRCE light curve classifier described in \citet{Sanchez-Saez20LC}. This classifier uses variability features from the ZTF alert stream ( i.e., only flux variations detected at greater than  $5\sigma$ significance in a reference-subtracted image); using both $g$ and $r$-band light curves whenever possible, and single-band light curves otherwise; colors from ZTF and WISE, and a morphology score from \citet{Tachibana18}, where values close to 1 indicate a point source and values close to 0 indicate an extended source. We included objects classified as AGN (host-dominated), QSO (core-dominated), and Blazar (jet dominated) in our candidate selection. This sample covers the entire ZTF sky, namely declination larger than $-30 \rm deg$. We will refer to this set as the ‘Alerts’ set. For this set, we use data taken between March 2018 and November 2022.

The second and third sets used a similar classifier but trained on the full point spread function photometry (PSF) light curves provided by ZTF in their data releases \citep{masci18}. The ZTF DR11 used here contains data from March 2018 to March 2022. These light curves are built from PSF-photometry on the science images of all epochs. The benefit of these data sets is that they include photometry for all the sources present in the reference catalogs, with detection in the ZTF science images. This means that it includes objects of lower variability amplitudes and more data points per light curve than the Alerts. The drawback is that the PSF photometry on these non reference-subtracted images is not ideal for tracking the flux of a variable point source ( i.e., the AGN) superimposed on the extended image of their host galaxies. This combination results in light curves with uncertain errors and modulations caused by seeing and weather conditions. This classification was made for the southern portion of the ZTF sky, namely $-30<\text{Dec}<+7$, to produce candidates for the Chilean AGN and Galaxy Evolution Survey (ChANGES; \citealt{Bauer2023}) project of 4MOST \citep{2022SPIE12184E..14D}, avoiding the Galactic plane. The classification was carried out independently for the $g$ and $r$-band light curves using data from the ZTF DR11. The classifier is described in \citet{Sanchez-Saez23}, and our selection includes objects classified as AGNs: lowz-AGN ($z\leq0.5$), midz-AGN ($0.5<z\leq3$), highz-AGN ($z>3$), and Blazar (jet dominated). We will refer to these sets as ‘DR-g’ and ‘DR-r’, respectively. Additionally, \citet{Sanchez-Saez23} discussed several caveats in the selection of candidates using ZTF DRs (see Section 7.1 in \citealt{Sanchez-Saez23}). In particular, for the $r$-band, there are regions of the sky with exceptionally large densities of epochs, which affect the computation of features and produce an over-density of highz-AGN candidates in the Galactic plane. Moreover, \citet{Sanchez-Saez23} also demonstrated that much purer candidate lists are obtained from the ZTF DRs in both $g$ and $r$ bands when the candidates are filtered by probability. Therefore, in order to have a clean sample of AGN candidates, we filtered the AGN sample from \citet{Sanchez-Saez23} as described below.

For ZTF DR11 $g$-band: 
\begin{itemize}
\item Probability of variability $\text{pred\_init\_class\_prob}>=0.9$ and 
\item $\text{abs(gal\_b)}>=20$
\end{itemize}

For ZTF DR11 $r$-band:
\begin{itemize}

\item$\text{pred\_init\_class\_prob}>=0.9$, 
\item$\text{abs(gal\_b)}>=20$, 
\item Significance of the Gaia DR3 proper motion $\text{pmsig}<=3\sigma$, 
\item Gaia proper motion $\text{PM}<3$, 
\item $\text{IAR\_phi}>=0.8$, 
\item $\text{GP\_DRW\_sigma}>=0.0001$,
\item $\text{GP\_DRW\_tau}>=5$ 
\item number of epochs $\text{nepochs}<=400$. 

\end{itemize}
All the features are explained in \citet{Sanchez-Saez23} and \citet{Sanchez-Saez21} (most of the definitions are there). These filtering was made to improve the purity of the samples for the selection of targets of the 4MOST ChANGES program and will be described in detail in Bauer et al. (in prep.). In particular, a number of Galactic sources were incorrectly identified as high-redshift AGNs in the original selection, especially in the $r$-band selection. Applying the previously mentioned restrictions on proper motion, as well as on the structure and timescales of the variations (using IAR\_phi and the DRW parameters), eliminated the majority of these misidentifications.

The fourth set is made using Forced Photometry light curves. However, given the large number of objects (on the order of tens of millions), the ZTF forced photometry service \citep{Masci2023}, which provides light curves on a request basis, was not adequate for our purposes. Therefore, we performed new aperture photometry on the {reference-subtracted science images} (i.e., difference images), provided by ZTF for all good quality epochs (i.e., with ZTF metadata $\text{infobits}=0$, $\text{maglimit}>20 \rm mag$, $\text{seeing }< 4 \rm arcsec$) in the $g$-band only, using at most one observation per night, and using an aperture of four arcseconds. The $g$-band was selected because of the tendency of AGNs to show larger variability amplitude at bluer wavelengths \citep[e.g.,][]{2010ApJ...721.1014MacLeod_photoVar}, and the lower level of contamination from the host galaxy in bluer bands, which makes this band more sensitive to variability. The data used covered the period between March 2018 and September 2022. The photometry was forced on the location of all sources detected in the reference images. This experiment was conducted for the region of the sky with $-30<\text{Dec}<+15.5$, to cover the ChANGES region. The Galactic plane area was not included due to the large number of contaminating sources in those fields. The light curves obtained from the aperture photometry were built with the purpose of selecting AGN out of the tens of millions of detectable objects in this region of the sky. The variability features were measured from the light curves and we used a similar classifier as the one used for the DR light curves. For this classifier we include some additional features, namely PANSTARRS $i-z$ color, proper motion provided by Gaia, Mexican-Hat filtered variance at timescales of 45 and 450 days, the error on the excess variance and a flux asymmetry estimator ($lc-asymmetry=\frac{N_p-N_n}{N}$, where $N_p$ and $N_n$ are the number of epochs with flux higher and lower than the mean flux of the light curve, and $N$ is the number of total epochs). The processing of these light curves and the classifier with the additional features is described in Arévalo et al. (in prep.). We refer to this set as ‘Forced Photometry’.

ZTF provides unique object IDs for unique combinations of RA, Dec, filter, field, CCD, and CCD quadrant. Thus, a single object can have multiple IDs in a given band. Since the flux calibration is done on a quadrant basis, there can be small offsets between the mean fluxes of observations with different IDs, which could be mistaken for rapid fluctuations. Thus, for the DR light curves as well as for the Forced Photometry ones, we kept only the light curve (i.e., unique object ID) with the largest number of epochs associated with a single object. This can result in a reduced number of epochs for objects in the overlap regions between pointings. However, it does not adversely reduce the total baseline of the light curve or the sampling rate for the intermediate and long timescale fluctuations.

The sets described above compose our variability-selected AGN samples. To search for massive BHs, we cross-matched these samples to the population of galaxies with low stellar masses retrieved from the NASA-Sloan Atlas v1.0.1 \footnote{https://www.sdss4.org/dr17/manga/manga-target-selection/nsa/} (NSA) catalog. The NSA v1.0.1 contains a catalog of all low redshift galaxies ($z<0.15$) inside the footprint of SDSS DR8, identified in images of this survey. The catalog provides, among many other derived quantities, measurements of the total, K-corrected stellar mass, which we used to select our parent sample. For our work, we selected galaxies by SERSIC-MASS, which is listed in units of $M_{\odot} h^{-2}$. We adopt the value of $h=0.7$ for the conversion into masses in $M_{\odot}$. We set a limit to search for AGN candidates in LSMGs at $\text{SERSIC-MASS}<10^{10}M_{\odot} h^{-2}$,  i.e., $M_*<2\times 10^{10}M_{\odot}$. The NSA catalog contains $220,830$ objects that meet the mass condition. All these objects lie within the sky region covered by the Alerts set ($\text{Dec}>-30$). The smaller Forced Photometry sky area ($-30<\text{Dec}<+15.5$) contains $103,054$ ($47\%$) of the NSA galaxies, while the DR-g and DR-r sky region ($-30<\text{Dec}<+7$) contains 73,964 ($33\%$) of the NSA objects. We cross-matched the positions of the four variable AGN samples with the LSMGs using 
%the right ascension (RA) and declination (Dec) coordinates and 
a radius of $1.5$ arcseconds. The total number of AGN candidates matched to low mass galaxies is 383 in the Alerts set, 215 in the Forced Photometry set, 71 in the DR-g set, and 69 in the DR-r set. Since one candidate can be part of one or more sets, counting unique objects the match produced a total of 506 AGN  variability-selected candidates. 

\begin{figure}[htpb]
    \centering
    \includegraphics[width=0.48\textwidth]{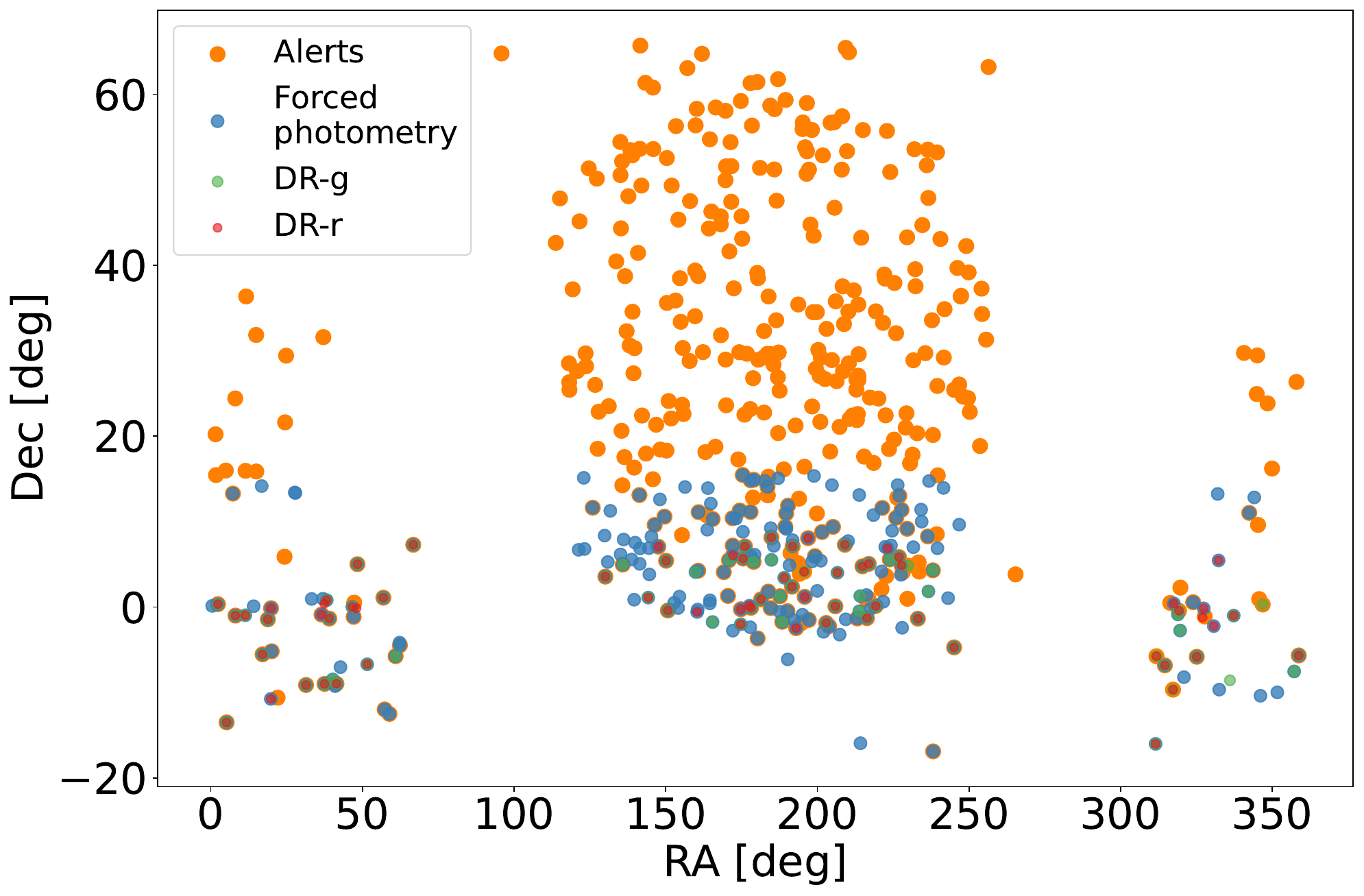}
   
    \caption{Sky distribution of AGN candidates in low stellar-mass galaxies. The different selection sets are marked by different colors, whereas AGN candidates selected in multiple sets have their symbols overlaid.}
    \label{fig:candidates}
\end{figure}

\begin{figure}[t]
    \centering
    \includegraphics[width=0.48\textwidth]{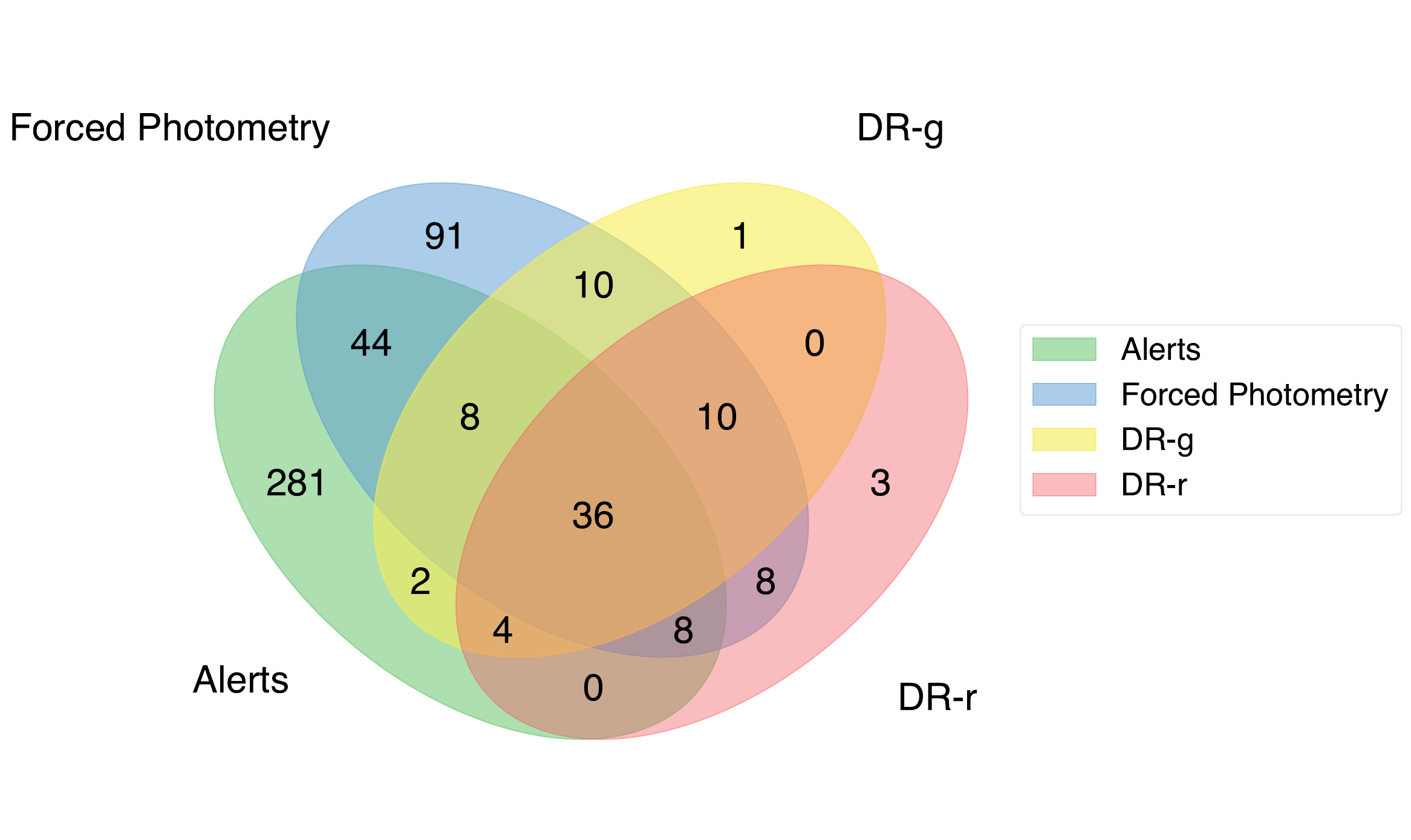}
    \caption{Venn diagram to show the number of elements within each intersection between the different sets. Note that the number of objects included only in the Alerts set is higher than in the other sets due to the larger sky region coverage of the Alerts set.}
    \label{fig:set_intersections}
\end{figure}

Figure \ref{fig:candidates} shows the sky locations of the matched sample. The Alerts set is marked in orange, the Forced Photometry set in blue, the DR-g in green, and DR-r in red. Evidently, most of the sky of the NSA sample is only covered by our Alerts set. The Forced Photometry set covers the area below $\text{Dec}=+15.5$ and the DR-g and DR-r sets below $\text{Dec}=+7$. We note that in the sky area covered by both sets, most of the Alerts candidates also appear in the Forced Photometry set (96 out of 137), while a majority of Forced Photometry candidates are not selected in the Alerts set (120 out of 216 are not in the Alerts set). More strikingly, almost all DR-g candidates are also selected in the Forced Photometry set, but about half of the Forced Photometry candidates are not selected in the DR-g set, and similarly for the DR-r set. These differences happen, in the case of the DR sets, because the DR light curves of extended objects are too noisy to allow the detection of low-amplitude variations, and for the case of the alert light curves, because the 5-sigma threshold prevents the identification of low-amplitude variations with respect to the reference image. The number of sources in each set and the coincidences in the regions of overlap between Forced Photometry and the other sets are summarized in Table \ref{tab:candidates}.  Additionally, Fig.~\ref{fig:set_intersections} shows a Venn diagram for the four sets to visualize the number of elements within each intersection between sets.

\begin{table}
    \centering
     \caption{Number of AGN candidates in low mass galaxies in the overlap regions between different sets.}
    \begin{tabular}{c|c|c}
        Alerts & Forced Photometry & Coincidences\\
         \hline
         137 & 216 & 96\\
        \hline
        \hline
        DR-g & Forced Photometry & Coincidences\\
         \hline
         71 & 143 & 64\\
        \hline
        \hline
        DR-r & Forced Photometry & Coincidences\\
         \hline
         69 & 143 & 62\\
        \hline
        \hline
        DR-g & DR-r & Coincidences\\
         \hline
         71 & 69 & 50\\
        \hline
    \end{tabular}
\begin{tablenotes}
   \item {\bf Note.} The overlap regions correspond to $\text{Dec}< 15.5 deg$ for the top row and $\text{Dec}< 7 deg$ for the next two rows. The coincidences column refers to the objects selected in both sets in each row. 
\end{tablenotes}
   
    \label{tab:candidates}
\end{table}

\begin{figure}[htpb]
    \centering
    \includegraphics[width=0.48\textwidth]{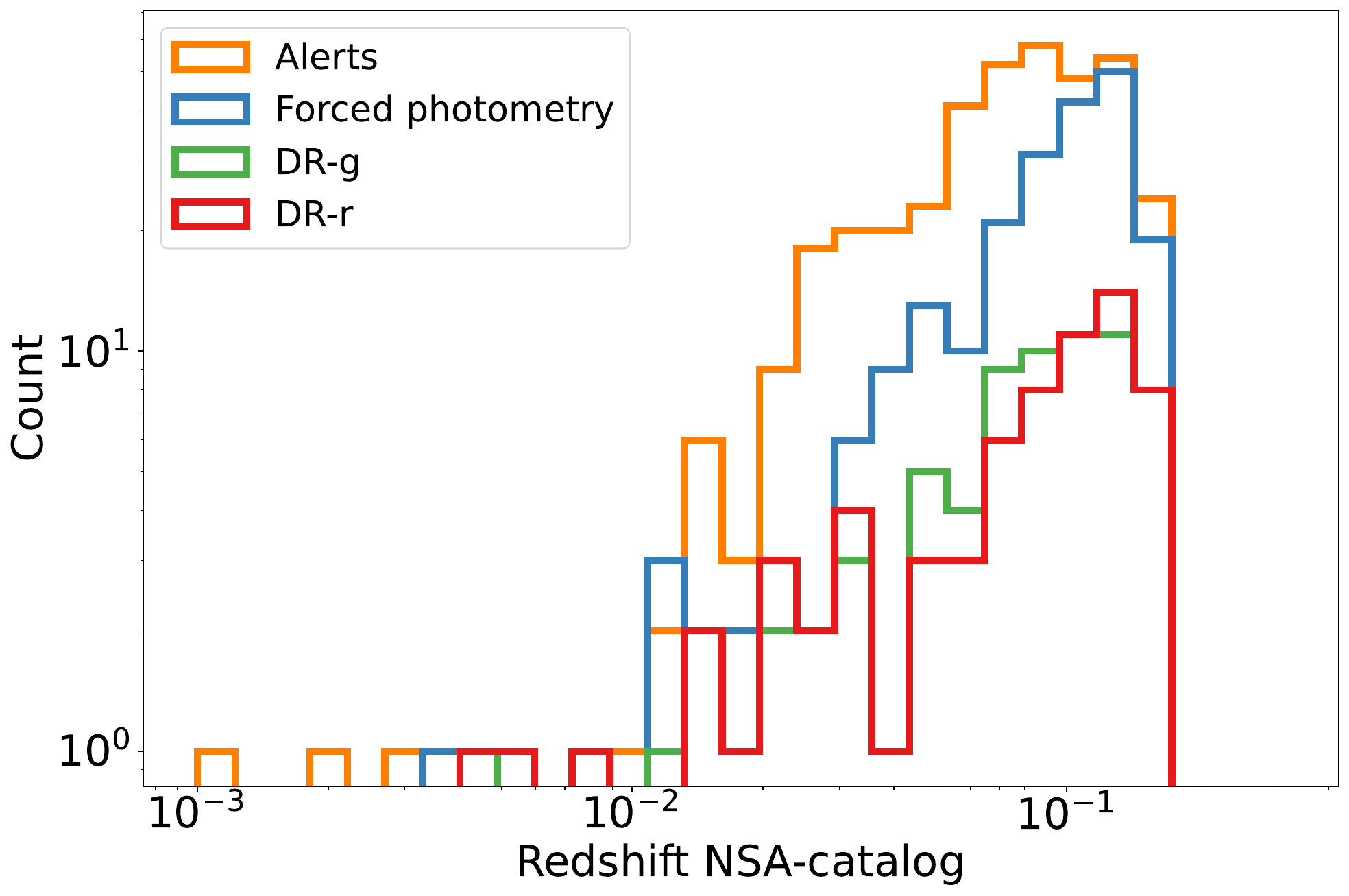}
    \caption{NSA v1.0.1 catalog redshift distributions for all the objects in each of the four sets: orange for Alerts, blue for Forced Photometry, green for ZTFDR11 $g$-band, and red for ZTFDR11 $r$-band.}
    \label{fig:z_nsa_candidates}
\end{figure}

Additionally, we present some properties of the selected objects using the data from the NSA v1.0.1 catalog. Fig.~\ref{fig:z_nsa_candidates} shows the distribution of the redshift for each set, which appear similar, with median redshift values of 0.076 for the Alerts, 0.099 for the Forced Photometry, 0.089 for DR-g, and 0.092 for DR-r. As we note in the next section, we find some objects with different redshift to the ones listed in the NSA catalog. However, this number is small and does not affect the redshift distribution and the magnitude and color distributions shown below.

\begin{figure}[htpb]
    \centering
    \includegraphics[width=0.48\textwidth]{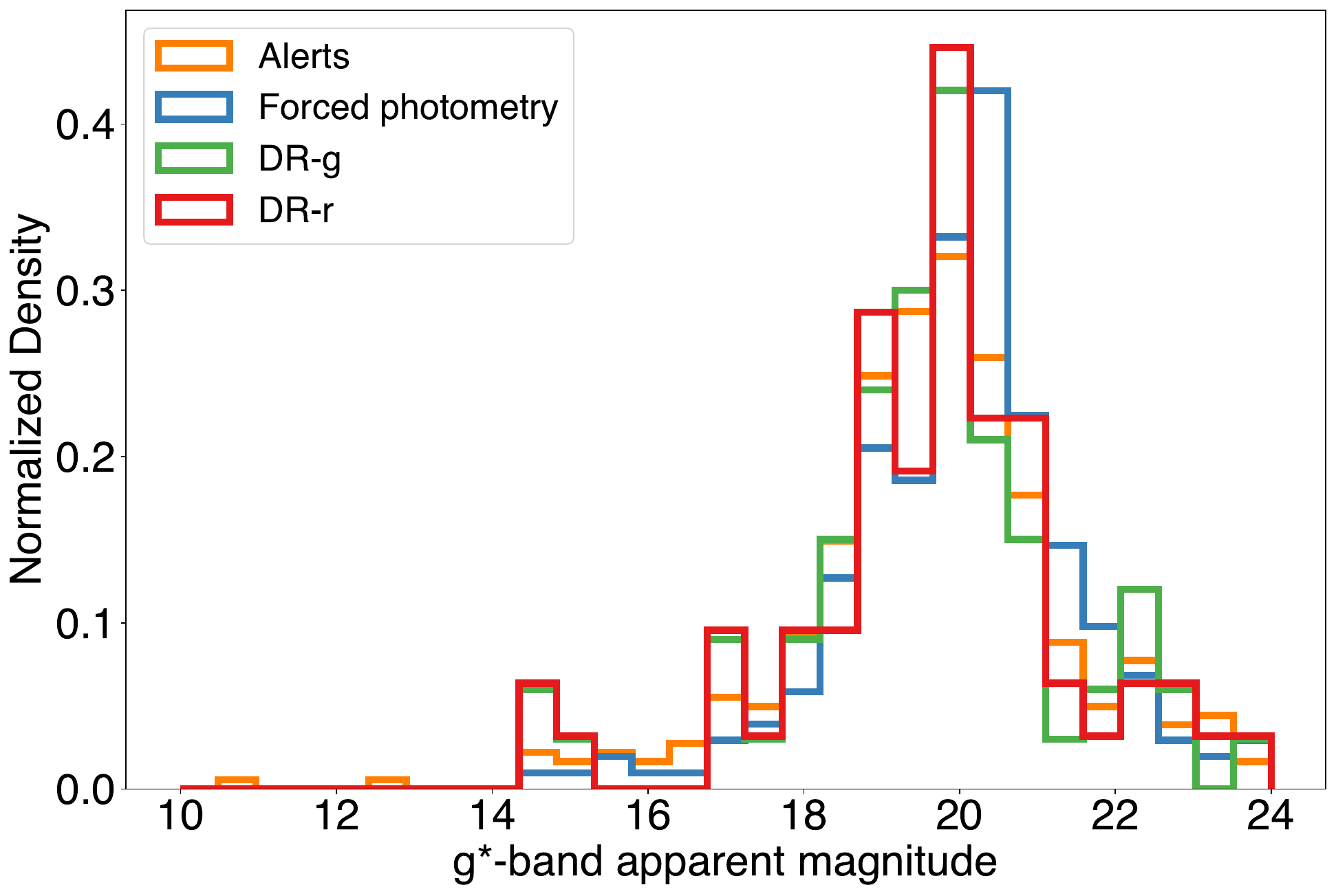}
    \\
    \includegraphics[width=0.48\textwidth]{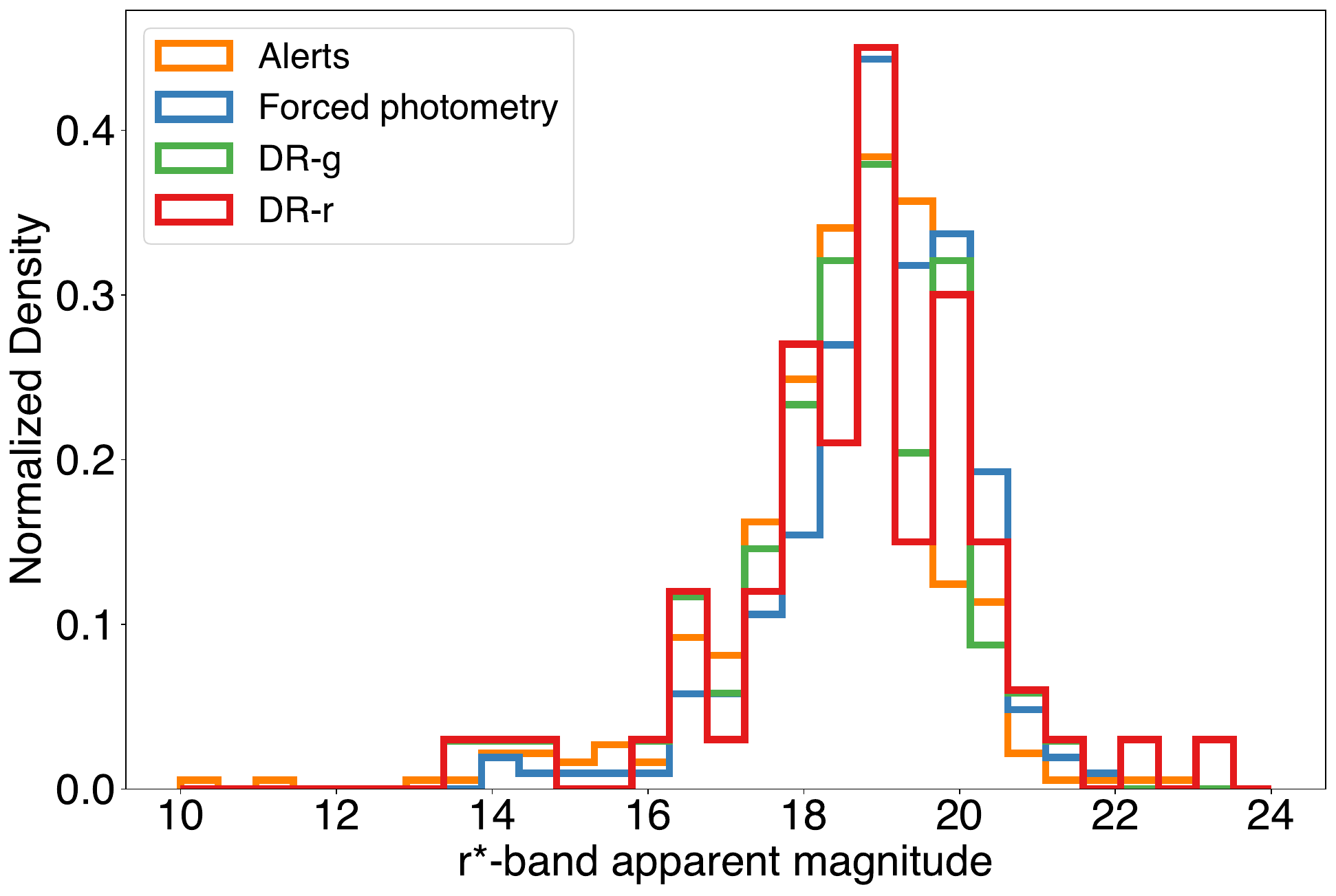}
    \\
    \includegraphics[width=0.48\textwidth]{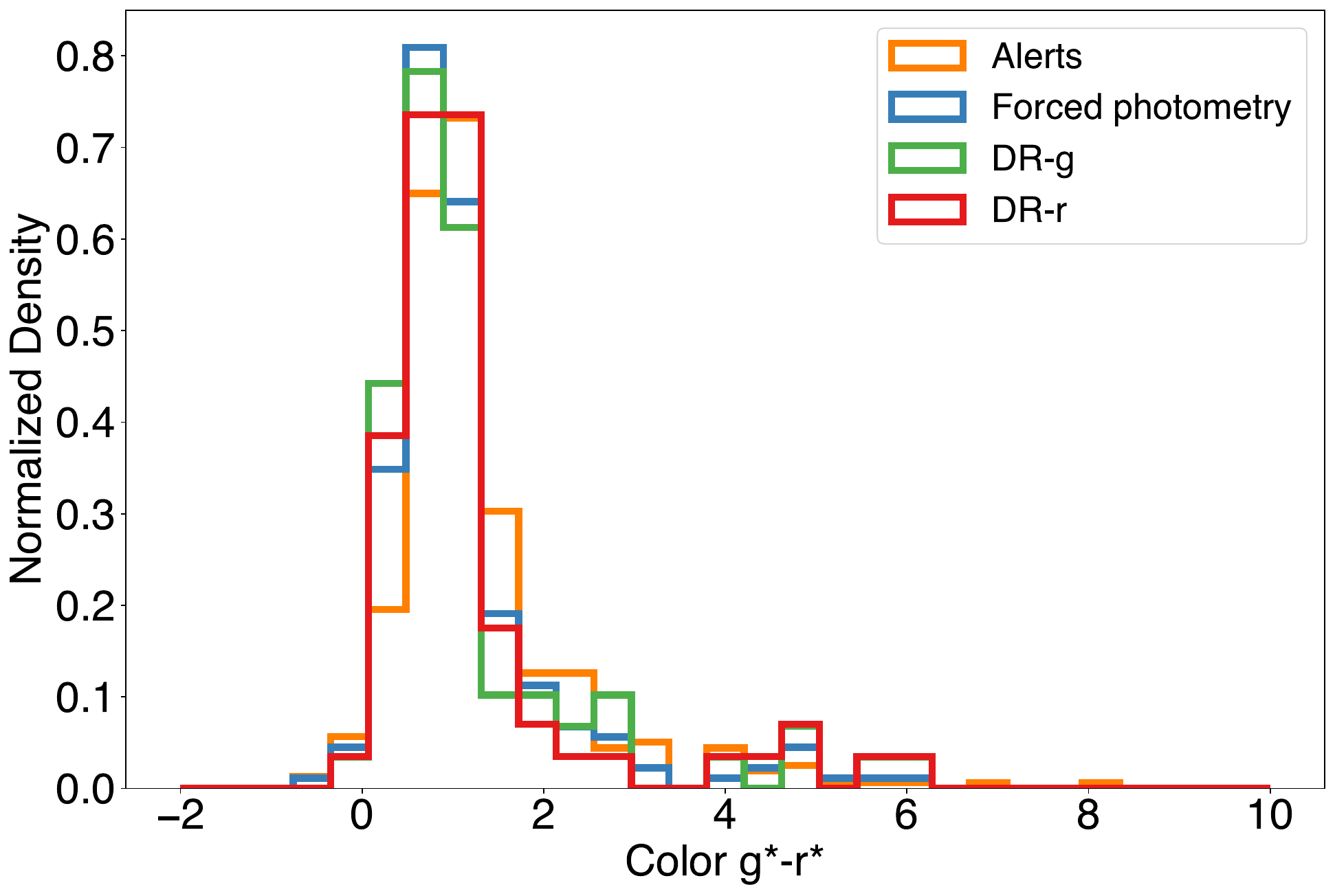}
    \caption{Apparent rest-frame magnitude distribution of AGN candidates in low stellar-mass galaxies for the $g^{*}$ (top) and $r^{*}$ (middle) bands. The bottom panel shows the color $g^{*}-r^{*}$ distributions. Each set of variability-selected AGN candidates is plotted in different colors, orange for Alerts, blue for Forced Photometry, green for ZTFDR11 $g$-band, and red for ZTFDR11 $r$-band.}
    \label{fig:appmag_candidates}
\end{figure}

Figure \ref{fig:appmag_candidates} shows the normalized distribution of the apparent rest-frame $g^{*}$ and $r^{*}$ magnitudes using the NSA v1.0.1 Sersic absolute magnitude (these magnitudes are computed using the flux from a Sersic profile modelling in each SDSS bands; we use the $^{*}$ symbol to note the rest-frame bands)  and redshift. The median $g*$-band AB magnitudes for the Alerts, Forced photometry, DR-g and DR-r sets are $19.73$, $20.15$, $19.72$, and $19.86$, respectively, while for the $r^{*}$-band they are $18.67$,  $19.09$, $18.71$, and $18.82$ for the same sets. The Forced photometry set contains objects with higher median $g^{*}$-band and $r^{*}$-band magnitudes than the other sets. The color $g^{*}-r^{*}$ distributions are also shown in Fig.~\ref{fig:appmag_candidates}, with median ABmag colors of $1.02$, $0.88$, $0.82$, $0.96$ for the Alerts, Forced Photometry, DR-g, and DR-r sets.

\section{Selection and analysis of spectroscopic data}\label{Sec:Spectro data and alysis}

\subsection{Selection of archival SDSS spectra}\label{Sec:Validation selection}

Spectroscopic characteristics in AGN candidates, like broad emission lines (BELs), are strong confirmation of nuclear activity. As a consequence, AGN candidates selected by different approaches are normally confirmed from their optical spectra (e.g.: \cite{Sanchez-Saez19}, \cite{2024AJ....167..169H}). Here, we searched for optical spectra of our variability selected sets using the tools available in the SDSS web-page\footnote{https://skyserver.sdss.org/} using the ZTF RA and Dec coordinates from the Data Release 17 catalog \citep[DR17;][]{2022ApJS..259...35AsdssDR17} with a matching radius of 1.5 arcsec. From the total sample, we found 454 objects with SDSS spectra, of which 450 were of good quality (i.e., SN-median-all > 2). In Sec.\ref{sec:bias} we show the main differences between the samples with and without spectra.

The NSA parent sample used spectroscopic redshifts from SDSS, NASA Extragalactic Database, the Six-degree Field Galaxy Redshift Survey, the Two-degree Field Galaxy Redshift Survey, and ZCAT (the CfA Redshift Survey), to limit the study to the local universe. However, we found that 16 of our matched AGN candidates have a higher spectroscopic redshift according to the SDSS pipeline and subsequent visual verification, in cases where the redshifts listed in the NSA catalog are from a source different of SDSS. Fifteen of these 16 objects are classified as QSO by SDSS spectral classification. The SDSS classification uses a least-squares minimization performed by the comparison of each spectrum to a full range of templates spanning galaxies, quasars, and stars, all in a range of redshifts (See \citealt{2012AJ....144..144B} for full description). These objects are listed in Table \ref{tab:wrong_z} in Appendix \ref{ap:redshifts} and have spectroscopic redshifts ranging from 0.225 to 2.427, whereas the NSA catalog redshifts were all below 0.15. 

Additionally, visual inspection showed that a further seven candidates present BELs consistent with an AGN, but lie at higher redshifts than the value reported by the SDSS pipeline. All these higher redshift AGN were classified as such in the SDSS quasar catalog with visual inspections produced by \citet{Lyke2020} and are listed in Table \ref{tab:wrong_z2} of Appendix \ref{ap:redshifts}, showing spectroscopic redshifts in the range $z=0.47942$ to $3.650$. In all these cases, although the AGN classification was correct, the stellar mass estimate is most likely wrong. %Therefore, these objects were removed from the optical spectral analysis below. 
Visual inspection of the remaining spectra revealed that six appear like galaxies without emission lines, three have blue, featureless continua consistent with either Blazars or stars, and two were not readily identifiable as AGN and could correspond to Broad Absorption Line (BAL) quasars. The spectra of these last five objects are shown in Appendix \ref{ap: suspected AGN}. Finally, one spectrum shows no data in the H$_\alpha$ region, flat, featureless continuum, and bad pixels in the H$_\beta$ region, preventing us from obtaining a useful fit. All the aforementioned 35 candidates with spectra were excluded from further consideration.  

Whit this reduction, we are left with 415 AGN candidates at low redshift ($z<0.15$) with good spectra, we will refer to these objects as the visually cleaned spectra (VCS) sample. For this sample, spectral fitting was performed following the methodology described below in Sec. \ref{sec:spectral-fitting}. 

\subsection{Assessment of biases in the spectroscopic sample}
\label{sec:bias}
\begin{figure}[htpb]
    \centering
    \includegraphics[width=0.48\textwidth]{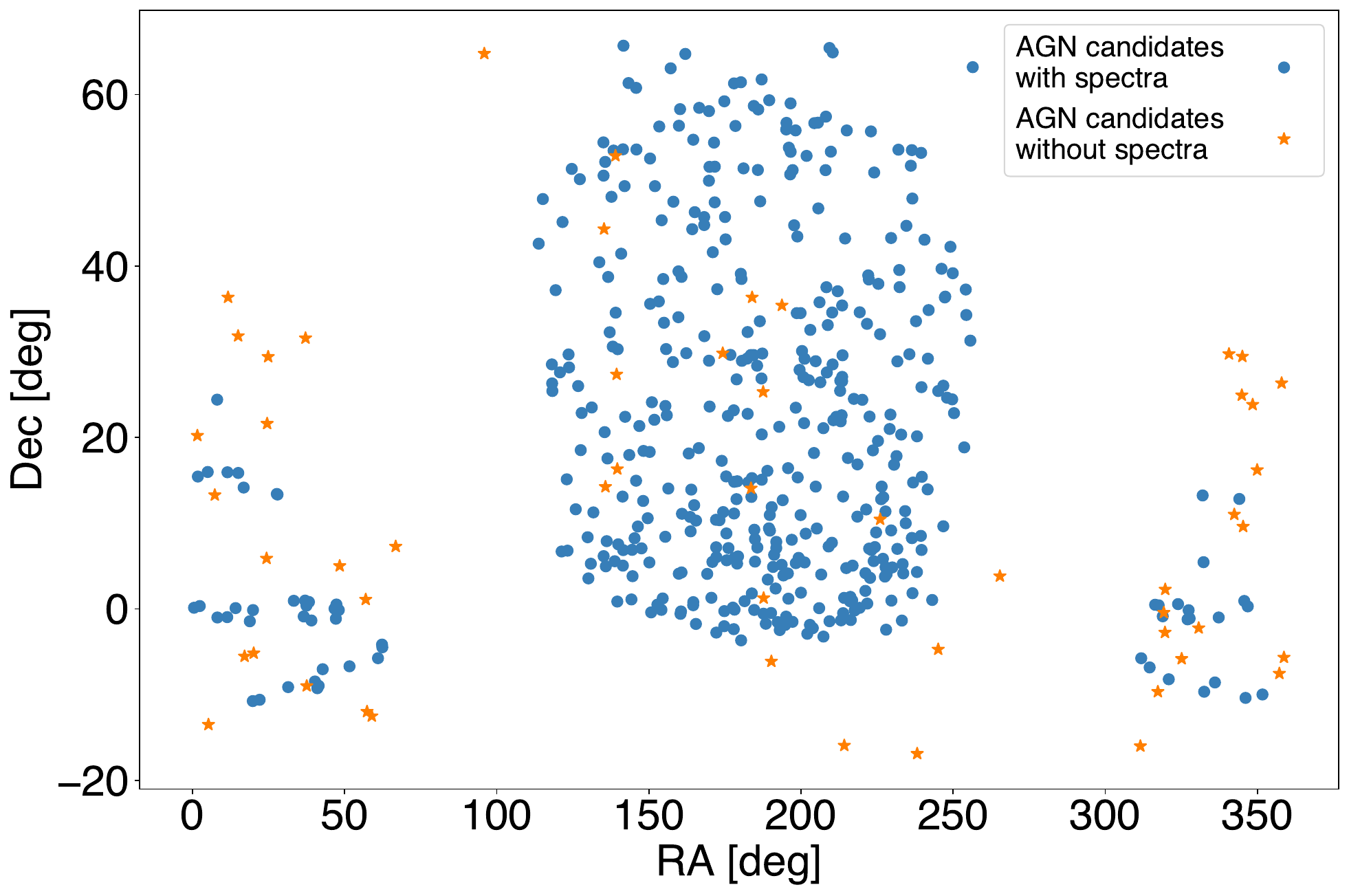}
      \includegraphics[width=0.48\textwidth]{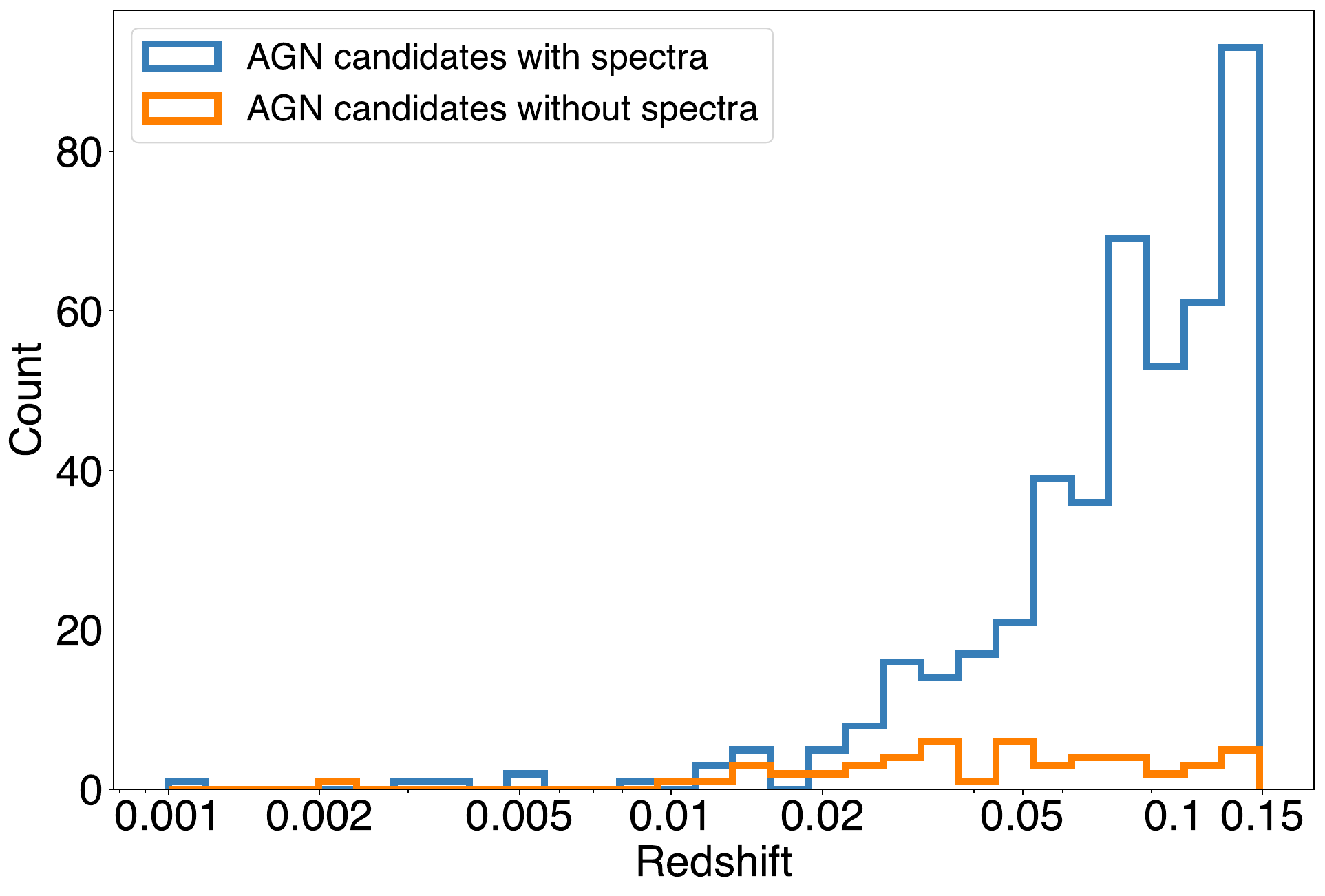}
    
    \caption{Comparison of AGN candidates with SDSS spectra (blue) and without spectra (orange). The top panel shows the distribution of these samples in the sky and the bottom panel shows the distribution of their redshifts from the NSA catalog.}
    \label{fig:sky_w_wo_spectra}
\end{figure}

Here, we investigate how the missing candidates differ from those with spectra. We note that of the 506 AGN candidates selected by variability features, only 454 have archival spectra of any quality from SDSS.   The top panel in Fig.~\ref{fig:sky_w_wo_spectra} shows the distribution in the sky of all candidates with and without available SDSS spectra. Some candidates fall outside the main areas of SDSS spectroscopic surveys, in particular targets below $\text{Dec}{=}-10$. This explains why the DR-g and DR-r sets have lower fractions of candidates with spectra than the other two sets, since they are restricted to regions of the sky with lower declination and therefore also lower SDSS coverage (see Fig.~\ref{fig:candidates}).  The bottom panel in Fig.~\ref{fig:sky_w_wo_spectra} shows the distribution of redshifts taken from the NSA catalog, for the candidates with and without spectra, which is the quantity where the samples differ more strongly. Evidently, a larger fraction of the lowest redshift (and also lowest-mass) galaxies do not have an archival spectrum. A consequence of this bias in the availability of spectra is that some of the lowest-mass black holes will not be included in the sample with mass measurements, which limits the ability of our method to confirm low-mass black holes. We stress, however, that the missing spectra only represent about 10\% of the AGN candidate sample and hence this limitation does not strongly bias the mass distribution of confirmed AGN candidates and other properties reported in the next sections. In any case, the low-mass AGN candidates without spectra should be confirmed or rejected with future spectroscopic campaigns to refine the selection process.     

\subsection{Spectral fitting}\label{sec:spectral-fitting}

 For the spectral fitting, we used the pPXF software (\citealt{pPXF2004PASP..116..138C}, \citealt{pPXF2017MNRAS.466..798C}) with a model that includes the stellar populations from the E-MILES library \citep{EMILES2016MNRAS.463.3409V}, a set of Gaussian profiles for the NELs with a wavelength shorter than $6300\AA$ and another set for NELs with wavelength longer than $6300\AA$, a set of Gaussians with four Gauss-Hermite moments for the BELs, templates for the Balmer higher order emission and Balmer continuum, templates for the FeII pseudo-continuum, and a set of power laws for the accretion disc continuum emission. The power laws are defined as $f_\lambda=(\frac{\lambda}{N})^{\alpha}$, with $\lambda$ being the wavelength, $N$ the normalization factor, and $\alpha$ the slope of the power law, which can take values in the range $-3\leq \alpha \leq 0 $. We will refer to this combination of components as Model 1. Each component has its own radial velocity and velocity dispersion. For the set of NELs with wavelengths shorter and longer than $6300\AA$, the kinematic moments were not tied to allow for deviations in wavelength calibrations. However, the differences between them, as determined from the best-fit model, are small and within the range of the spectral resolution, except for a few specific objects as discussed below. We highlight this here but did not conduct further analysis.
 
 We note that 46 objects show more complex spectra profiles for the broad H$_\alpha$ and H$_\beta$, and/or offset wings on the [O{\sc iii}]5007 emission lines. For this reason, a second model was fitted (Model 2) for 39 of the 46 objects. This model adds to Model 1 an extra Gaussian profile for both broad H$_\alpha$ and H$_\beta$ emission lines, as well as an additional Gaussian profile for the [O{\sc iii}]5007 doublet, to model possible winds. For the remaining seven objects, Model 2 was used with an adjustment: the H$_\beta$ narrow emission line Gaussian had its radial velocity and velocity dispersion allowed to vary independently, leading to an improved fit. This modified approach is referred to as Model 3.
 
 We also note that the method implemented by pPXF is sensitive to the initial values of velocity and velocity dispersion. In some cases, the best-fit was only achieved when one or both of these parameters were fixed. For 53 objects, when all parameters were left free, visual inspection revealed unsatisfactory fits, particularly in the velocity shifts of the narrow line templates. As a result, for these 53 cases, the most problematic parameter was manually adjusted and fixed for subsequent fitting. The fixed parameters were either the line-of-sight velocity (nV), the velocity dispersion (nVd), or both, of the narrow line templates only. The broad line kinematic parameters and the flux normalization for all templates were always free to vary. The specific model used, along with any fixed parameters (i.e., nV and nVd), are detailed for each object in the complete online version of Table \ref{tab:results_short}.

 In order to account for the errors on the fitted parameters, we followed the advice in \citet{pPXF2004PASP..116..138C} and performed Monte Carlo simulations. We created simulated spectra by combining the best-fit model for each object with different realizations of the observational noise.  Below we describe the process we followed. First, we calculated the standard deviation of the residuals from the best-fitting model. In the next step, we used the standard deviation of the residuals to scale a Gaussian deviation to produce random values for each wavelength bin, thereby generating the simulated noise. This provides an empirical measure of the error in the flux. Finally, we combined the best-fit model with the simulated noise and then used the same models , and the same fixed parameters if needed, as explained above to fit this simulated spectra, with the difference that in this case, we included only the stellar templates selected in the best-fitting model. This approach assumes that the stellar template is well-determined in the best-fit model of the observed spectrum. The highest uncertainty is assumed to come from the normalization of the stellar templates in the final fit, and this normalization is always free during the fitting process. This procedure is repeated 100 times for each spectrum. Finally, we computed the 16th and 84th percentiles of the fitted parameters to report the lower and upper estimations of each parameter.
 Note that use the residuals to generate the simulated noise instead of errors provided by SDSS, which allows us to test the robustness of the fit by evaluating not only the impact of systematic errors but also how specific deviations between the best-fit model and the observed spectrum influence the final results. As the average deviations are slightly larger than the SDSS errors, the use of residuals is a more conservative approach for the estimation of errors on the fitted parameters.

\subsection{Estimating black hole masses and Eddington ratios}\label{sec:mass_REdd_eq}

Estimates of the black hole masses were made using the relation obtained by \citet{MejiaRestrepo16}: 
\begin{equation}
M_{BH}=K(L_{\lambda})^{\alpha}FWHM^2,
\label{eq:M}
\end{equation}

\noindent where the values of $K$ and $\alpha$ depend on combinations of the monochromatic luminosity of the AGN continuum $L_{\lambda}$ and Full-Width-Half-Maximum ($FWHM$) of the BEL used (see Table 7 of \citealt{MejiaRestrepo16}). For $L_{\lambda}=L_{5100}$ and the $FWHM$ of H$_\beta$, the values are $K=10^{6.864}$ and $\alpha=0.568$. For $L_{\lambda}=L_{5100}$ and the $FWHM$ of H$_\alpha$, the values are $K=10^{6.958}$ and
$\alpha=0.569$. The monochromatic luminosity is defined as $L_{\lambda}=4\pi r^2 F_{\lambda}$, where $F_{\lambda}$ is the monochromatic flux at $\lambda$, and $r=z\times c/H_0$, with $z$ being the redshift, $c$ the speed of light, and $H_0=70 kms^{-1}Mpc^{-1}$. 
For the Eddington ratio, we used the relation:

\begin{equation}
L_{REdd}=C_{Bol}\frac{L_{5100}}{1.5\times10^{38}(M_{BH}/M_{\odot})},
\label{eq:LREdd}
\end{equation}

\noindent with $C_{Bol}=9.26$ (see \citealt{2008ApJ...680..169S}) and $M_{BH}$ being the mass calculated with the previous equation. To estimate the lower limits in Mass and Eddington ratio, we calculated these values for each Monte Carlo simulated spectra. Then, we obtain the 16th  and 84th percentiles, for the lower and upper limits of the mass and Eddington ratio. To report the errors for black hole mass and Eddington ratio we add in quadrature the uncertainty associated with the equations \ref{eq:M} and \ref{eq:LREdd}. For the mass the uncertainty is 0.19 dex \citep{MejiaRestrepo16} and for the $C_{Bol}$ value 0.1 dex \citep{2008ApJ...680..169S}.

\section{Results}\label{sec: results}
\subsection{Confirmation of type I AGN}

\begin{figure}[htpb]
    \centering
    \includegraphics[width=0.48\textwidth]{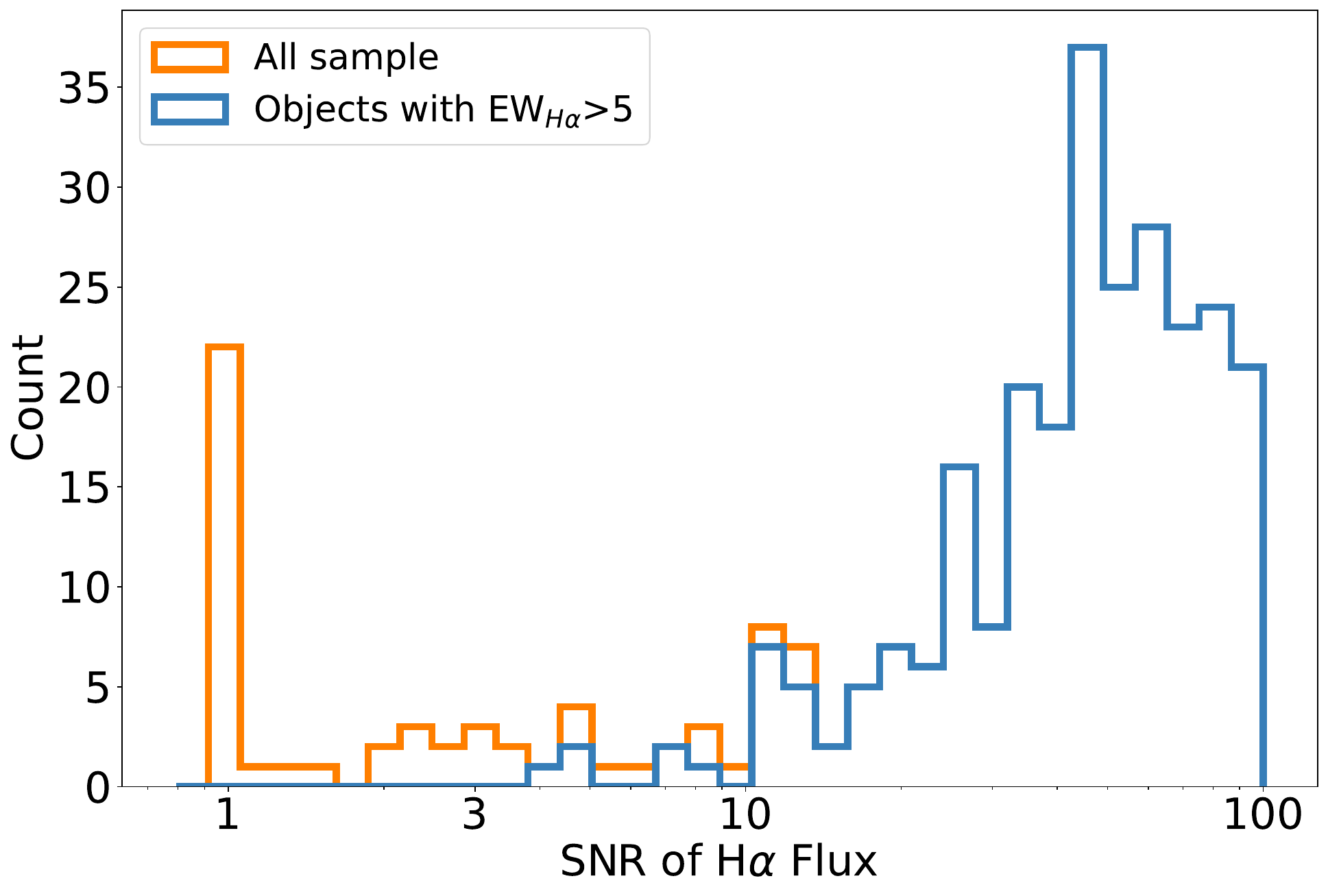}
    \caption{Distribution of the SNR of the flux of the broad component of H$_\alpha$. All the variability-selected AGN candidates in low mass galaxies are plotted in orange and, of these, only the ones with  $EW_{\rm H\alpha} > 5\AA$ in blue.  }
    \label{fig:Ha_flux_SNR}
\end{figure}

Type I AGNs are characterized by the presence of BELs in their spectra. We used the criteria of the equivalent width of the broad H$_\alpha$ emission line to be $EW_{\rm H\alpha} > 5\AA$ and the signal-to-noise ratio (SNR) of broad H$_\alpha$ flux to be larger than three to consider a BEL detection on the fitting of the spectra. {The SNR is defined using the simulation results as the ratio between the 50th percentile ($p50$) and the difference between the 50th percentile and the 16th percentile ($p16$), expressed as SNR = $\frac{p50}{p50 - p16}$.} Additionally, for objects at the low end of the FWHM (FWHM=940km/s) and high end (FWHM>9400km/s), we visually inspect the reliability of the measured BEL. As a result, of the 415 AGN candidates in the VCS sample, the fits returned 355 objects that met the $EW_{\rm H\alpha}$ and broad H$_\alpha$ flux criteria, of which 323 also had  $EW_{\rm H\beta}>5$ \AA. Fig.~\ref{fig:Ha_flux_SNR} shows the distribution of the SNR of H$_\alpha$ flux for all the candidates and for those with  $EW_{\rm H\alpha} > 5\AA$.
We additionally include two objects, where the spectra show no data in the H$_\alpha$ range, but show  $EW_{\rm H\beta}>5\AA$ and SNR of broad H$_\beta$ flux larger than 3. Therefore, the fits returned 357 objects that met the criteria for the presence of BELs, either in H$_\alpha$ or in H$_\beta$. 

Table \ref{tab:confirmation} shows the number of candidates in each set that had available good spectra and the number of these spectra classified as AGN, either by the presence of BELs described here for the LSMG in column LSMG-EW, or by the higher-redshift QSO classification criteria described in the previous section. We also include a "suspected AGN" category after a further visual inspection of the spectra that were not selected for fitting. These objects have mainly either a featureless blue continuum, which might correspond to a Blazar classification, or an uncommon spectrum that might correspond to broad absorption line quasars. The number of suspected AGNs objects are included in Table \ref{tab:confirmation}. The spectra of these objects are described in Appendix \ref{ap: suspected AGN} and shown in Fig.~\ref{fig:suspected-agn}. Given their uncertainty, we do not include them in the further analysis, and stress that their small numbers do not affect the percentage of confirmed candidates.

\begin{table*}[t]
    \centering
    \caption{Number of confirmed AGN}
    \begin{tabular}{l|c|cc|ccc|c}
        Set & candidates& \multicolumn{2}{c|}{spectra}& \multicolumn{3}{c|}{confirmed AGN} & suspected AGN \\
             &           & all & VCS      & LSMG-$EW$ & higher-z & Total & \\
         \hline
        Alerts & 383 & 336 & 314  & 258& 13 & 271 & 4\\ 
        
        Forced phot. &215 & 188 & 170 & 168 & 14 & 182 & 3\\
        
          DR-g&71 & 54 & 40 & 40 & 11 & 51 & 3\\ 
          
         DR-r &69 & 54 & 40 & 40 & 11 & 51 & 1 
         
    \end{tabular}
    \label{tab:confirmation}
        \begin{tablenotes}
        \item {\bf Note.} The column "candidates" refers to the total 506 variability-selected candidates, while column "spectra" refers to the candidates with good quality SDSS spectra in "all", and those with consistent redshift related with LSMGs in the "VCS". Confirmation is based on the detection of significant broad  H$_\alpha$ emission lines (column LSMG-EW) or other permitted lines for the objects at higher redshift (column higher-z). Suspected AGN are those with either featureless blue continua, consistent with Blazars but that might in fact be stars, and unusual spectra suggestive of BAL. Plots of suspected AGN spectra are presented in Appendix \ref{ap: suspected AGN}.
    \end{tablenotes}

\end{table*}

For 54 candidates in the VCS sample, a BEL was not detected in the spectra above  $EW_{\rm H\alpha} = 5\AA$, all of these arise from the Alerts set. In addition, four objects were classified as unreliable detections after visual inspection of the best-fit model, as the resulting $EW_{\rm H\alpha}>5$ was due to a large FWHM and relatively low peak flux, indicating that the feature was not a true BEL, two of them in the Alerts set and two in the Forced Photometry set. Furthermore, five objects in the Alerts set and two in the DR-r set, which have spectra but are not in the VCS sample (see Sec. \ref{Sec:Validation selection}), are also not confirmed or suspected AGN. In total, 65 objects, representing 12.8\% of the 506 variability-selected candidates, are not confirmed as AGNs according to their optical spectra.

We visually inspected the light curves and image stamps for the 65 candidates described above, using the ALeRCE ZTF explorer\footnote{\url{https://alerce.online}} with the options "Apparent Magnitude" and "Toggle DR" activated, which allows the user to visualize the alert light curve together with the ZTF DR 6 light curve. We found that 15 of the 65 corresponded to a bad subtraction in the ZTF difference images,\footnote{\url{https://alerce.online/object/ZTF18aaounyj}} 30 had a nuclear transient in the ZTF template,\footnote{\url{https://alerce.online/object/ZTF18aabxrxc}} two candidates corresponded to bright and seven to weak nuclear transients\footnote{\url{https://alerce.online/object/ZTF18aaqjyon}} including potential tidal disruption events (TDEs).\footnote{\url{https://alerce.online/object/ZTF18abtizze}} The last 11 candidates correspond to sources with an unclear nature, but whose alert light curves seem to show real stochastic variations.\footnote{\url{https://alerce.online/object/ZTF19aangwsm}} The footnotes provide one example of each mentioned case. We summarize all these cases in Table \ref{tab:class_objects_no_BELs}. The misclassifications due to bad subtractions or transients in the templates are hard to correct from the point of view of a light curve classifier. For the case of nuclear transients, a model that includes the class nuclear transient would solve the issue. Currently, the ALeRCE broker team is working on a new model that includes the class TDE, which will prevent issues like this in the future (F. F{\"o}rster, private communication).

\begin{table}
    \centering
    \caption{Not confirmed AGN candidates}
    \begin{tabular}{l|c }
    Case& Number of objects\\
\hline
Transient in template & 30\\ 
Transient & 2 \\ 
Weak transient & 7\\ 
Bad subtraction & 15\\ 
Other & 11\\ 
\hline
Total & 65  
         
    \end{tabular}
    \label{tab:class_objects_no_BELs}
    \begin{tablenotes}
        \item {\bf Note.} Case names and number of selected candidates that were not confirmed from their optical spectra. The Transient cases correspond to nuclear variability. Bad subtraction cases are those with poor subtraction in the ZTF difference images. The Other cases denote light curves where the measured variability appears to be real, based on visual inspection. 
    \end{tablenotes}
\end{table}

On the other hand, considering the candidates in the VCS sample as reliable LSMGs, we identified 258 out of 314 objects in the Alerts set with significant BELs. In the Forced Photometry set, 168 out of 170 objects in the VCS exhibit the same characteristic, while all 40 objects in the DR-g and 40 in DR-r subsets display significant BELs. Additionally, high-redshift objects identified through their spectra are also consistent with AGN BELs, with 13 in the Alerts set, 14 in the Forced Photometry set, 11 in the DR-g set, and 1 in the DR-r set.

\begin{figure}[htpb]
    \centering
    \includegraphics[width=0.48\textwidth]{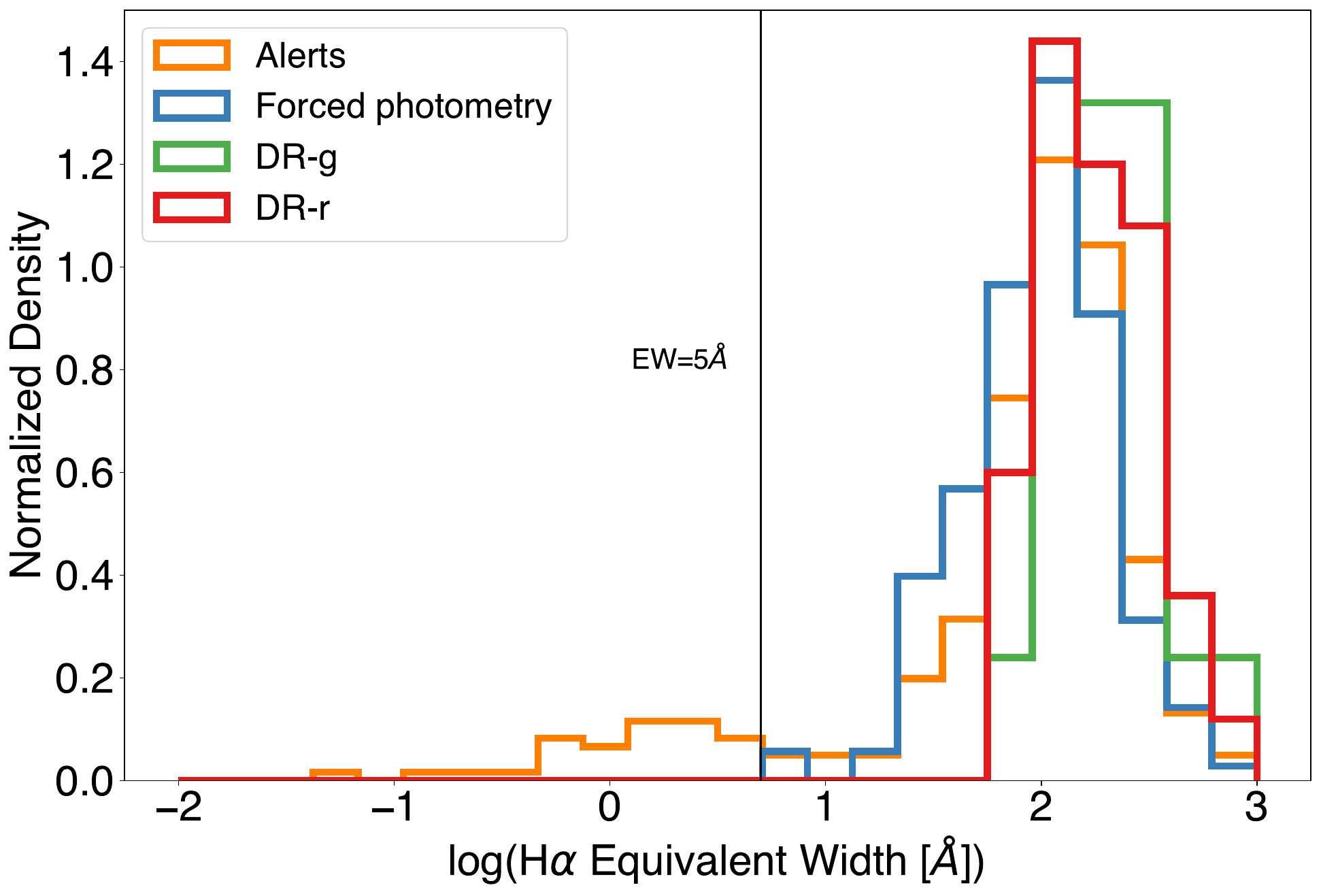}
    \\
    
    \includegraphics[width=0.48\textwidth]{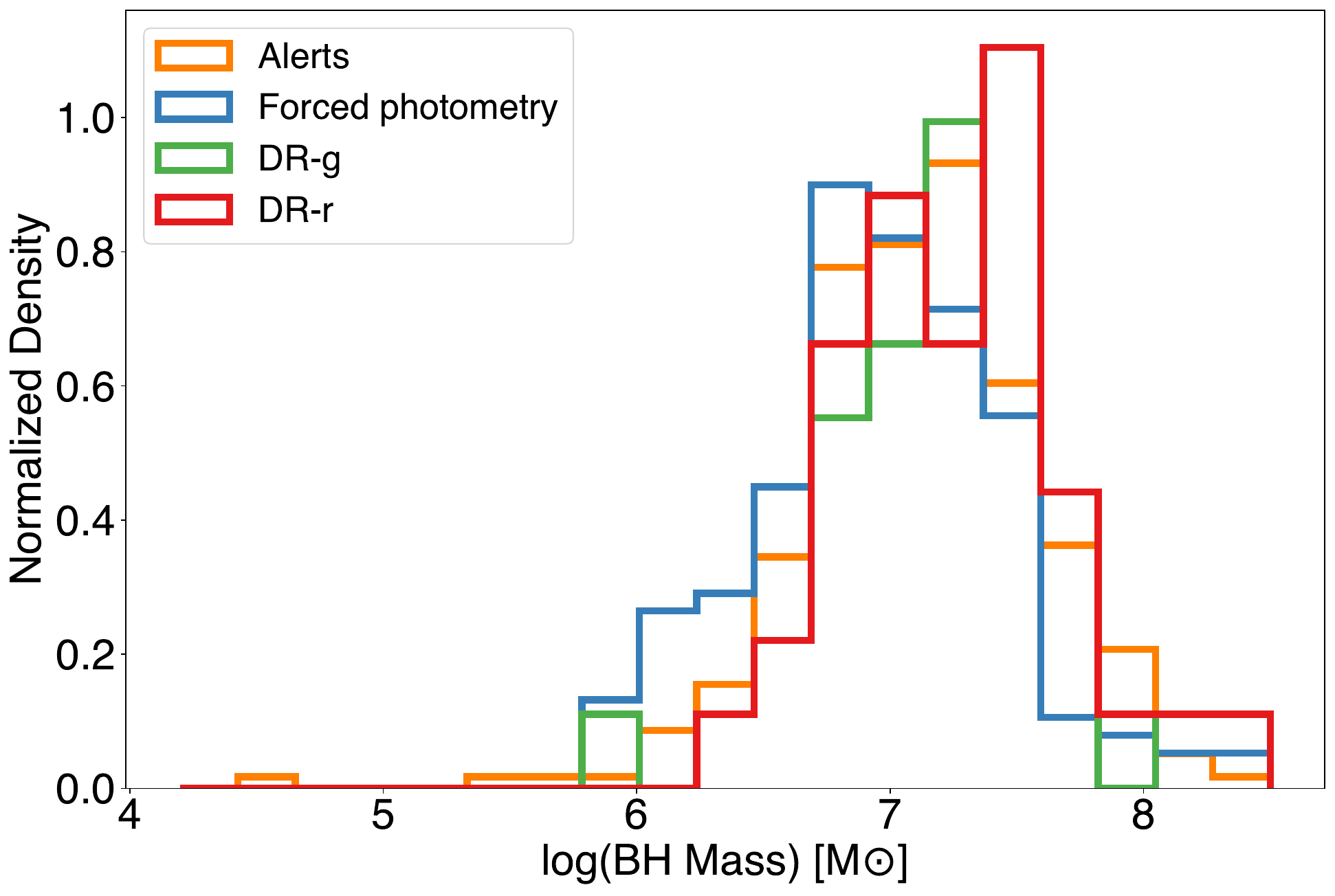}
    \caption{Distribution of the equivalent width of the broad component of H$_\alpha$ (top) and of black hole masses derived using Eq. \ref{eq:M} (bottom) for objects with  $EW_{\rm H\alpha}>5\AA$. In both histograms, each set of variability-selected AGN candidates is plotted in different colors, orange for Alerts, blue for Forced Photometry, green for ZTFDR11 $g$-band, and red for ZTFDR11 $r$-band. In the top panel, the vertical line indicates $EW_{\rm H\alpha}=5\AA$.}
    \label{fig:histogram}
\end{figure}

For all the fitted objects (objects in the VCS sample) the distribution of H$_\alpha$ equivalent widths is shown as a histogram in the top panel in Fig.\ref{fig:histogram}, where we have plotted each variability set separately, even though many objects belong to more than one set. The equivalent widths cluster around $100-200 \AA$, with the Alerts set peaking at a lower $EW$ than the DR-g and DR-r sets and the Forced Photometry set peaking at even lower values. This shift demonstrates the ability of the Forced Photometry selection (and to a lesser extent the Alerts selection) to detect weaker AGN, embedded in stellar continua than the data release light curves. Only the Alerts set shows a tail toward very low equivalent widths, below our threshold of $5\AA$. As described above, these AGN candidates are probably produced by spurious variations such as bad image subtractions and transient events appearing in the reference images, which affect the alerts-based classification much more strongly than those based on full light curves.

\subsection{Black hole masses and Eddington ratio} \label{sec: mass REdd results}
We estimated black hole masses for all objects with sufficiently significant broad Balmer lines, selecting only galaxies where the  $EW_{\rm H\alpha} > 5 \AA$ and the SNR of broad H$_\alpha$ flux is larger than 3. The fitted widths, equivalent widths, fluxes, and mass estimates of this sample are presented in Table~\ref{tab:results_short}.

\begin{table*}[t]
\small
    \centering
    \caption{Values of the different estimated magnitudes of each object. The checkmarks indicate in which sets the object was included (short version).}
    \begin{tabular}{lccccccccc}
\toprule
        IAUNAME &         FWHM &       $EW_{\rm H\alpha}$ &      Flux H$_\alpha$ &      BH Mass &    LREdd &       Alerts & Forced  & ZTFDR11& ZTFDR11\\
            &($km\ s^{-1}$)&(\AA) &($erg\ cm^{-2}\ s^{-1}$)& ($Log[M_{\odot}$])&
            ($erg\ s^{-1}$)& & photometry&$g$-band & $r$-band\\
\midrule

J092547.31+050231.6 &   940.0 &  111.2 &    555.0 &      6.3 &  0.259 &          X &        \checkmark &             X &             X \\
J003238.20-010035.2 &  2110.0 &   61.6 &   1351.0 &      7.1 &  0.065 & \checkmark &        \checkmark &    \checkmark &    \checkmark \\
J090148.19+203632.0 &  2682.0 &   94.6 &   1323.0 &      7.1 &  0.029 & \checkmark &                 X &             X &             X \\
J121754.80+583926.9 &  2268.0 &   80.1 &  10414.0 &      6.9 &  0.034 & \checkmark &                 X &             X &             X \\
J142138.63+173536.2 &  5362.0 &   39.0 &   1208.0 &      7.5 &  0.004 & \checkmark &                 X &             X &             X \\
J162445.70+393949.6 &  4426.0 &   53.1 &   2570.0 &      7.7 &  0.014 & \checkmark &                 X &             X &             X \\
J024052.19-082827.4 &   940.0 &  211.4 &    538.0 &      5.9 &  0.126 &          X &        \checkmark &    \checkmark &             X \\
J122548.86+333248.7 &  1403.0 &  166.5 &   6616.0 &      4.6 &  0.003 & \checkmark &                 X &             X &             X \\
J135510.15+330556.8 &  1745.0 &   98.8 &    817.0 &      7.0 &  0.103 & \checkmark &                 X &             X &             X \\
J152534.58+174900.0 &  1548.0 &  350.3 &   5114.0 &      7.1 &  0.172 & \checkmark &                 X &             X &             X \\
J123748.50+092323.1 &  1597.0 &   56.6 &   1438.0 &      7.1 &  0.155 & \checkmark &        \checkmark &             X &             X \\
J135321.45+373053.9 &  5497.0 &  152.4 &   3040.0 &      8.0 &  0.010 & \checkmark &                 X &             X &             X \\
J155229.98+200721.0 &  3044.0 &  152.9 &   3631.0 &      7.4 &  0.029 & \checkmark &                 X &             X &             X \\
J081259.96+064711.0 &  2261.0 &   50.4 &    488.0 &      6.9 &  0.033 &          X &        \checkmark &             X &             X \\
J230427.08-102318.5 &  1683.0 &   59.8 &    608.0 &      6.8 &  0.079 &          X &        \checkmark &             X &             X \\
J144034.34+242250.3 &  6980.0 &  136.1 &   4710.0 &      7.9 &  0.004 & \checkmark &                 X &             X &             X \\
J140130.21+283026.7 & 11750.0 &    6.3 &    142.0 &      8.1 &  0.001 & \checkmark &                 X &             X &             X \\
J124726.37+070525.0 &  1866.0 &  239.0 &   7336.0 &      7.2 &  0.118 & \checkmark &        \checkmark &    \checkmark &    \checkmark \\
    \end{tabular}
    
    \label{tab:results_short}

    \begin{tablenotes}
        \item {\bf Note.} The complete table is available at CDS.
    \end{tablenotes}
\end{table*}

The bottom panel in Fig.~\ref{fig:histogram} shows the distribution of the black hole masses derived using Eq.\ref{eq:M} for all cases where  $EW_{\rm H\alpha} > 5\AA$, except for the object J102530.29+140207.3. This object shows an absorbed (red) spectrum profile so the fitted model only used the stellar populations to fit the continuum and it is therefore not possible to estimate $L_{5100}$ for the AGN continuum. Similarly to the top panel in this figure, the distribution is shown independently for each variability set. The mean black hole mass for the Forced Photometry set is $1.7\times 10^7M_\odot$, for the Alerts set is $2.2\times 10^7M_\odot$, for DR-g is $2.7\times 10^7M_\odot$, and for DR-r is $2.8\times 10^7M_\odot$. Almost all black hole masses below $3\times 10^6M_\odot$ are only found in the Alerts and Forced Photometry sets. To estimate the significance of the differences in the black hole mass  distributions of the four sets we performed an Anderson-Darling test.\footnote{\href{https://docs.scipy.org/doc/scipy/reference/generated/scipy.stats.anderson_ksamp.html}{The Anderson-Darling test for k-samples}} Based on the p-values obtained, the DR-r, DR-g, and Alerts sets are all consistent with each other (p-value $\ge$ 0.17--0.25). The Forced Photometry set, however, has a different distribution  compared to any of the other sets (p-value < 0.001). Table \ref{tab:mass_quantiles} summarizes the distribution of black hole masses obtained using Eq. \ref{eq:M} for each set of AGN candidates. The lowest black hole mass obtained corresponds to the well-known low mass AGN NGC\,4395 included in the Alerts set, and our result of $M_{BH}=3.9^{+0.9}_{-1.2} \times 10^4 M_{\odot}$ is in agreement with the reverberation mapping mass $M_{BH} = (4.9\pm 2.6) \times 10^4M_{\odot}$ estimated by \citet{2012ApJ...756...73EmassNGC}. We note that the masses derived using the FWHM of H$_\beta$ in Eq. \ref{eq:M} are systematically offset and 30\% lower compared to those derived using the FWHM of H$_\alpha$. Since they are measured using the same continuum flux and the fitted  width of H$_\alpha$ and H$_\beta$ are the same, the difference is only set by the parameters used in Eq. \ref{eq:M}.

\begin{table}
    \centering
    \caption{Quantiles of the distribution of black hole masses in units of $M_\odot$.}
    \begin{tabular}{l|c c c }
    &\multicolumn{3}{c}{Quantile}\\
        Set & 10 &50&90 \\
\hline
Alerts & 3.9E6& 1.3E7& 5.1E7\\
Forced phot.& 1.8E6& 9.1E6& 3.1E7\\
DR-g& 6.3E6 &1.8E7 & 4.7E7\\
DR-r& 6.4E6& 1.6E7& 5.2E7\\
         
    \end{tabular}
    \label{tab:mass_quantiles}
\end{table}

\begin{figure}[htpb]
    \centering
    \includegraphics[width=0.48\textwidth]{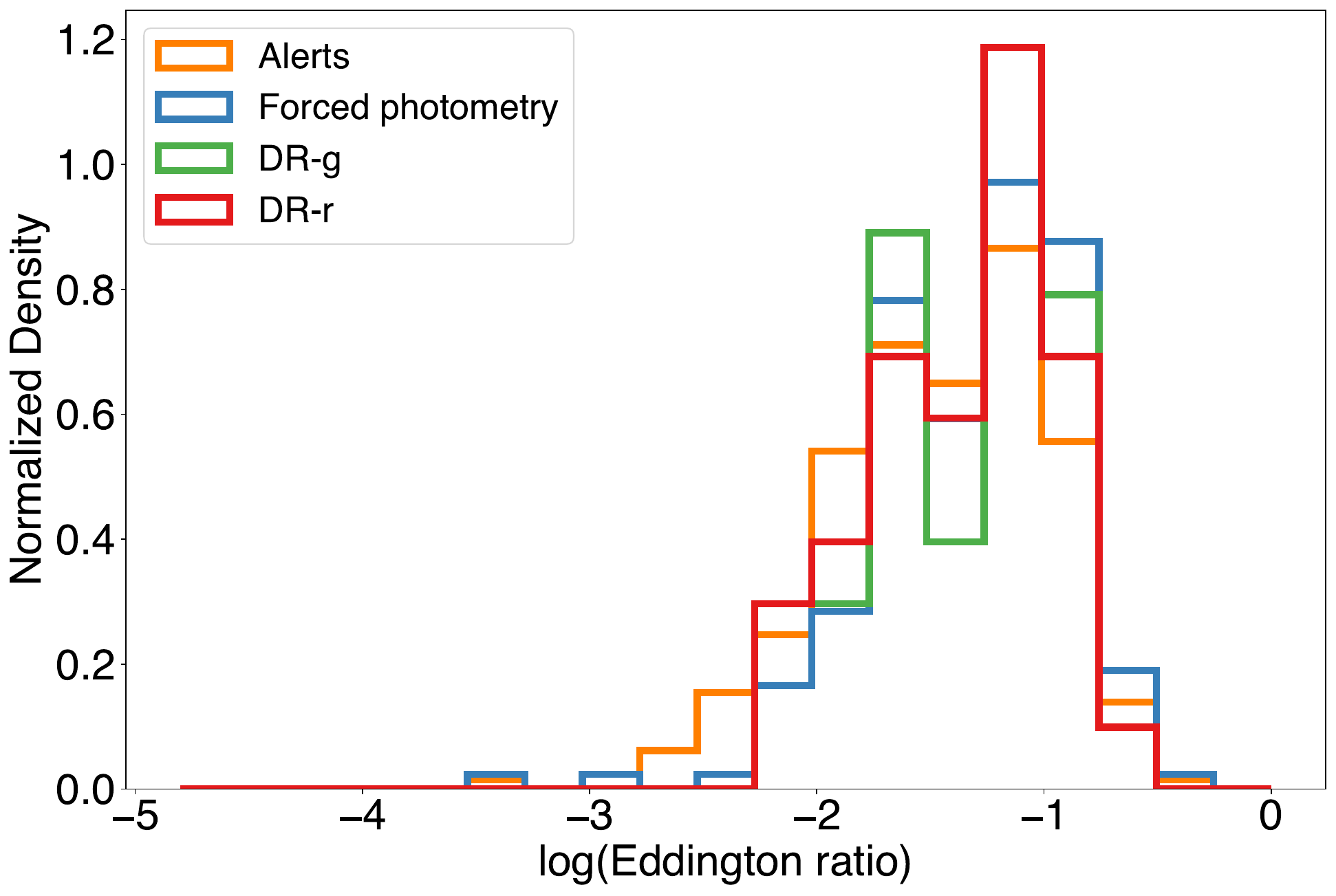}
    \caption{Distribution of the Eddington ratios derived for each set of AGN candidates. Each set is plotted in different colors, orange for Alerts, blue for Forced Photometry, green for ZTFDR11 $g$-band, and red for ZTFDR11 $r$-band.}
    \label{fig:histogram_REdd}
\end{figure}

We estimated the luminosity at $5100\AA$ from the measured flux from the best-fit power law continuum component in $erg\ s^{-1}$, and a bolometric correction factor of $C_{Bol}=9.26$ as mentioned in Sec. \ref{sec:mass_REdd_eq}. Combining these luminosities with the previously estimated mass, we calculated the Eddington ratio R$_{\rm Edd}$, and show them in the same table (Table \ref{tab:results_short}). The distribution of the Eddington ratio is shown in Fig.\ref{fig:histogram_REdd}. We notice that, unlike the mass distribution, the Eddington ratio for all sets is similarly distributed. The mean Eddington ratio in each set are $7.1\times 10^{-2}$ for the Forced Photometry, $5.7\times 10^{-2}$ for the Alerts, $6.4\times 10^{-2}$ for DR-g and $6.5\times 10^{-2}$ for DR-r. Performing an Anderson-Darling test to estimate the difference between the Eddington ratio distributions, we find a p-value of 0.23 for a simultaneous comparison of the four sets, confirming their consistency.

\subsection{Galaxy classification by narrow emission-line ratios}
\label{sec:BPT}
The spectral fitting described above also produced fluxes for the narrow permitted and forbidden emission lines. The relative fluxes of these lines can differentiate between different excitation mechanisms, such as star formation (SF), Seyfert activity, and LINERs, ( i.e., Baldwin, Phillips \& Terlevich (BPT) diagnostics, \citealt{BPT_Baldwin81}). We state whether each object falls in the SF, Seyfert, or LINER portions of each BPT diagram, following the relations in  \citet{Kewley2001ApJ...556..121K}, \citet{Kauffmann2003MNRAS.346.1055K}, \citet{Kewley2006MNRAS.372..961K} and \citet{Schawinski2007MNRAS.382.1415S}. 

Figure \ref{fig:BPT diagrams} shows the BPT diagrams for all the objects where the emission lines were measured. The color bar indicates the black hole mass on a logarithmic scale. Objects with no BH mass estimate, mainly because they lacked BELs, are plotted with black triangles. In the top left of each panel we show the average error bars of both line ratios. These are estimated from the percentiles of the distributions of {line ratios} for each object resulting from the Monte Carlo simulations described in Sec. \ref{sec:spectral-fitting}. In these diagrams objects with lower black hole mass are preferably found in the star-forming region but the distribution of masses largely overlaps. {For example, the mean, median, and standard deviation of the mass in logarithm scale and solar masses ($log(M/M_\odot)$) for objects in the star-forming class are 6.83, 6.90, and 0.49 respectively; for the Composite class are 7.01, 7.05, and 0.50; and for the Seyfert class are 7.08, 7.09, and 0.48.  
From the 415 objects in the VCS sample, 20 have no BPT classification on the [O{\sc iii}]/H$_\beta$ vs. [N{\sc ii}]/H$_\alpha$ diagram: for three of them, the spectrum has no data in the H$_\alpha$ region (note that these objects are different from the one excluded in Section \ref{Sec:Validation selection} because of no data in the H$_\alpha$ region, as we were able to fit the H$_\beta$ region for the three mentioned here, and in two cases, we found a BEL); in three cases the best-fit model does not fit a narrow H$_\beta$ line; and in 14 cases the best-fit model could not separate the overlapped [N{\sc ii}] and H$_\alpha$ narrow emission lines. All these 14 objects are classified as QSO by the SDSS pipeline and all have a significant broad H$_\alpha$ detection.

\begin{figure}[htpb]
    \centering
    \includegraphics[width=0.49\textwidth]{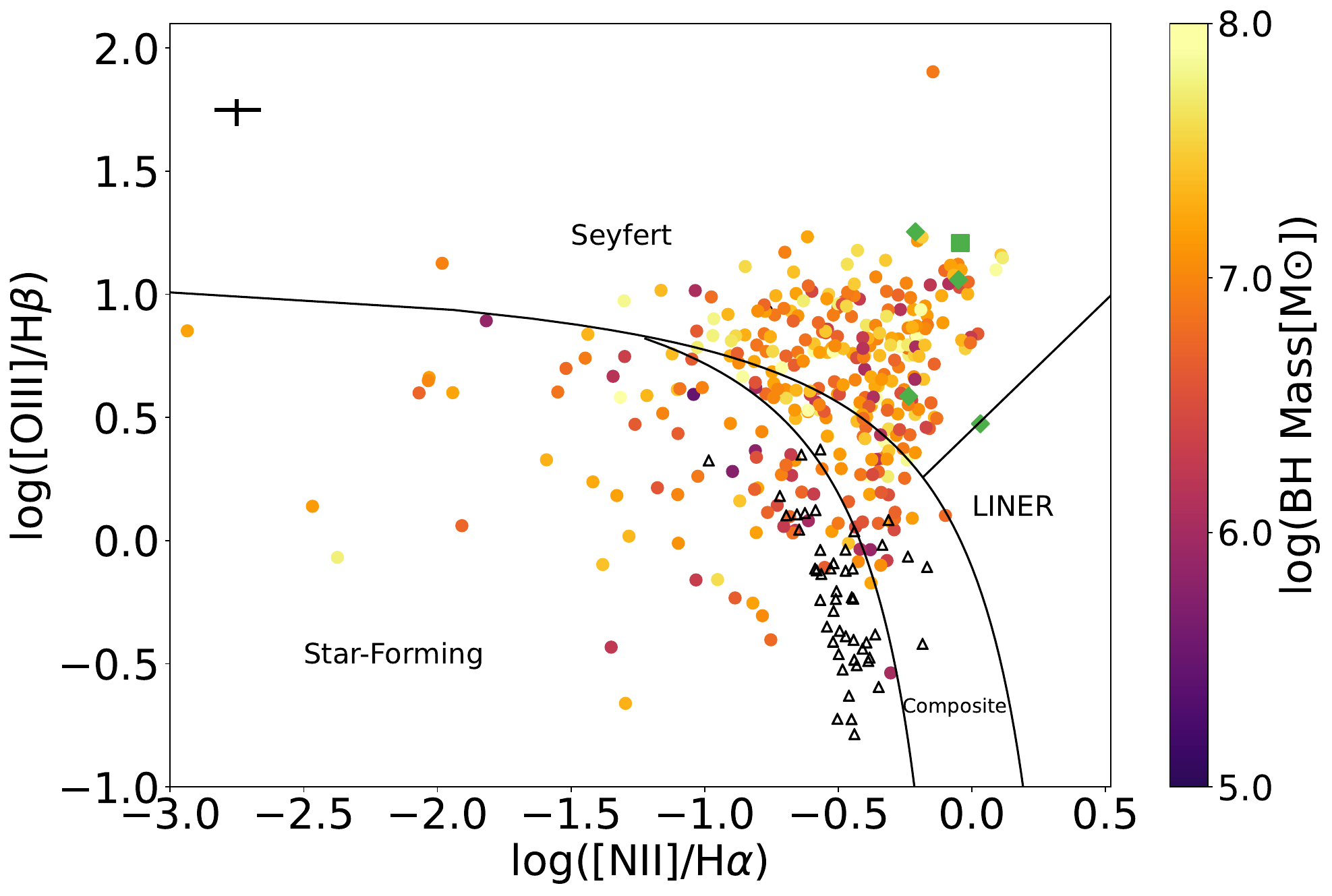}
    \includegraphics[width=0.49\textwidth]{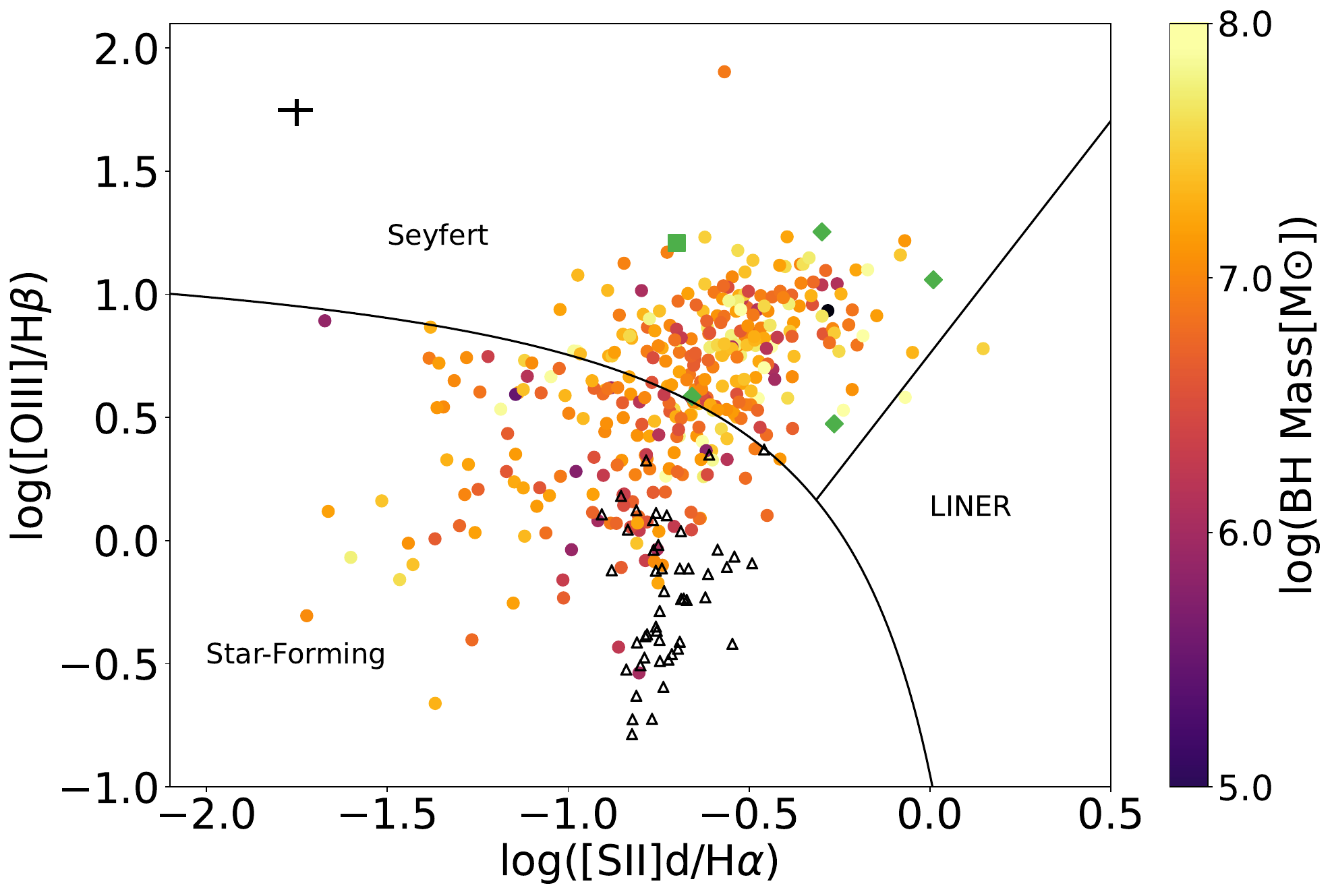}
    \includegraphics[width=0.49\textwidth]{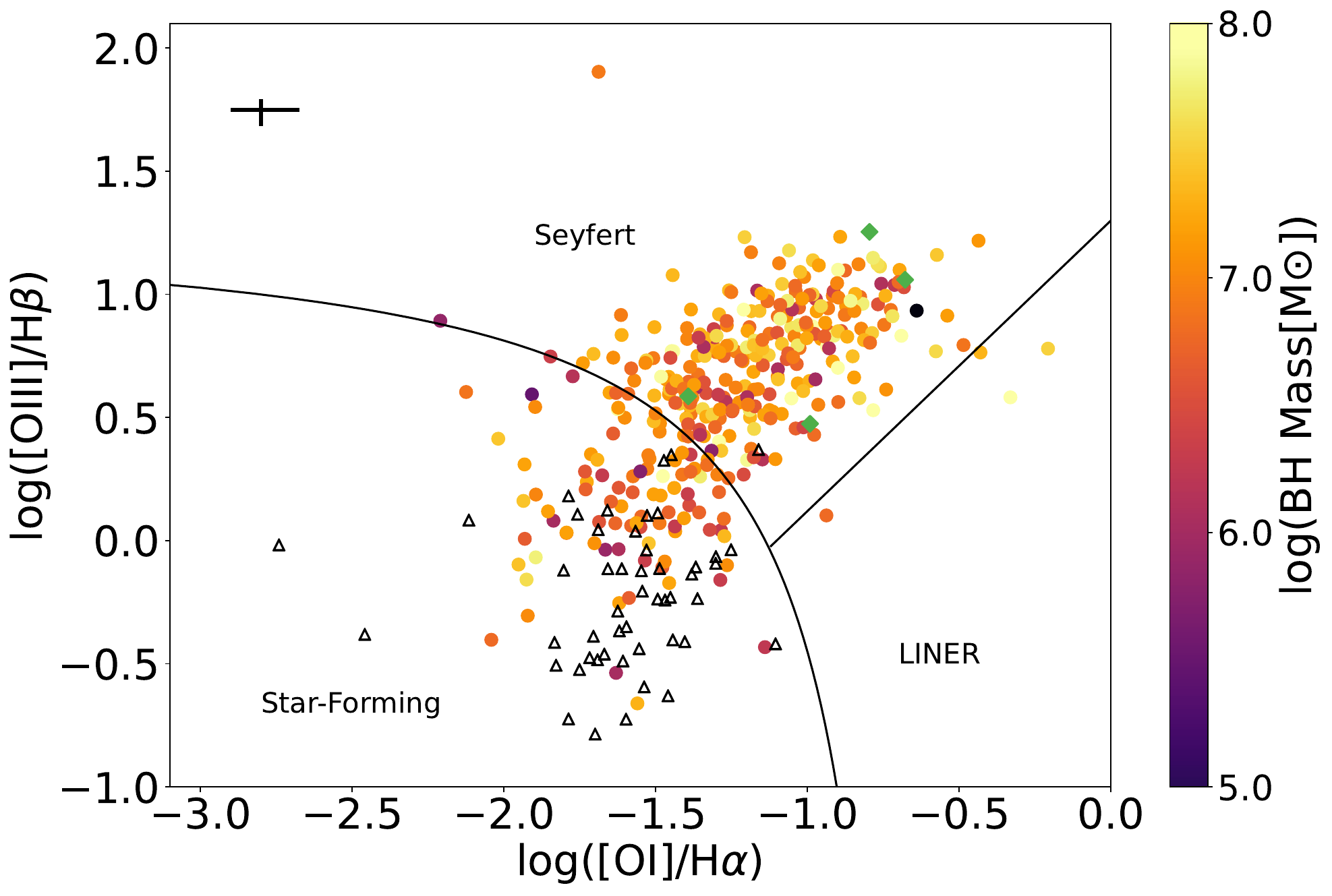}    
    \caption{BPT diagrams from the measured flux of narrow emission lines. The color bar indicates the black hole mass on a logarithmic scale. And, the black triangles represent the objects with no mass estimation. In the top left corner, black lines represent the mean errors. We notice that there are four objects classified as Seyfert (LINER) in all three BPT diagrams with no mass estimation (green diamonds). For these objects the WHAN classification is presented in Table \ref{tab:WHAN}. Furthermore, the object represented as a green square has an X-ray counterpart in the eRASSv1.1 catalog.}
    \label{fig:BPT diagrams}
\end{figure}

In Fig.~\ref{fig:BPT diagrams}, the objects represented as green diamonds, J124438.47+061804.6, J120141.43+382821.5, J131305.85+232733.6, J215055.73-010654.1, and a square (differentiated because this object has an X-ray counterpart; see Sec. \ref{sec:xrays} and \ref{sec:consistency}), J121736.78+293628.8, fall on the AGN region. All of them show an $EW$ of broad H$_\alpha$ and H$_\beta$ lower than $5\AA$, which we classify as having no BELs. 
For these objects, we used the equivalent width of the narrow H$_\alpha$ emission line versus the [N{\sc ii}]/H$_\alpha$ (WHAN) criteria introduced by \citep{Cid2011}. The WHAN diagnostic helps to distinguish between LINER spectra excited by an AGN and similar spectra produced instead by old stellar populations. Table \ref{tab:WHAN} shows the classification according to the WHAN diagram. As shown in Fig.~\ref{fig:WHAN diagram} the five objects fall on the AGN classification.

\begin{figure}[htpb]
    \centering
    \includegraphics[width=0.49\textwidth]{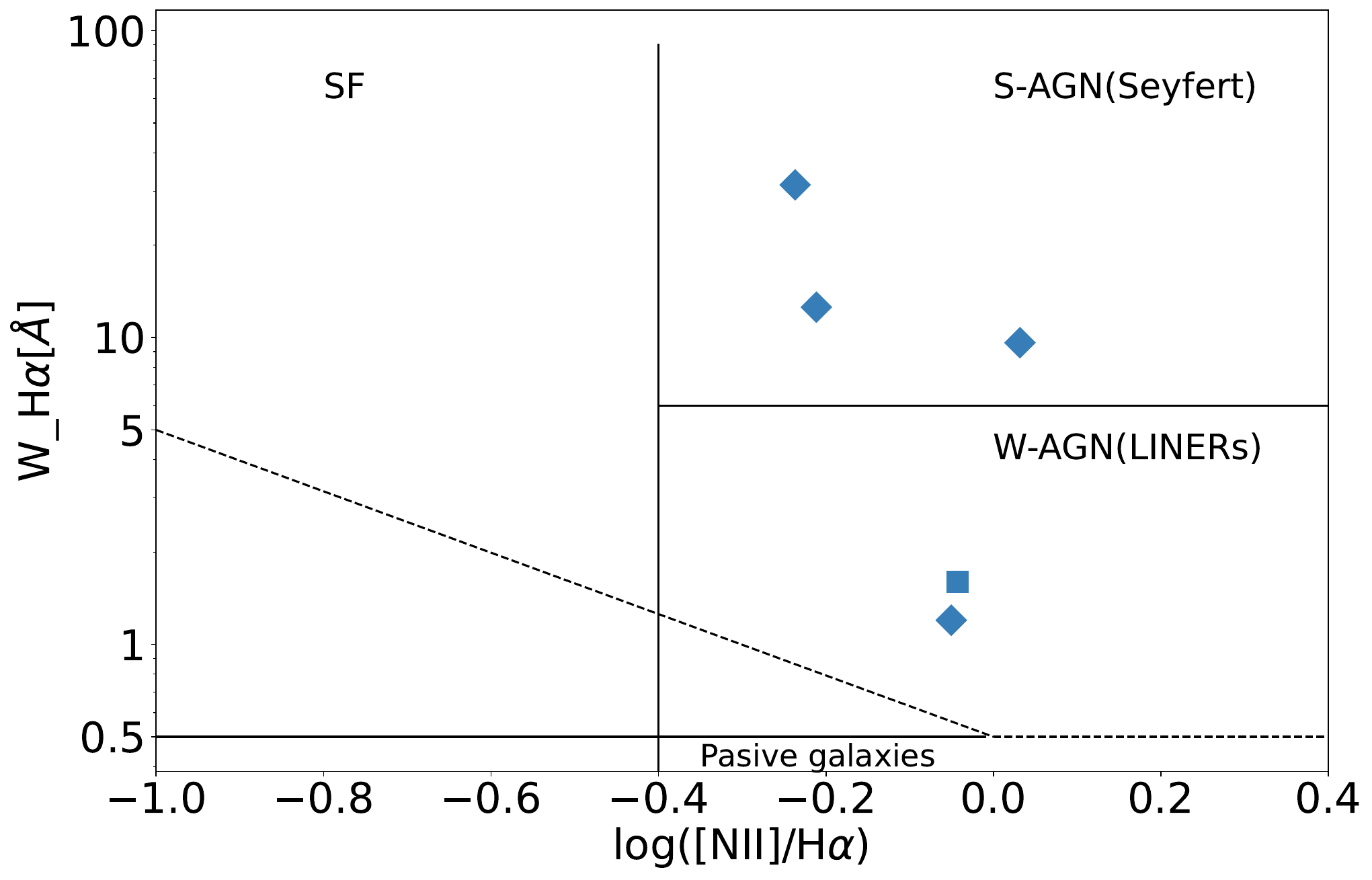}
    \caption{WHAN diagram for the five objects resulting in the AGN region on BPT diagrams but showing no BELs. Blue markers show the position of these objects in the diagram. Tree of them fall in the S-AGN region and two in the W-AGN region.}
    \label{fig:WHAN diagram}
\end{figure}

\begin{table}
    \centering
    \caption{WHAN diagram classification. }
    \begin{tabular}{l|c c  }
    
        IAUNAME &  $EW_{\rm H\alpha}$-narrow & WHAN Class \\
\hline
         J124438.47+061804.6&9.62&S-AGN\\
         J120141.43+382821.5&31.44&S-AGN\\
         J131305.85+232733.6&1.20&W-AGN\\
         J215055.73-010654.1&12.56&S-AGN\\
         J121736.78+293628.8&1.60 &W-AGN
    \end{tabular}
    \label{tab:WHAN}
    \begin{tablenotes}
        \item {\bf Note.} S-AGN means strong AGN, and W-AGN means weak AGN into the WHAN classification criteria.
    \end{tablenotes}
\end{table}

There are 121 objects classified as Star-Forming by the BPT diagnostic diagram  [O{\sc iii}]/H$_\beta$ vs. [N{\sc ii}]/H$_\alpha$. From these, 75 have a significant $EW$ of the broad H$_\alpha$ emission line, with 55 objects in the Alerts set, 37 in the Forced Photometry, 13 in the DR-g, and nine in the DR-r set. We note again that one object can be in one or more sets. Furthermore, 62 of the 70 objects that were classified as Composite also show significant $EW$ of broad H$_\alpha$. Of these 62, 40 are in the Alerts set, 30 in the Forced Photometry set,  four in the DR-g and three in the DR-r set.

We used the publicly available measurements provided by SDSS to compare our BPT classification with those obtained by other groups. We utilized the catalog from the MPA-JHU group, whose method for spectral fitting and measurements is described on the SDSS webpage\footnote{https://www.sdss4.org/dr17/spectro/galaxy\_mpajhu} and based on \citet{2004MNRAS.351.1151B}, \citet{2003MNRAS.341...33K} and \citet{2004ApJ...613..898T}. We found a total of 410 objects after cross-matching the VCS sample with the MPA-JHU catalog, using the coordinates provided in the NSA catalog, with a radius of 1.5 arcsec. In addition, we used the Portsmouth Stellar Kinematics and Emission Line Fluxes Value Added Catalog from SDSS for the same comparison. The description and methodology are also in the SDSS webpage \footnote{http://www.sdss.org/dr14/spectro/galaxy\_portsmouth/\#kinematics} and in \citet{2013MNRAS.431.1383T}. After a cross-match between the Portsmouth catalog and our VCS sample, we found 116 objects, from which 17 are also in the MPA-JHU catalog. In table \ref{tab:BPT_class} we show the comparison of the different BPT classes between this work and the MPA-JHU and Portsmouth catalogs. We notice that our results for BPT classification are more consistent with the results obtained by the Portsmouth group. This could be a consequence of the differences in the models used to fit the spectra. In the case of the MPA-JHU method, the goal had been to separate star-forming galaxies from AGNs, so the model does not distinguish broad from narrow emission lines. As a consequence, the fluxes of the narrow H$_\alpha$ and H$_\beta$ lines are overestimated in the spectra that contain a broad component. This explains why many objects classified as Seyfert and Composite in our work appear as star-forming in the MPA-JHU catalog, while almost all the star-forming classifications in our work are also classified as star-forming in the MPA-JHU catalog (see Table \ref{tab:BPT_class}). We explore these flux differences due to the differences in model fitting in the Appendix \ref{ap: BPT comparison}.

\begin{table*}[t]
    \centering
    \caption{BPT classification comparison between this work and the MPA-JHU and Portsmouth groups.  }
    \begin{tabular}{l|c|ccccc|ccccc}
        BPT class & This work& \multicolumn{5}{c|}{MPA-JHU Group}& \multicolumn{5}{c}{Portsmouth Group}  \\
        & &Seyfert&LINER&Comp.&SF&N-C&Seyfert&LINER&Comp.&SF&N-C\\
         \hline
        Seyfert &203 &\boxed{44}&0&37&117&2& \boxed{12} & 2 & 10 & 0 & 0\\ 
        LINER &1& 1& \boxed{0} & 0 & 0 & 0 & 0 & \boxed{1} & 0 & 0 & 0\\
          Composite&70 &  4 & 0 & \boxed{26} & 39 & 0& 1 & 3 & \boxed{15} & 11 & 0\\
         Star-Forming& 121 &  0 & 0 & 2 & \boxed{118} & 0 & 0 & 0 & 1 & \boxed{57} & 0\\
         No-class& 20 & 3 & 1 & 0 & 14 & 2 & 0 & 1 & 0 & 0 & 2
    \end{tabular}
    \label{tab:BPT_class}
            \begin{tablenotes}
        \item {\bf Note.} The column 'This work' accounts for all the 415 objects in the VCS sample with their BPT classification on the first column named BPT class. Table blocks for the MPA-JHU and Portsmouth groups show the classification of the cross-matched objects. Here the column 'N-C' is the No-class category. The numbers marked in boxes are the coincidences in classification between our results and those of the MPA-JHU and Portsmouth groups.
    \end{tablenotes}

\end{table*}

 \subsection{Contribution of the AGN component to the continuum}\label{sec: AGN contribution}

The continuum emission in the spectra of low redshift (z<0.15) AGNs at optical wavelengths is mainly composed of the stellar light of the host galaxy and the continuum from the accretion disk; the relative contribution of these two components is dependent on the properties of the host galaxy and the AGN. For objects where the $EW_{\rm H\alpha}>5\AA$, we measured the relative contribution of the AGN component at $5100\AA$ as the ratio of the monochromatic flux of the AGN to the total continuum flux ($f_{\rm 5100,AGN}/f_{\rm 5100,cont.}$). To compare the emissivity profile of the AGN component obtained from pPXF, we fit this component as one power law using the model $f_\lambda=(\frac{\lambda}{N})^{\gamma}$, and in Fig.~\ref{fig:AGN contribution} we compare the slope $\gamma$  and the fractional contribution of the AGN. For bluer (meaning more negative values of $\gamma$) AGN continuum, the relative contribution of the AGN to the total continuum shows a tendency to increase.

{There are 108 objects with $-0.1<\gamma\leq0.0$. In most cases, this flat AGN component is expected due to the limit in the $\alpha$ parameter in the model introduced in section \ref{sec:spectral-fitting} ($f_\lambda=(\frac{\lambda}{N})^{\alpha}$, with $-3\leq \alpha \leq 0 $) when fitting the spectra that show flat profiles, possibly dominated by the host galaxy. However, we identified 10 objects with $\gamma=0$ that exhibit spectra dominated by the AGN component, with  $0.51<f_{\rm 5100,AGN}/f_{\rm 5100,cont.}<0.76$; the spectra of these objects show flat profiles and do not display very strong stellar absorption features. For these cases, the normalization of the stellar component and the AGN component could be degenerate (see spectral decompositions in Appendix \ref{ap:AGN contribution}). However, note that the black hole mass measured using the AGN continuum (Eq. \ref{eq:M}) and using the H$_\alpha$ luminosity are in agreement as shown in Sec. \ref{sec: BH vs M*}}. In addition, in the top panel of Fig.~\ref{fig:AGN contribution} we show the histograms of the AGN relative contribution of the different sample sets. The distributions show that Alerts and Forced Photometry sets find more objects with lower ($<0.4$) relative AGN contribution than the DR.  In the right panel of Fig.~\ref{fig:AGN contribution} we show the distribution of the power law slope, where the Alerts and Forced Photometry sets tend to find more objects with redder AGN continua or host dominated. 

When comparing the distribution of the AGN relative contribution and power law slope $\gamma$ for different BPT-classes, namely Star-Forming, Composite, and Seyfert/LINER (Figure \ref{fig:BPT-AGN contribution}), we do not see strong distinctions;  i.e., objects classified as Star-Forming by the BPT diagrams show similar characteristics of their continuum emission to the ones classified as AGN. The difference between Star-Forming and AGN classes comes from the strength of the narrow emission lines. {As shown by \citet{Mezcua17}, and in more recent work by \citet{2024MNRAS.528.5252Mezcua}, using IFU data, dwarf galaxies can have narrow-line characteristics of AGN in small spatial regions, while the rest of the galaxy is dominated by star formation. In our case, it is possible that the broad-line AGN with star-forming classification are similarly affected by contamination from the rest of the galaxy within the fiber aperture. This contamination appears to influence the line emission and continuum differently, potentially depending on the level of star formation activity, which alters the narrow-line fluxes relative to the stellar continuum. Consequently, we do not find a correlation between BPT type and AGN continuum dominance.}

\begin{figure}[t]
    \centering
    \includegraphics[width=0.48\textwidth]{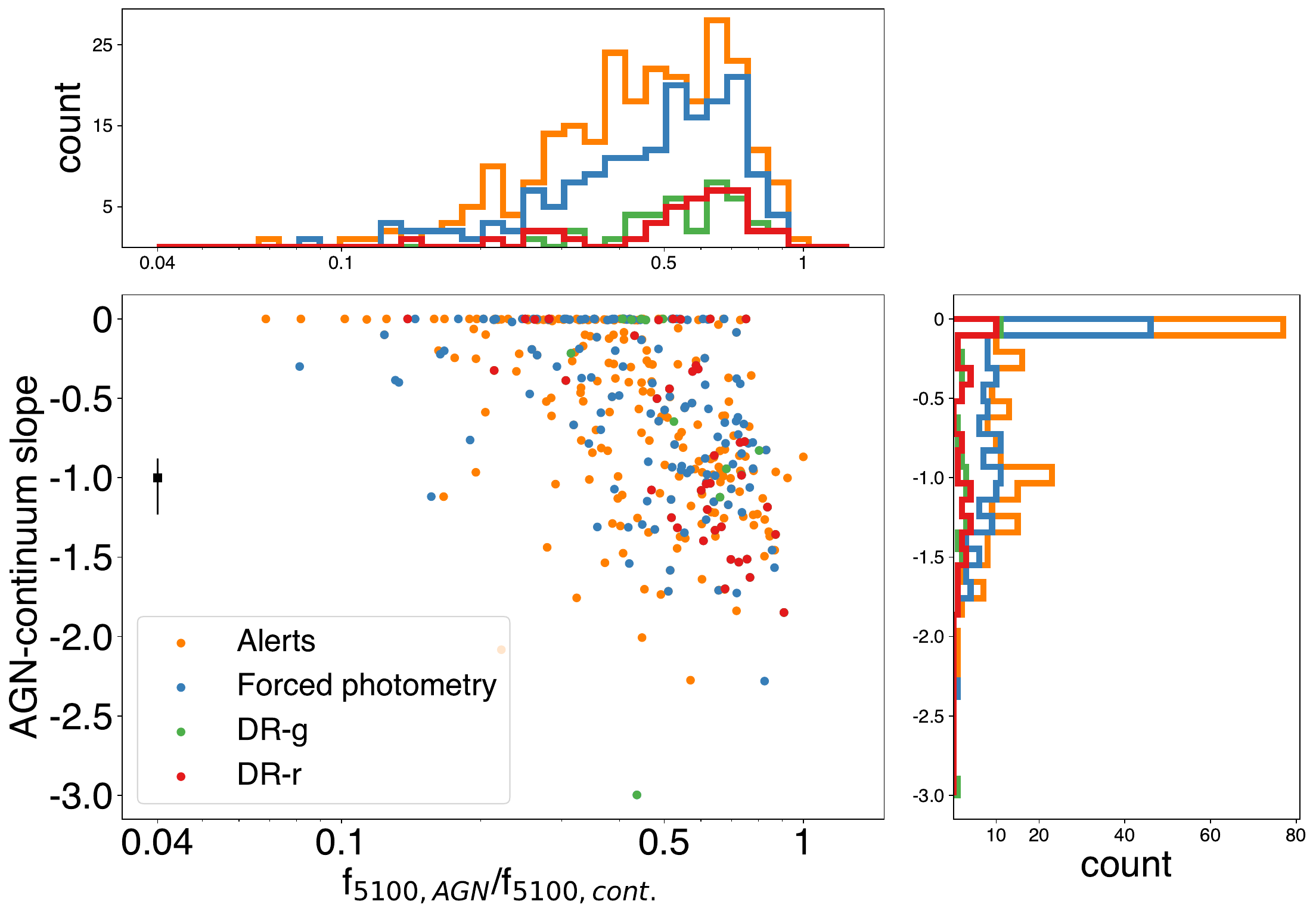}
    \caption{In the central plot, we present the comparison of the AGN-continuum slope from the best fit with the ratio of the AGN component and total continuum at $5100\AA$. Different colors represent different sets: orange for Alerts, blue for Forced Photometry, green for DR-g, and red for DR-r. The black square and error bars represent the typical error in the AGN-continuum slope constrained by the 16th and 84th percentile values from simulations. In top panel, the distribution of the AGN relative contribution ($f_{\rm 5100,AGN}/f_{\rm 5100,cont.}$) is plotted for each set. Additionally, in the right panel, the power law slope ($\gamma$) distribution of each set is shown.}
    \label{fig:AGN contribution}
\end{figure}  

\begin{figure}
    \centering
    \includegraphics[width=0.45\textwidth]{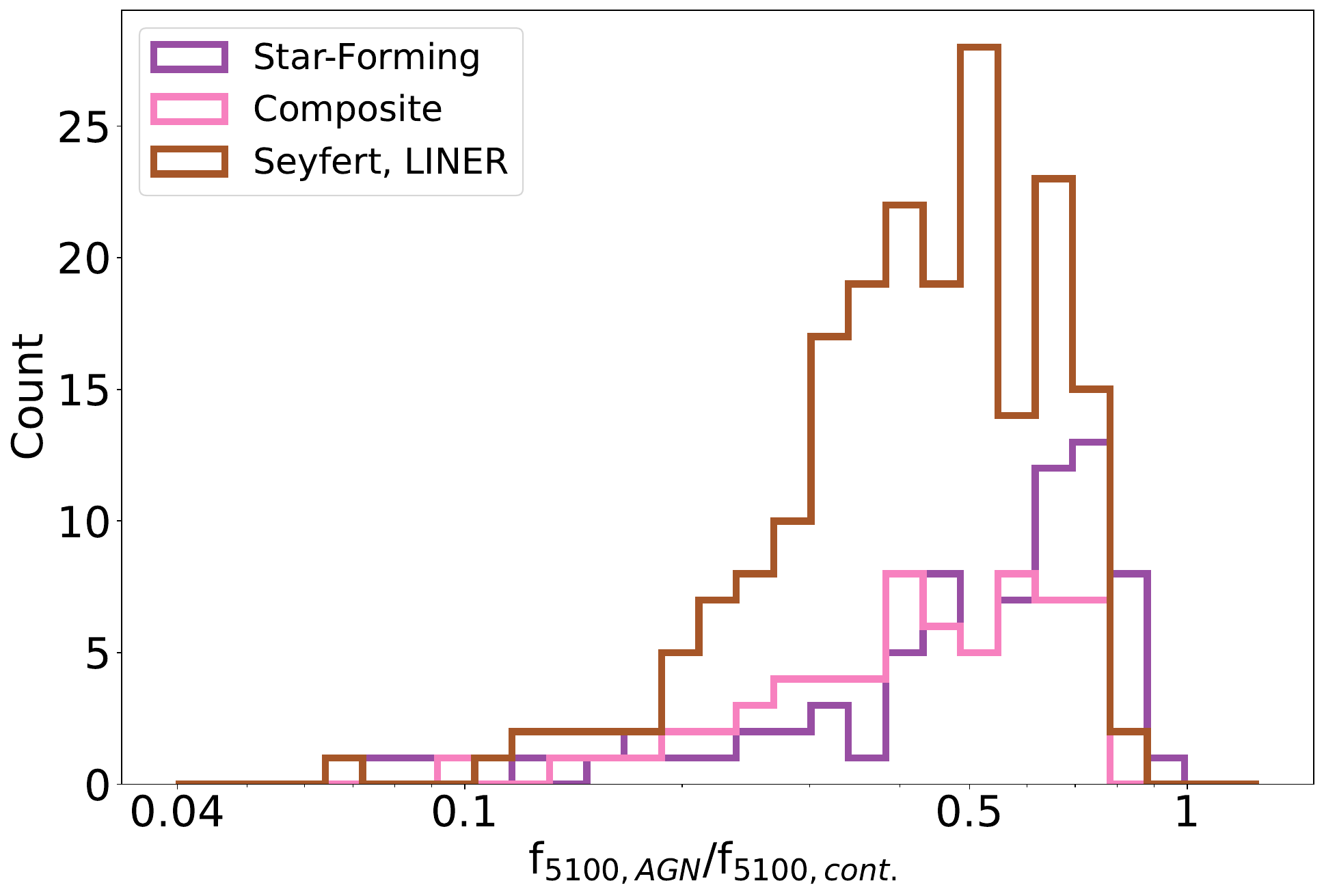}
    \includegraphics[width=0.45\textwidth]{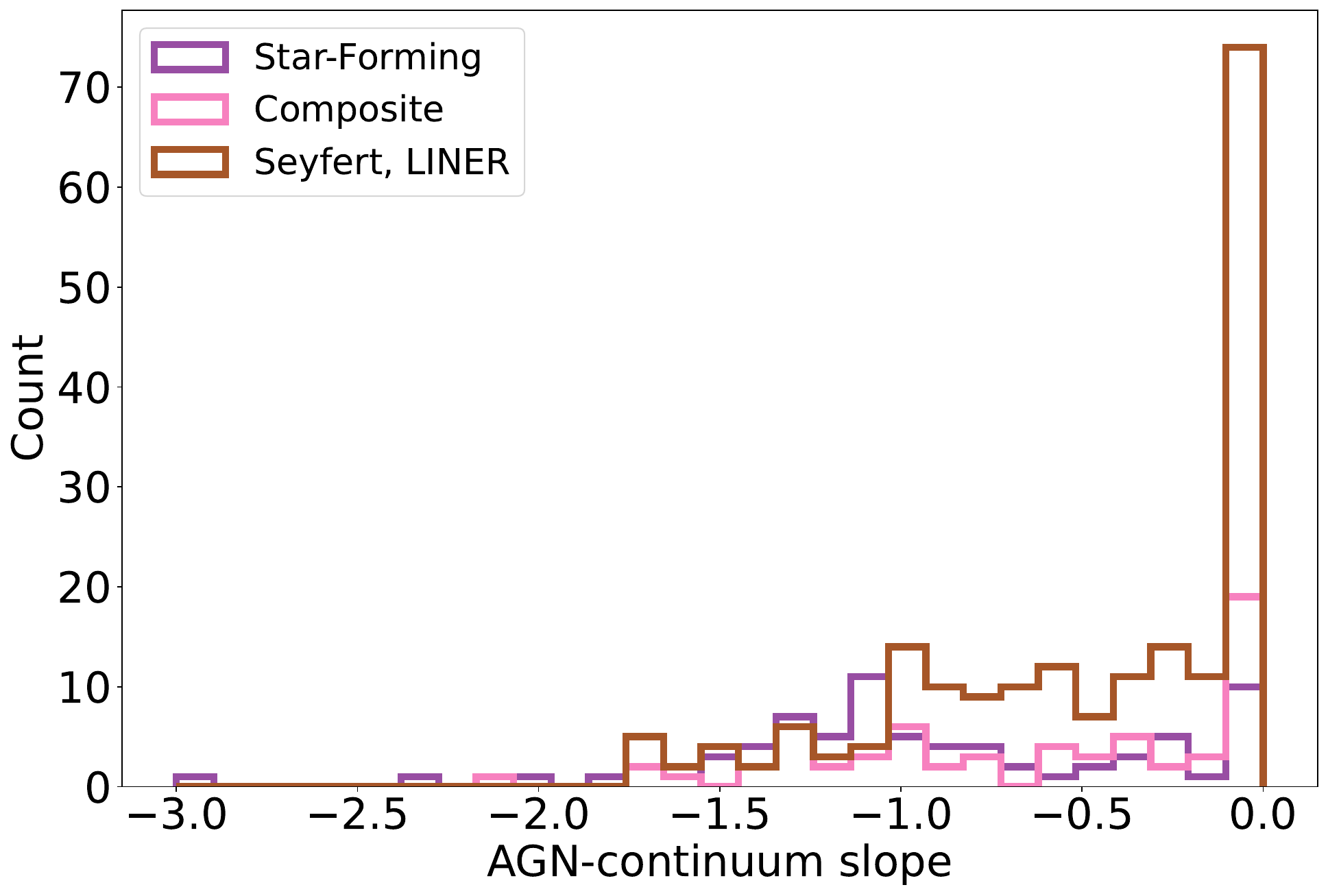}
    \caption{(Top) Distribution of the AGN relative contribution ($f_{\rm 5100,AGN}/f_{\rm 5100,cont.}$). For different BPT classes, according to the [O{\sc iii}]/H$_\beta$ vs. [N{\sc ii}]/H$_\alpha$ diagram. The distributions for different classes, Star-Forming (purple), Composite (pink), and Seyfert/LINER (brown) are similar with the Seyfert/LINER objects concentrated between $0.3<f_{\rm 5100,AGN}/f_{\rm 5100,cont.}<0.75$. (Bottom) Distribution of the fitted power slope for the same BPT classes.}
    \label{fig:BPT-AGN contribution}
\end{figure}

\subsection{Black hole vs stellar mass}\label{sec: BH vs M*}
In the top panel in Fig.~\ref{fig:mass_ratio} we show a histogram of the ratio between our estimate of the black hole mass and the stellar mass of its host galaxy given in the NSA catalog. For all samples, the ratio is distributed mainly between values of $M_{*}/M_{BH} = 300$ to 3000, with the DR sets distributed almost uniformly in this range and the Alerts and Forced Photometry sets peaking around $M_{*}/M_{BH} = 1000$.  The median value of the ratios $M_{*}/M_{BH}$ for the Alerts, Forced Photometry, DR-g, and DR-r sets are 906, 1252, 592, and 787, respectively. An Anderson-Darling test results in a p-value lower than 0.05 for the comparison between the Forced Photometry set with the other sets. While for the comparison between any pair of the Alerts, DR-g, and DR-r sets the p-value is always higher than 0.05. This shows that the mass ratio distribution of the Forced Photometry set is significantly different from the other sets.

\citet{Reines2015} studied the relation between black hole mass and total stellar mass for a sample of 341 galaxies with stellar masses in the range $10^{8}-10^{12}$ $M_\odot$, although only about 40 of them with masses below $10^{10}$ $M_\odot$. They find a linear relation between black hole mass and total stellar mass with a ratio $M_{*}/M_{BH} \sim 4000$, i.e., with black hole masses four times less massive than those found here. In the bottom panel in Fig.~\ref{fig:mass_ratio} we show the relation between black hole mass and total stellar mass in our sample, together with the best-fitting line obtained by \citet{Reines2015} of $M_{\rm BH}=0.00025M_*$. Although some of our sources reach this level, the majority of our sample lies above. We also plot for reference the relation $M_{\rm BH}=0.001M_*$ \citep{2013ARA&A..51..511K}, found for generally more massive elliptical galaxies, which is closer to our data. 

One key difference to examine is how black hole mass is estimated. In the Eq.~\ref{eq:M} in this work we use a single-epoch mass estimation from \citet{MejiaRestrepo16}, that uses the FWHM of the BELs and the AGN luminosity at $5100\AA$. While in \citet{Reines2015} (eq.~1) the mass is estimated, also using a single-epoch, but considering only the H$_\alpha$ properties (FWHM, Luminosity). Fig.~\ref{fig:mass_comparison} shows the comparison of the masses estimated by both methods for our sample, which give very consistent results, with only a slight discrepancy towards higher masses at the level of $\sim 0.3$ dex. {As a consequence, this cannot account for the difference in the $M_{*}/M_{BH}$.}

{Another key difference to investigate is the adopted samples}. Although the parent sample in \citet{Reines2015} and this work is the same, i.e., the NSA catalog of local galaxies, the criteria for the selection of AGN differs. In our work, we select {506} candidates based on the optical variability features, of which {357} have a significant broad Balmer line detection (i.e., a flux SNR$>$3 and EW${\ge}$5\AA\ in broad H$_\alpha$ or H$_\beta$). In \citet{Reines2015}, {341} AGN candidates were selected based on the simultaneous identification  of broad Balmer lines and a Seyfert or AGN classification in all three BPT diagnostic diagrams using the optical spectra; by contrast, a broad line was considered detected in their work if it led to a 50\% improvement in the $\chi^2$ of the fitted model. As shown in Secs. \ref{sec:BPT} and later in \ref{sec:consistency}, 169 of our variability-selected AGN are not classified as AGN in all three BPT diagrams in Fig.~\ref{fig:BPT diagrams}, despite having significant broad Balmer lines and in many cases an X-ray detection as well. The {222} galaxies classified as Star-Forming or Composite (including mixed BPT classifications) in our sample appear in all stellar mass bins, but are more frequent for lower masses and comprise almost all the AGN candidates in galaxies with stellar masses below $10^{9}$ $M_\odot$. Therefore, the removal of candidates with non-AGN BPT classification in \citet{Reines2015} at least partially explains why our sample has many more galaxies with stellar masses below $10^{10}M_\odot$ (i.e., 161 AGN in galaxies with stellar mass below $10^{10}M_\odot$ in our work vs. 41 in \citealp{Reines2015}, with 15 objects in common in this mass range).

The difference in the {confirmation of AGN} criteria, however, does not explain why our objects have, on average, larger black hole masses for a given stellar mass. The difference is probably produced by the limitations of the variability selection: although it can find AGN in lower mass galaxies, it preferentially detects the galaxies with the largest black holes. This supposition is supported by the median of the mass ratios detected by the different sets, where the more sensitive sets, i.e., Alerts and Forced Photometry, detect AGN up to larger ratios of stellar mass to black hole masses than the DR sets (see the top panel in Fig.~\ref{fig:mass_ratio}).  This limitation might be reduced with even more sensitive, higher SNR light curves that will be produced in the future by the Vera Rubin LSST \citep{Ivezic2019}. 

{Finally, we note that the stellar masses calculated via SED fitting, such as those in the NSA catalog, can be overestimated in active galaxies, as demonstrated in \citet{Buchner2024}. These authors show that the bias factor is a function of AGN bolometric luminosity and stellar mass. For the majority of the bolometric luminosities (i.e., log L$_{bol}=41-44$) and stellar masses (i.e., log$M_*=9-10$) estimated for our sample, the mean and standard deviation values of this bias are in the range $0.1\pm0.2$ to $0.2\pm0.4$, which should not affect considerably our results.} 

\begin{figure}[htpb]
    \centering
    \includegraphics[width=0.49\textwidth]{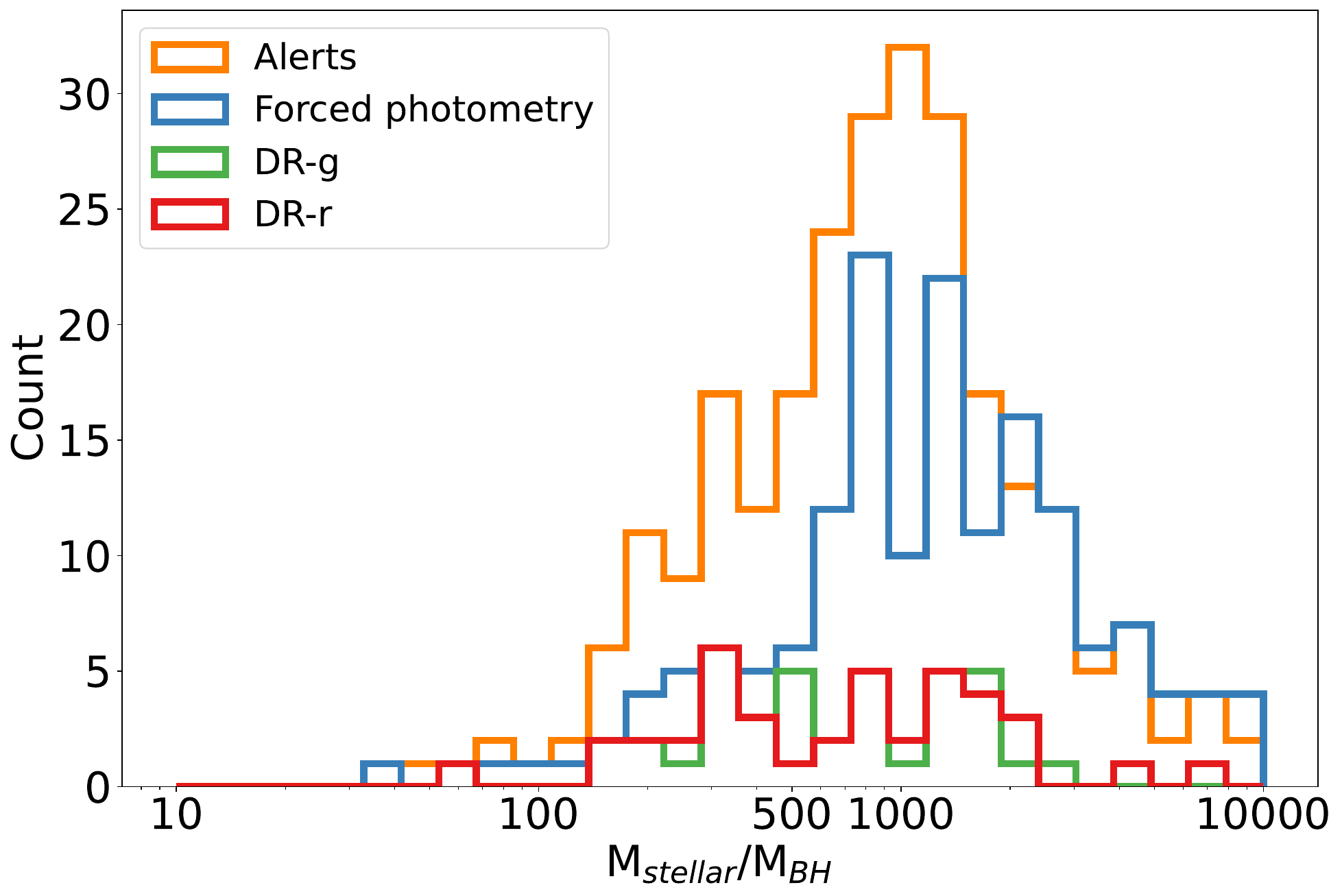}
    
    \includegraphics[width=0.49\textwidth]{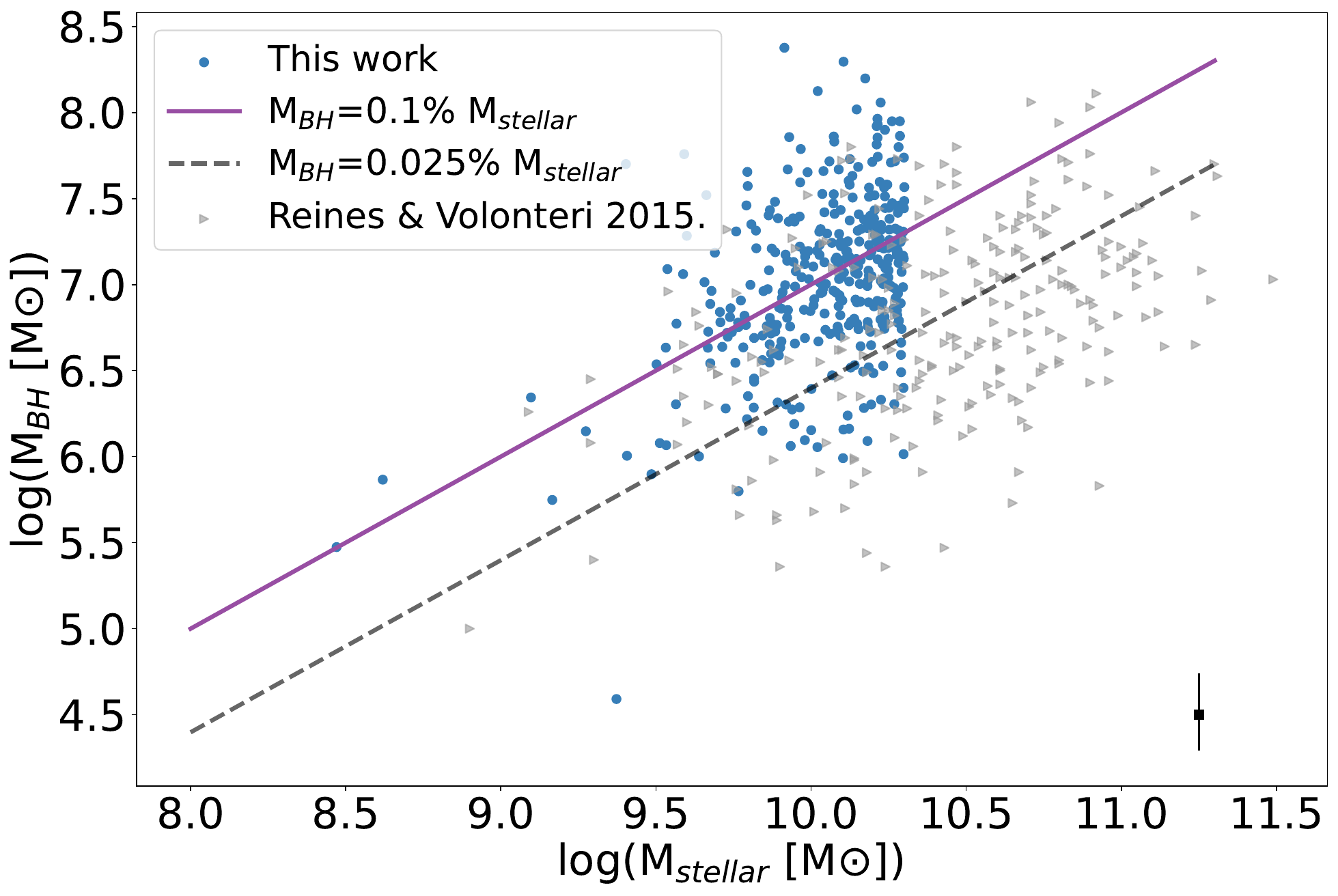}
    
    \caption{Top: Distribution of the ratio between our estimates of the black hole mass to the stellar mass of the host galaxies listed in the NSA catalog, separated by selection set. Bottom: Black hole mass estimated using Eq. \ref{eq:M} versus the stellar mass of the host galaxy. {The results from \citet{Reines2015} are shown in grey for comparison. The purple line represents the $M_{\rm BH}=0.001M_*$ relation \citep{2013ARA&A..51..511K}, while the black dashed line indicates the relation obtained in \citet{Reines2015}}. Mean errors are shown with error bars in the bottom right corner. }
    \label{fig:mass_ratio}
\end{figure}

\begin{figure}[htpb]
    \centering
    \includegraphics[width=0.48\textwidth]{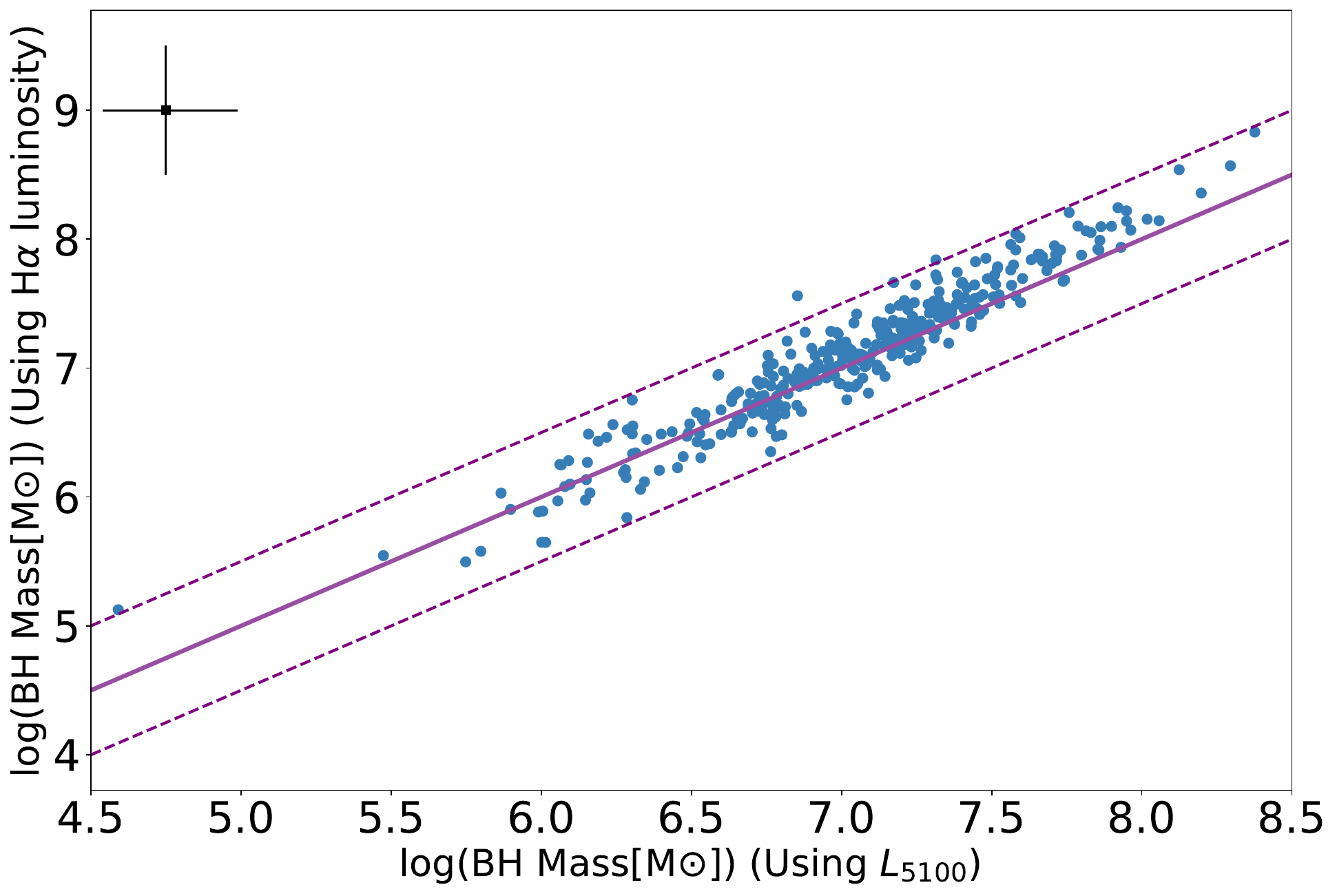}
    \caption{Comparison of the Black hole mass computed by Eq.~\ref{eq:M} in this work (i.e., using the FWHM of BELs and AGN luminosity at 5100\AA) versus the mass computed using eq.~1 of \citealt{Reines2015} (i.e., using the H$_\alpha$ FWHM and its luminosity).   The purple line shows the 1:1 relation in the logarithmic scale. {The dashed purple lines indicate the $0.5\rm dex$ uncertainty from the eq.~1 of \citealt{Reines2015}. Mean errors are shown in the top left with error bars. Since both mass estimates use the FWHM of H$_\alpha$ the errors are partly correlated}}
    \label{fig:mass_comparison}
\end{figure}

\section{Frequency of X-ray counterparts of the variability-selected AGN in low mass galaxies}
\label{sec:xrays}
AGN activity often results in the emission of X-rays and conversely, sufficiently high X-ray luminosity in galaxies is often considered proof of black hole activity \citep{1994MNRAS.267..193P}. We explored the incidence of X-ray detections in our sample of variability-selected AGN candidates in the interest of establishing whether X-ray emission is also ubiquitous in low mass galaxies/low mass SMBHs, in particular those selected by optical variability. \citet{Arcodia2024} carried out a similar analysis on several literature samples, which we discuss in Sec. \ref{Sec:Comparison}. 

We used the recent publication of the first data release (DR1) of the X-ray source catalog of the SRG/eROSITA all-sky survey from the German Consortium \citep{Merloni2024}, to search for soft X-ray counterparts of our variability-selected AGN sets.
We will refer to the soft X-ray counterparts simply as X-ray counterparts in the following text.

The eRASSv1.1 catalog covers half the sky, specifically the Galactic 
longitudes in the range 180 to 360 degrees, to a depth of about $5 \times 10^{-14} $ \ergs\ $cm^{-2}$ in the 0.2--2.3 keV band. Of our 506 AGN 
candidates, 230 fall in this region of the sky. Performing a simple cross-match by sky coordinates between both samples with a matching radius of 5 (10) $\arcsec$ results in 123 (150) AGN candidates with an X-ray 
counterpart,  i.e., 53(65)\% of the candidates have a match. We note that the source density of the eRASSv1.1 catalog in this area of the sky is about 30 points per deg$^2$ or equivalently $0.00073$ sources in a circle of $10\arcsec$ radius. Therefore, even using the larger cross-matching radius the probability of chance alignments is small. Increasing the matching radius to $15\arcsec$ only results in an addition of five matches, from which we conclude that the counterparts are most likely related to the AGN candidates and not to chance alignments.  

The discussion above includes all variability-selected AGN in the low stellar mass sample of the NSA catalog. As discussed in Sec. \ref{Sec:Validation selection}, some of these galaxies are indeed QSOs but have higher redshifts than expected for the NSA sample, so their stellar masses are incorrect. Therefore, their spectra were not modeled. If we consider only the 415 VCS sample of galaxies with good SNR spectra, that are indeed LSMGs at $z<0.15$ and show at least narrow emission lines, we find 195 in the portion of the sky included in the eRASSv1.1 catalog. Of these, 130 have an X-ray counterpart within $10\arcsec$, or 67\% of the sample, a similar fraction as in the total variability-selected AGN sample in the eROSITA-DE sky. 

Table \ref{tab:xrays} shows the number of AGN candidates in the eROSITA-DE sky in each set, and the number of those candidates with an X-ray counterpart in the eRASSv1 catalog. We note that the ratio of matched candidates in each individual set is larger than the ratio of matches for the combined sets in this area of the sky (150/230). This counter-intuitive result is caused by the larger probability of X-ray matches for sources that appear in multiple sets, while sources that appear in only one set have a lower ratio of matches.  The total number of unmatched candidates in the combined set is 80, while the sum of the unmatched candidates in each individual set is 100, showing that most of the unmatched candidates only appear in one set while only a few appear in two or more sets.

\begin{table}[]
    \centering
    \caption{AGN candidates with X-ray counterparts separated by selection set.}
    \begin{tabular}{c|cc}
    Set & \multicolumn{2}{c}{ in eRASSv1.1/in eROSITA-DE sky} \\
     & in VCS & all candidates \\
 \hline
Alerts &  87/130=67\% & 105 / 154 = 68\% \\
Forced phot.&  92/120=77\% &105 / 142 = 74\% \\
DR-g& 23/25=92\% &30 / 38 = 79\%\\
DR-r&  23/25=92\% &29 / 35 = 83\%\\ 
    \end{tabular}
    \label{tab:xrays}
     \begin{tablenotes}
        \item {\bf Note.} Here we show the number of Variability-selected AGN candidates in the VCS sample, meaning LSMGs at $z<0.15$ in the column called "in VCS", and in column "all candidates" all variability-selected AGN candidates galaxies of the NSA v1.1 catalog that fall in the eROSITA-DE sky regardless of whether they have spectra or not or if they are higher redshift objects erroneously included in the NSA catalog.
    \end{tablenotes}
\end{table}

In Fig.~\ref{fig:ew_Ha_erass} we show the distribution of equivalent widths of broad H$_\alpha$ obtained for all the AGN candidates in bona fide LSMGs that are in the eRASSv1.1 sky (orange), the ones with X-ray counterparts (blue) and, of these, the ones where the BPT diagnostic [O{\sc iii}]/H$_\beta$ vs. [N{\sc ii}]/H$_\alpha$ returns a star-forming type (red).{ We added 0.6 to objects with  $EW_{\rm H\alpha}=0$\AA\ (12 in total), so they can be shown in the plot with a logarithmic scale.} There are 56 AGN candidates with fitted spectra in this region of the sky which have a Star-Forming classification according to the [O{\sc iii}]/H$_\beta$ vs. [N{\sc ii}]/H$_\alpha$  BPT diagnostic. From these, 27 have an X-ray counterpart and all have equivalent widths of H$_\alpha > 5\AA\ $.

{We note that, only one of the 25 objects with an equivalent width $EW_{\rm H\alpha}<5\AA$ in the eRASS sky has an X-ray counterpart. The object J121736.78+293628.8, which has an X-ray counterpart, shows a spectrum profile dominated by the stellar light and is classified as Seyfert in the [O{\sc iii}]/H$_\beta$ vs. [N{\sc ii}]/H$_\alpha$  BPT-diagram and it is also classified as W-AGN in the WHAN diagram.} 
From these numbers, we conclude that the majority of AGN candidates with  $EW_{\rm H\alpha}>5\AA$ have X-ray counterparts, even if they are classified as Star-Forming by the BPT diagnostics.

\begin{figure}
    \centering
    \includegraphics[width=0.45\textwidth]{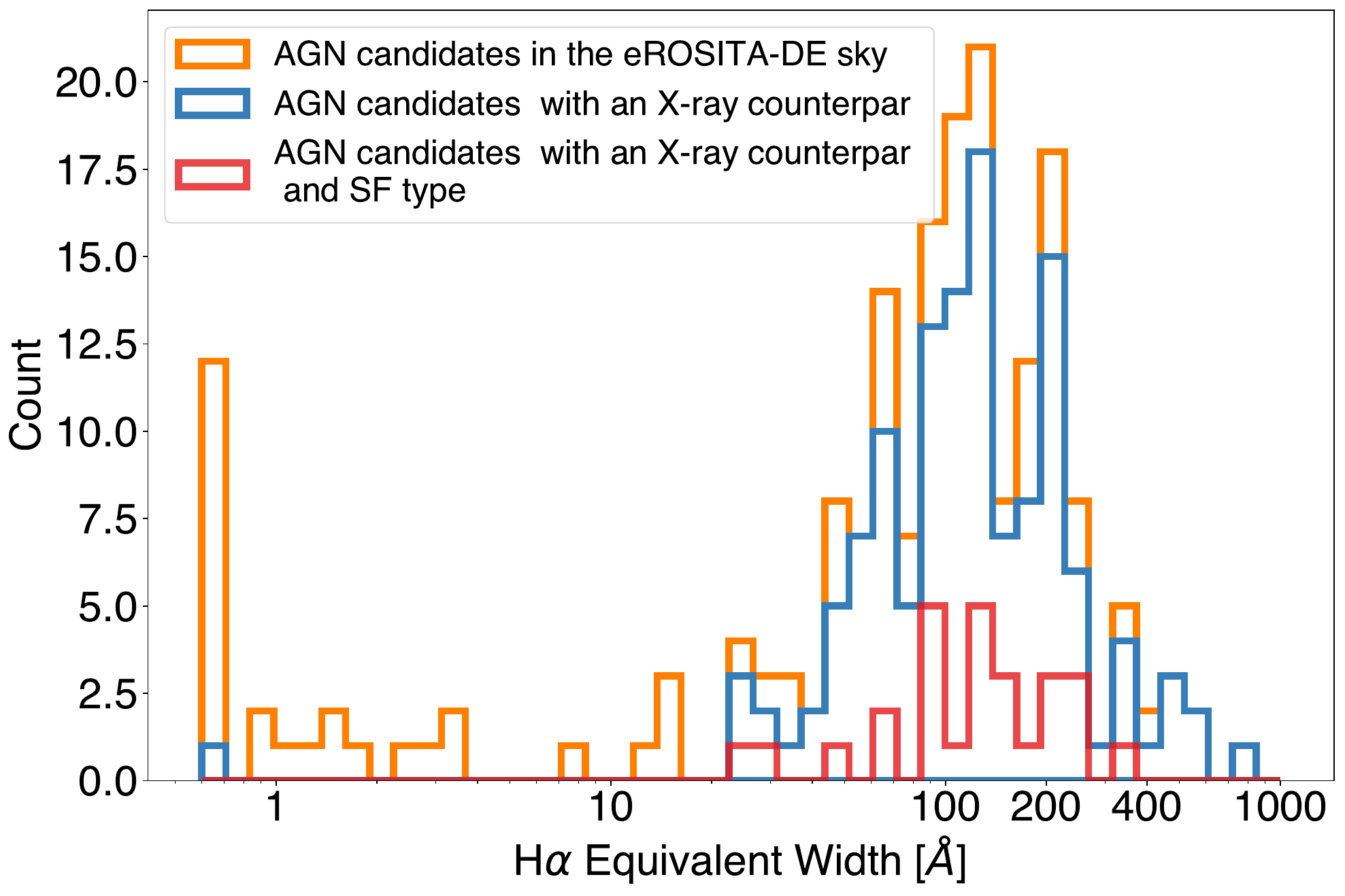}
    \caption{Distribution of H$_\alpha$ equivalent widths for the AGN candidates with fitted spectra in the eROSITA-DE sky (grey), of these the ones with X-ray counterparts (blue), and of these, the ones classified as star-forming by the BPT diagnostics (red). Objects with  $EW_{\rm H\alpha}=0\AA$, 12 in total, are counted using an added $EW$ value of 0.6. }
    \label{fig:ew_Ha_erass}
\end{figure}

{The relationship between X-ray emission and H$_\alpha$ luminosity has been studied by various authors (e.g., \citealt{2006A&A...455..173Panessa_X-ray}, \citealt{2001ApJ...549L..51Ho_Xray}, \citealt{2010ApJ...714..115Shi_Xray}). \citet{2006A&A...455..173Panessa_X-ray} demonstrates the relationship between X-ray luminosity in the 2–10 keV band and H$_\alpha$ luminosity for Type 1 Seyferts, Type 2 Seyferts, mixed Seyferts, Compton-thick candidates, and low-redshift quasars. This relationship is consistent with the findings of \citet{2001ApJ...549L..51Ho_Xray}, where the best-fit equation is given by $log(L_X)=(1.11 \pm 0.054) log L_{H\alpha}-(3.50 \pm 2.27)$. For comparison, in Fig.~\ref{fig:Xray_Halpha}, we plot the X-ray luminosity in the 2–8 keV band in the observed frame of reference from the eRASSv1.1 catalog and the H$_\alpha$ luminosity measured from our best-fit models, including both the broad and narrow components, for objects with $EW_{\rm H\alpha}>5 \AA\ $. We find that the data are well distributed around the relationship from \citet{2001ApJ...549L..51Ho_Xray} (purple line in the plot). To contrast, in Fig. \ref{fig:Xray_Halpha}, we also plot the relation between X-ray luminosity in the 2–10 keV band and H$_\alpha$ luminosity for low mass Star-forming galaxies, given by $log(L_X)= log L_{H\alpha}-1.40$ as presented in \citet{2009MNRAS.399..487RG} (red line in the plot).

As expected, objects without X-ray detection in eRASSv1.1 have a lower value of H$_\alpha$ (narrow + broad) flux. For these objects, the median is $\rm Flux_{H\alpha}= 988\; erg\ cm^{-2}\ s^{-1}$, while for those with an X-ray counterpart, it is $\rm Flux_{H\alpha}=2395\; erg\ cm^{-2}\ s^{-1}$. 
We intend to conduct further comparisons and analyses of X-ray properties in future work, as this is beyond the scope of this paper.}

\begin{figure}
    \centering
    
    \includegraphics[width=0.45\textwidth]{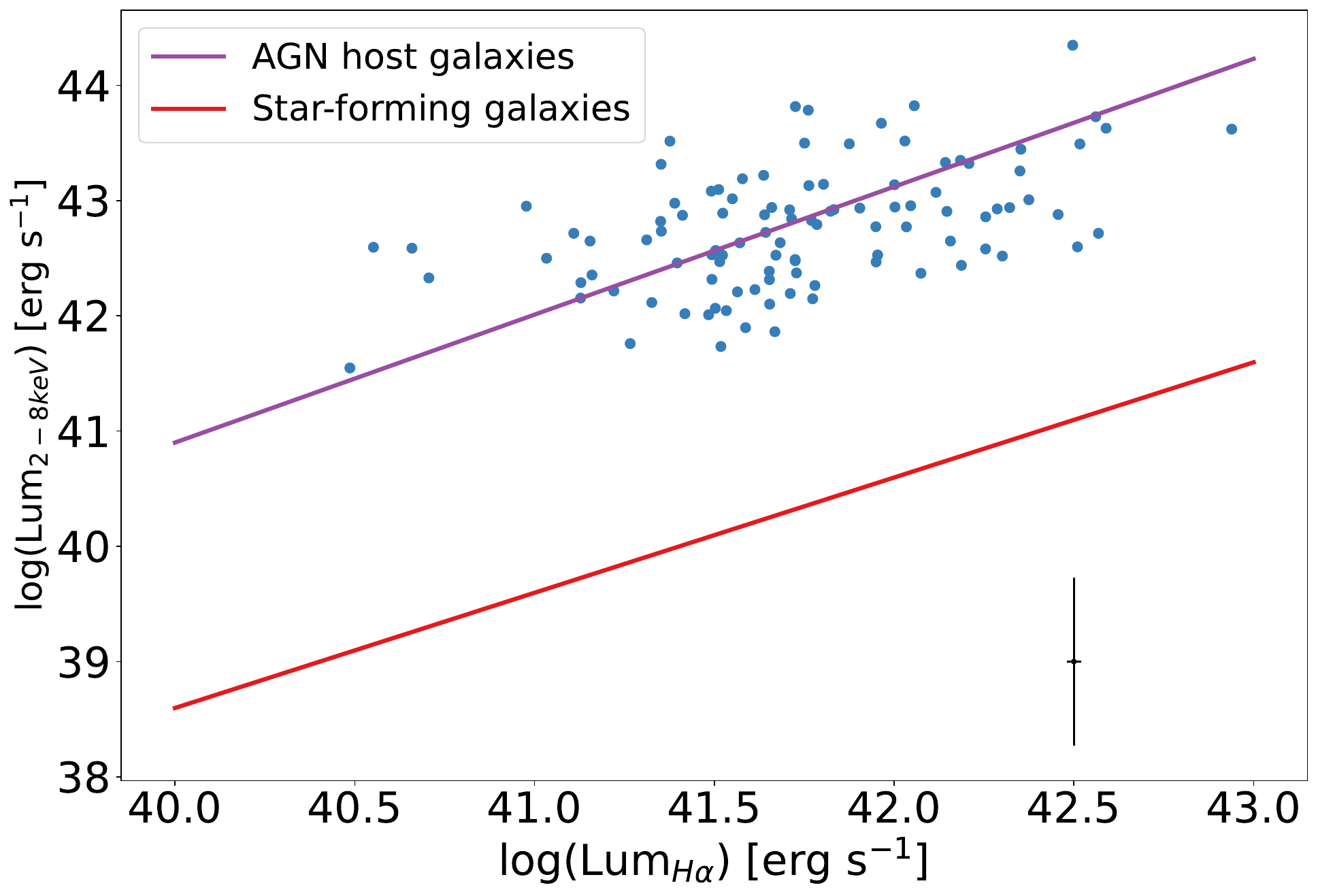}
    \caption{X-ray luminosity (2-8 keV band) and H$_\alpha$ (narrow+broad) luminosity comparison. The purple line plots the relation found by \citet{2001ApJ...549L..51Ho_Xray} for AGN host galaxies, while the red line plots the relation for Star-forming galaxies presented in \citet{2009MNRAS.399..487RG}. Mean errors are shown in the bottom right with error bars.}
    \label{fig:Xray_Halpha}
\end{figure}

\section{Discussion}\label{sec: Discussion}
\subsection{Consistency between different indicators of activity}
\label{sec:consistency}
We consider the NEL diagnostics, the presence of broad permitted lines, and the detection of an X-ray counterpart as indicators of black hole activity. In this section we quantify the candidates that fulfill one or more of the criteria, limiting the discussion to the visually cleaned 415 objects with good quality optical spectra that are indeed low redshift, low stellar-mass galaxies, and that show at least narrow emission lines.  

Regarding classification based on NEL ratios, we note that four candidates without BELs are classified as Seyfert/AGN or LINER types in all three BPT diagrams. One additional candidate without BELs has the same classification in two BPT diagrams but lacks classification in the [O{\sc iii}]/H$_\beta$ vs. [O{\sc i}]/H$_\alpha$ diagram due to the non-detection of the [O{\sc i}] line. Of these, three are classified as S-AGN and two as W-AGN in the WHAN diagram, and hence consistent with black hole activity, of either Seyfert II or LINER type. Three of the five are in the eROSITA-DE sky and one has an X-ray counterpart in eRASSv1.1. The BPT and WHAN classification of the object with X-ray counterpart is highlighted by using a different (square) marker in Figs.~\ref{fig:BPT diagrams} and \ref{fig:WHAN diagram}. The other two of these galaxies were re-observed after they showed alerts in ZTF and therefore became candidates of changing state AGN (CSAGN): SDSS J215055.73-010654.1/ZTF18abtizze by \citet{Lopez22} and J120141.43+382821.5/ZTF18aaqjyon by \citet{Lopez2023CLAGN}, but neither had developed BELs. 
Therefore, they represent either true type II AGN or transient flaring events in AGN that have switched off. Apart from these five galaxies, all objects classified as AGN/Seyfert by all three BPT diagnostics have significant BELs. 

Of the 415 galaxies with fitted spectra, 185 have an AGN/Seyfert, Composite, or LINER classification in all three BPT diagrams. Of these, 181 also have significant BELs. Of the 185 galaxies with consistent AGN classification, 86 broad-line and two narrow-line AGN fall in the eROSITA-DE sky. Of the broad-line objects, 68 have an X-ray counterpart (i.e., 68/86=79\%) while only one of the two narrow-line objects has an X-ray counterpart. We note here that the spectrum of the source with an X-ray counterpart was taken in 2006 and the X-ray observation published in the eRASSv1.1 started on December 13, 2019 \citep{2021A&A...647A...1PeROSITA}. Thus, considering the existence of CSAGNs (e.g., \citealt{Lopez2023CLAGN}), new spectroscopic observations can help to unveil the presence or absence of BELs together with the X-ray emission.

At the other extreme, there are 104 galaxies classified as SF in all three BPT diagrams. Of these, 46 do not have significant broad H$_\alpha$. Of the 58 galaxies with consistent SF classification that do show BELs, 29 are in the eROSITA-DE sky and 23 have X-ray counterparts ( i.e., 23/29=79\%), while none of the ones without BELs have X-ray counterparts. For completeness, we note that there are 118 galaxies with mixed (meaning different classification in one or two BPT diagrams, or no classification in the [O{\sc iii}]/H$_\beta$ vs. [N{\sc ii}]/H$_\alpha$ diagram) BPT classifications, and of these 111 have BELs.

Therefore, the likelihood of finding broad lines is much higher for consistent AGN/Seyfert/Composite classifications (181/185=98\%) than for consistent SF classifications (58/104=56\%), while mixed types also have a higher likelihood (111/118=94\%). 
The probability of finding X-ray counterparts is much higher for galaxies that show BELs: of 169 galaxies with significant BELs in the eROSITA-DE sky 129 have X-ray counterparts (129/169=76\%), while for the 26 galaxies in this region of the sky that do not have BELs only one has an X-ray counterpart (1/26=4\%). 

Finally, the probability of finding X-ray counterparts in SF vs AGN/Seyfert/Composite BPT types only depends on whether they have BELs or not. For galaxies with BELs, the fraction of X-ray matches is 79\% for both, galaxies with consistent SF or with consistent AGN/Seyfert/Composite classifications, while almost all galaxies without broad lines have no X-ray counterparts, evidently regardless of where they appear in the BPTs.

In summary, Fig.~\ref{fig:Comparison_ANG_indicators} presents a Venn diagram illustrating the distribution of 195 VCS objects that are located within the eROSITA-DE sky. These candidates are grouped based on the following criteria: objects with BEL detections, classified as either 'Seyfert' or 'LINER' in at least two BPT diagrams (this criterion was chosen due to the number of candidates showing mixed BPT classifications but displaying BELs); objects with X-ray counterparts; and the various intersections between the categories.

\begin{figure}[htpb]
    \centering
    \includegraphics[width=0.48\textwidth]{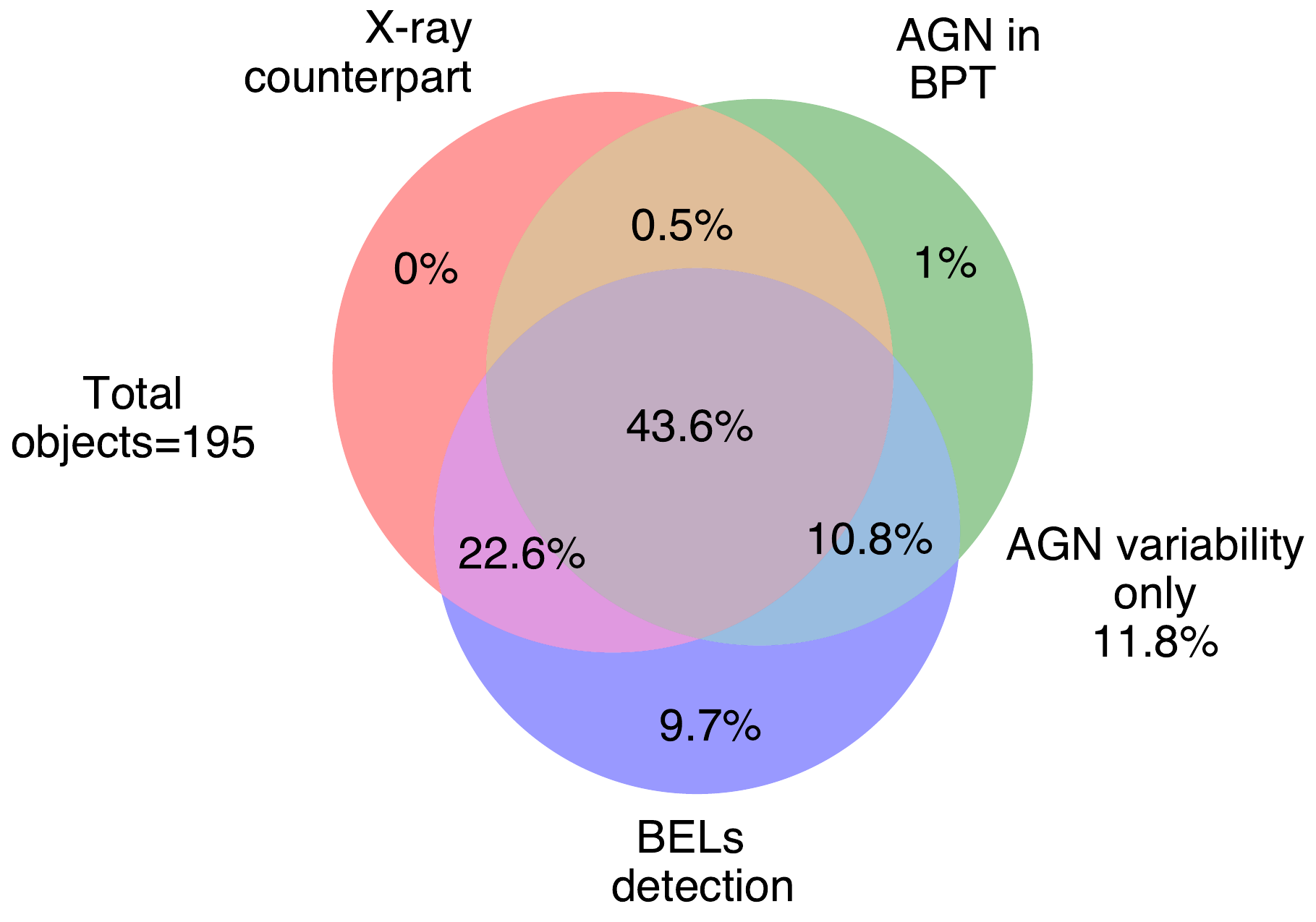}
    
    \caption{Venn diagram of the 195 candidates that are in the VCS sample and that lie in the eROSITA-DE sky. The groups are: objects with BEL detections (purple); objects classified as AGN (Seyfert or LINER) in at least two BPT diagrams (green); objects with X-ray counterparts (red); and the intersections. The "AGN variability only" refers to the candidates in the eROSITA-DE sky falling outside the three mentioned groups.}
    \label{fig:Comparison_ANG_indicators}
\end{figure}

\subsection{Comparison with previous variability-based selections}

\subsubsection{Comparison by optical properties}
\label{Sec:Comparison}
Several works have used optical variability to search for AGN in LSMGs, finding candidates with different methods and validation criteria.
Here, we compare AGN candidates selected by variability, confirmed either through the detection of BELs or by classification as AGNs in BPT diagrams, with large optical samples. Specifically, we relate our findings to the results from \citet{Baldassare18}, \citet{Baldassare2020}, and \citet{Ward2022}.

In \citet{Baldassare18}, AGN candidates were selected from the SDSS spectroscopy within Stripe 82, covering approximately 300 deg$^{2}$. To identify AGN candidates, the full sample of around 28,000 objects was cross-matched with the NSA v1.0.1 catalog, and difference-image photometry light curves were used to assess variability, yielding 135 AGN candidates with redshifts $z<0.15$ and galaxy stellar masses in the range of $2\times10^8<M_*<5\times 10^{11}M_{\odot}$.  
When considering only objects with stellar mass $M_*<10^{10}M_{\odot}$, the number of candidates reduces to 35, of which 10 (29\%) are classified as AGN, 7 (14\%) as composite and 16 (46\%) as star-forming. Of these 35 candidates, 16 (46\%) exhibit H$_{\alpha}$ BELs, with estimated black hole masses ranging from $10^{6.1}M_{\odot}$ to $10^{7.9}M_{\odot}$. In our study, when limiting the host galaxy mass to $M_*<10^{10}M_{\odot}$, the VCS sample includes 161 candidates. Of these, 60 (37\%) are classified as AGN based on BPT diagrams, 24 (15\%) as composite, and 69 (43\%) as star-forming. Among the 161 candidates, 127 show BELs, and the estimated black hole masses range from $10^{5.3}M_{\odot}$ to $10^{8.0}M_{\odot}$. In this comparison, the success rate of confirmed AGNs is higher in our study when considering the detection of BELs in LSMGs ($M_*<10^{10}M_{\odot}$): 46\% in \citet{Baldassare18} vs 79\% in the VCS sample. Further, restricting our sample to the SDSS Stripe 82 sky area we obtain 13 candidates with $M_*<10^{10}M_{\odot}$.
Of these, 12 (92\%) exhibit BELs, 7 (54\%) are classified as AGN by BPT diagrams, and 6 (46\%) as star-forming.

Our sample is more easily compared to the low-redshift sample of \citet{Baldassare2020}, who investigated variability-selected AGNs in low-mass galaxies by analyzing low-amplitude variability in the
r-band light curves from the Palomar Transient Factory (PTF). They identified 417 candidates using the NSA v0 catalog, selecting objects with $z<0.055$ and stellar mass $M_*<10^{12}M_{\odot}$.
When constraining the candidates to $M_*<10^{10}M_{\odot}$, they have 237 candidates, all with SDSS spectral data. Of these 25\% are AGN/Composite, and 12 (5\%) exhibit BELs. 
If we restrict the redshift of our VCS sample to $z<0.055$, and the stellar mass to $M_*<10^{10}M_{\odot}$, we have 55 candidates, of these 25 (45\%) are AGN/Composite, and 30 (54\%) exhibit BELs.

The third set for comparison is the one studied by \citet{Ward2022}. These authors identified 44 AGN candidates selected by optical variability from ZTF light curves in LSMGs from the NSA v1.0.1 catalog, with a stellar mass limit set at $M_{\rm Petrossian}< 3\times 10^{9}M_{\odot}/h^2$. Because of computational limitations, only about a third of the available galaxies were studied (25,714 out of 81,462). These 44 candidates have $z<0.15$ and $M_*<10^{9.75}M_{\odot}$, with 4 (9\%) classified as AGN by BPT diagrams, 4 (9\%) as composite, and 35 (81\%) as star-forming. Six (14\%) of the 44 candidates exhibited BELs and the estimated black hole masses range from $10^{6.3}M_{\odot}$ to $10^{7.5}M_{\odot}$. Applying the same stellar mass restriction, $M_*<10^{9.75}M_{\odot}$, to our VCS sample, we obtain 57 candidates, of which 17 (30\%) are classified as AGN, 5 (9\%) as composite, and 33 (58\%) as star-forming in BPT diagrams, while 37 (65\%) exhibit BELs, with black hole masses estimated to range from $10^{5.3}M_{\odot}$ to $10^{7.7}M_{\odot}$. Comparatively, our success rate in confirming AGNs in LSMGs is significantly higher than that reported in \citet{Ward2022}, particularly regarding the detection of BELs (i.e. 6/44 in \citet{Ward2022}, 37/57 in the present work).

Since the \citet{Ward2022} selection was also based on ZTF light curves in the $g$ band, we directly compare our samples further. Cross-matching the 44 AGN candidates selected from optical variability by \citet{Ward2022} to our AGN candidate list, we find only five matches. These all belong to the Alerts set and the only one of the five that falls inside our forced photometry footprint is also in the Forced Photometry set. All five have significant BELs in our analysis and are classified as QSO/broad-line by the SDSS pipeline, with four of the five carrying the additional label of starburst. Of the 44 AGN candidates identified by \citet{Ward2022}, 18 have $\rm Dec<15.5$ deg and we had produced forced photometry light curves for 11 of these. Our forced photometry classifier returned a classification of AGN for one of them (the one included also in the Alerts set as discussed above), a transient class for another one, and a non-variable galaxy-class for the remaining nine. We checked the SDSS spectra of these 11 galaxies, finding significant BELs in only two, the ones classified as AGN and transient above, while all the others only show narrow emission lines and BPT classifications of either star-forming type (8/9) or composite (1/9). In summary, the AGN candidates identified by both our classifier and \citet{Ward2022} are spectrally confirmed, while the ones identified by \citet{Ward2022} and where our classification based on forced photometry light curves produces a non-variable galaxy-class are not confirmed as AGN by their spectra. The one case where \citet{Ward2022} identified an AGN candidate that we classified as transient and not as AGN with our forced photometry classification is confirmed as an AGN by its spectrum. Therefore, it is possible that our classification has missed some AGN by assigning them a transient class rather than an AGN class.

\subsubsection{Comparison by X-ray detection}

We now consider the detection of X-rays in the AGN candidates as an independent validation criterion, following the work presented by \citet{Arcodia2024}. 
This study reported very low X-ray counterpart match rates among optical variability-selected AGN candidates in LSMGs, notably lower than the rates for our sample (Secs. \ref{sec:xrays} and  \ref{sec:consistency}). We compare our selection to the largest optical sets studied by \citet{Arcodia2024}, namely \citet{Baldassare2020, Kimura2020, Burke2022} and the optically-selected sample by \citet{Ward2022}. As we show below, the lack of X-ray counterparts in some cases is caused by shallow X-ray catalog sensitivity limits, and differences in selection in other cases.

\citet{Burke2022} searched for variable galaxies in Dark Energy Camera images, with a $\sim$7-day cadence and 6-year baseline in three different fields. These data are much deeper and also much more restricted in sky coverage than our sample. They selected low-mass black holes on the characteristic timescales of Damped Random Walk models fitted to the light curves, with shorter timescales expected for low-mass black holes. Of 46 AGN candidates in the eROSITA-DE sky with short variability 
timescales ( i.e., $\log (\tau /{\rm day}) <= 1.5$) 
\citet{Arcodia2024} only found one X-ray counterpart at the depth of eRASSv1.1. These AGN candidates are considerably dimmer than our sample, with a median $g$-band magnitude of 21.3 and a standard deviation of 1.5. Therefore, it is likely that the lack of X-ray counterparts is a product of the limited depth of the X-ray catalog. Indeed, the number of matches in \citet{Arcodia2024} rises to four when the deeper, stacked X-ray catalog is used. 

The sample by \citet{Kimura2020} is even more extreme, with fewer but deeper epochs restricted only to the COSMOS field. In their work, the identification of their variability-selected AGN with X-ray counterparts from deep observations of the Chandra COSMOS Legacy survey is 90\%, although no cross-matches are reported for the eROSITA data, pointing again to a limitation in the depth of eRASSv1.1.  

As mentioned, our sample is more readily comparable to the low-redshift sample of \citet{Baldassare2020}. As in our work, the parent sample was selected from the NSA catalog of local galaxies, and they found 237 AGN candidates in galaxies with stellar mass below $10^{10}M_{\odot}$. Of the 52 candidates in that sample that fall inside the eROSITA-DE sky, \citet{Arcodia2024} only found counterparts for three sources (i.e., 6\%) at the depth of eRASSv1.1. To make a fairer comparison with the sample in \citet{Baldassare2020}, we note that their parent sample was the v0 of the NSA catalog, reaching up to $z=0.055$, whereas we used version 1.0.1, including galaxies up to $z=0.15$. Applying these same cuts, we have 55 AGN candidates. Of these, 42 fall in the eROSITA-DE sky and 16 have X-ray counterparts within $10\arcsec$ (i.e., 38\%), a lower fraction than our total sample but still much higher than the fraction of X-ray matches found for the variability-selected PTF-NSA AGN sample studied by \citet{Arcodia2024}. For completeness, limiting our sample by redshift to $z<0.055$ but not by mass, i.e., allowing candidates up to $M_{*}<2\times10^{10}M_{\odot}$, results in 56 galaxies in the eROSITA-DE sky, of which 26 have X-ray counterparts (i.e., 46\%). This small increase in stellar mass changes the fraction of counterparts noticeably. 
In any case, the random forest algorithms used here, possibly together with the better sampled and higher cadence light curves from ZTF as compared to PTF, are significantly more successful at selecting AGN that do have X-ray counterparts.

Finally, \citet{Ward2022} found 44 AGN candidates selected by optical variability from ZTF light curves in LSMGs of the NSA catalog. Of these 44 candidates, seven were included in the analysis by \citet{Arcodia2024}, with no X-ray counterparts found. As mentioned, cross-matching the 44 AGN candidates selected based on optical variability by \citet{Ward2022} with our AGN candidate list yielded only five matches. Unfortunately, none of these matches fall within the eROSITA-DE footprint. However, based on their optical spectral classification, we regard them as correct AGN identifications.

\section{Conclusions}\label{sec:conclusions}

In this work, we have presented the performance of a variability, color, and morphology based random-forest classifier applied to ZTF data to search for IMBH candidates in low stellar mass galaxies (LSMGs) limited by $M_*<2\times 10^{10}M_\odot$ and redshift $z<0.15$ from the NSA v1.0.1 catalog. Here, four different data sets were used to select candidates, while archival spectra from SDSS were used to confirm the AGN activity in the sample. The analysis above shows that the selection based on the variability properties of type I AGN is effective in the majority of candidates selected from well-sampled light curves. The different selections used different light curves from ZTF data: the alert stream, the complete light curves available through data releases, in $g$ and $r$ band separately, and our custom-made forced photometry on the reference-subtracted images provided by ZTF. The Alerts and Forced Photometry light curves can be more accurate since they are extracted from difference images, while the data release and Forced Photometry light curves are more complete than the Alerts set because they include flux measurements for all epochs regardless of the variability amplitude. Our main conclusions are summarized as follows

   \begin{enumerate}
      \item The cross-match between the variability selection and the NSA v1.0.1 catalog produced 506 AGN candidates in LSMGs. From these, 450 have good quality SDSS archival spectra. After visual inspection, 415 objects were characterized by fitting the spectrum. Subsequently, applying the criteria of an  $EW_{\rm H\alpha}>5\AA$ and the SNR of H$_\alpha$ flux $>$ 3, a total of 357 candidates exhibited a significant BEL. The 357 objects include the two selected by the same criteria but for the H$_\beta$ emission line since the H$_\alpha$ region was missing in the spectra. Additionally, five objects that do not show BELs were classified as AGNs by the BPT and WHAN diagrams, bringing the total number of confirmed AGNs in the VCS sample to 362 out of 415. This means that $87\%$ of the LSMG candidates were confirmed as AGNs.
      Among the 35 objects with spectra that were not fitted, 22 are classified as AGN by the SDSS pipeline or are included in the quasar catalog of \citet{Lyke2020} but are at a higher redshift than expected from the NSA catalog.
      
      \item From the spectral analysis, 58 candidates in the VCS sample were not classified as AGN: 56 are from the Alerts set and two are from the Forced Photometry set. Additionally, eight objects that were not included in the candidates with good spectra, were also not classified as AGN. After visual inspection of the misclassified light curves, it was found that a few are the product of bad image subtraction, and most are produced by transients in the data. These problems in the candidate selection can be addressed by this analysis with the inclusion of the nuclear transient class in the light curve classifier. It is noteworthy that sets using complete light curves are minimally affected by transients or bad image subtraction. When comparing the performance of Forced Photometry and DR, both are equally pure, but Forced Photometry is capable of finding twice as many AGNs in LSMGs in the same sky region, selecting more objects with lower black hole mass.
     
      \item In the Forced Photometry set, 97\% of candidates with good quality spectra show significant BELs. The spectral analysis of these candidates results in equivalent widths between 20 and 800 Å and FWHM $>$ 1500 km/s. This means that all the broad lines are strongly evident and distinguishable from the narrow emission lines in the spectral profile. This strongly implies that the identification of lower-mass black holes ( $M_{\odot}<10^6$  ) is not limited by spectroscopic identification but by the variability selection from the available photometry data. New facilities with more sensitive light curves, such as the Vera C. Rubin LSST, will allow for the discovery of more low-mass black holes from variability selection.
      
      \item The black hole mass estimates were performed for 355 objects using Eq. \ref{eq:M}, with mean values of $1.6\times 10^7M_\odot$ for the Forced Photometry set, $2.1\times 10^7M_\odot$ for the Alerts set, $2.7\times 10^7M_\odot$ for DR-g, and $2.8\times 10^7M_\odot$ for DR-r. Almost all the black hole masses below $3\times 10^6M_\odot$ are in the Alerts and Forced Photometry sets, showing that the photometry performed on reference-subtracted images is more sensitive to lower black hole masses. This is clear when we note that the 10\% of the masses in the Forced Photometry set are below $1.8\times1^6 M_\odot$ while for the DR this limit is at $6.4-6.4 \times 10^6 M_\odot$.
      
      \item The Eddington ratios for the different sets are mainly distributed within the range of $10^{-2}$ and $10^{-1}$, showing that this is the range where the selection method is most sensitive.
      
      \item The classification of the 415 visually cleaned candidates by NEL ratios, i.e., BPT diagrams, results in 185 classified as AGN/Seyfert, composite, or LINER in all three diagrams, while 104 are classified as Star-Forming in all three BPT diagrams. The remaining show mixed classifications as AGN/Seyfert, composite, or LINER among one or more diagrams. Among the objects consistently classified as Star-Forming, 59 objects show a significant BEL, proving their type I AGN nature. As a consequence, methods searching for AGN candidates using the BPT AGN-class as an obligatory criterion will miss these objects.  The comparison with other methods for measuring the NEL ratios suggests that a more careful fitting provides better constraints for the BPT classification and can also improve the methods of AGN candidate selection.
      
      \item Cross-matching the eRASSv1.1 catalog of X-ray sources with the 415 variability-selected AGN candidates with good spectra and $z<0.15$, we find that $67\%$ of the latter have X-ray counterparts when restricting to the eROSITA-DE sky. Candidates that appear in more than one set are more likely to have an X-ray counterpart. Furthermore, we found that $75\%$ of candidates in the eROSITA-DE sky that show significant BELs have an X-ray counterpart. This result considers all BPT classifications, suggesting that the probability of finding an X-ray counterpart only depends on the presence of BELs and not on the BPT class. This also indicates that the SF classification of objects with X-ray counterparts is likely due to the dominant contamination of the host galaxy emission in the optical spectrum. And, as demonstrated by other studies (e.g., \citealt{Mezcua17}, \citealt{2020ApJ...898L..30M}, \citealt{2024MNRAS.528.5252Mezcua}), observations with higher spatial resolution have been shown to effectively resolve the narrow emission line region of the AGN.
      
      \item Variability-selected AGN candidates with higher AGN relative contributions tend to have bluer AGN continua. However, some objects (10 in total) with flat spectral profiles and evident BELs also show AGN relative contributions between 0.51 and 0.76. When comparing different sets, the Alerts and Forced Photometry find more objects with AGN relative contributions lower than 0.4. Additionally, we find no differences in the AGN relative contribution among different BPT-classes. This suggests that emission lines from the host galaxy can contaminate the AGN emission lines even when the AGN has a high relative contribution to the total continuum. Therefore, as mentioned in the previous conclusion, observations with higher spatial resolution are needed to reveal the AGN emission lines component. 
      
      \item  The ratio of the host galaxy stellar mass to the black hole mass for our sample shows a distribution closer to the relation found for massive elliptical galaxies ($M_{BH}=0.001M_*$) in strong contrast to the results of \citet{Reines2015}, for the same stellar mass range. We attribute this to the limitations of the variability selection in identifying black holes with the lowest masses for a given stellar mass host since the more sensitive selection sets ( i.e., Alerts and Forced Photometry) find AGNs with lower $M_{BH}/M_*$ ratios than the other sets. We expect that this limitation might be mitigated with more sensitive and higher SNR data by new observatories like the Vera Rubin LSST.
      
      \item  A comparison of results from different studies utilizing optical variability to find AGN in LSMGs reveals that the method used here produced a much higher fraction of AGN candidates with eROSITA-DE X-ray counterparts. We argue that, at least for works with similar input data, the main difference is the selection method, in our case using RF algorithms that include several classes of variable and non-variable objects.
     
   \end{enumerate}

   The results presented in this work, aimed at searching for IMBH in LSMGs showed that only a few where consistent with a mass lower than $10^6M_\odot$. However, the number of confirmed candidates selected by variability demonstrates the high performance of the applied selection method and offers insights on how to improve it. Notably, the number of candidates studied was made possible by the advantages of photometric and spectroscopic surveys. Therefore, future surveys and new generations of instruments will provide better and significantly more data for similar studies. These advancements, together with the refinement of selection techniques, will increase the number of IMBH detections, helping to reveal the history of AGNs in low stellar mass galaxies.

\section*{Data availability}

Table \ref{tab:results_short} is only available in electronic form at the CDS via anonymous ftp to \href{https://cdsarc.u-strasbg.fr/}{cdsarc.u-strasbg.fr} (130.79.128.5) or via \href{http://cdsweb.u-strasbg.fr/cgi-bin/qcat?J/A+A/}{http://cdsweb.u-strasbg.fr/cgi-bin/qcat?J/A+A/}.
 
\begin{acknowledgements}

      Part of this work was supported by the European Southern Observatory (ESO) project SSDF 28/23D. The authors acknowledge the National Agency for Research and Development (ANID) grants: Programa de Becas/Doctorado Nacional 21212344 (SB); Millennium Science Initiative Programs NCN$2023\_002$ (SB, PA, PL), ICN12\_009 (FEB) and AIM23-0001 (FEB, PA); and Fondecyt Regular 1241422 (PA, FEB, PL), 1240875 (PL, PA), and 1241005 (FEB, PA), and CATA-BASAL FB210003 (FEB). PA acknowledges support from Centro de Astrofísica de Valparaíso - CAV, CIDI N. 21 (Universidad de Valparaíso, Chile).

%SDSS
Funding for the Sloan Digital Sky 
Survey IV has been provided by the 
Alfred P. Sloan Foundation, the U.S. 
Department of Energy Office of 
Science, and the Participating 
Institutions. 

SDSS-IV acknowledges support and 
resources from the Center for High 
Performance Computing  at the 
University of Utah. The SDSS 
website is www.sdss4.org.

SDSS-IV is managed by the 
Astrophysical Research Consortium 
for the Participating Institutions 
of the SDSS Collaboration including 
the Brazilian Participation Group, 
the Carnegie Institution for Science, 
Carnegie Mellon University, Center for 
Astrophysics | Harvard \&
Smithsonian, the Chilean Participation 
Group, the French Participation Group, 
Instituto de Astrof\'isica de 
Canarias, The Johns Hopkins 
University, Kavli Institute for the 
Physics and Mathematics of the 
Universe (IPMU) / University of 
Tokyo, the Korean Participation Group, 
Lawrence Berkeley National Laboratory, 
Leibniz Institut f\"ur Astrophysik 
Potsdam (AIP),  Max-Planck-Institut 
f\"ur Astronomie (MPIA Heidelberg), 
Max-Planck-Institut f\"ur 
Astrophysik (MPA Garching), 
Max-Planck-Institut f\"ur 
Extraterrestrische Physik (MPE), 
National Astronomical Observatories of 
China, New Mexico State University, 
New York University, University of 
Notre Dame, Observat\'ario 
Nacional / MCTI, The Ohio State 
University, Pennsylvania State 
University, Shanghai 
Astronomical Observatory, United 
Kingdom Participation Group, 
Universidad Nacional Aut\'onoma 
de M\'exico, University of Arizona, 
University of Colorado Boulder, 
University of Oxford, University of 
Portsmouth, University of Utah, 
University of Virginia, University 
of Washington, University of 
Wisconsin, Vanderbilt University, 
and Yale University.

%ZTF
Based on observations obtained with the Samuel Oschin 48-inch Telescope at the Palomar Observatory as part of the Zwicky Transient Facility project. ZTF is supported by the National Science Foundation under Grant No. AST-1440341 and a collaboration including Caltech, IPAC, the Weizmann Institute for Science, the Oskar Klein Center at Stockholm University, the University of Maryland, the University of Washington, Deutsches Elektronen-Synchrotron and Humboldt University, Los Alamos National Laboratories, the TANGO Consortium of Taiwan, the University of Wisconsin at Milwaukee, and Lawrence Berkeley National Laboratories. Operations are conducted by COO, IPAC, and UW.

Based on observations obtained with the Samuel Oschin Telescope 48-inch and the 60-inch Telescope at the Palomar Observatory as part of the Zwicky Transient Facility project. ZTF is supported by the National Science Foundation under Grants No. AST-1440341 and AST-2034437 and a collaboration including current partners Caltech, IPAC, the Weizmann Institute for Science, the Oskar Klein Center at Stockholm University, the University of Maryland, Deutsches Elektronen-Synchrotron and Humboldt University, the TANGO Consortium of Taiwan, the University of Wisconsin at Milwaukee, Trinity College Dublin, Lawrence Livermore National Laboratories, IN2P3, University of Warwick, Ruhr University Bochum, Northwestern University and former partners the University of Washington, Los Alamos National Laboratories, and Lawrence Berkeley National Laboratories. Operations are conducted by COO, IPAC, and UW.

%eROSITA
Part of this work is based on data products from  eROSITA aboard SRG, a joint Russian-German science mission supported by the Russian Space Agency (Roskosmos), in the interests of the Russian
Academy of Sciences represented by its Space Research Institute (IKI),
and the Deutsches Zentrum f\"ur Luft- und Raumfahrt (DLR). The SRG
spacecraft was built by Lavochkin Association (NPOL) and its
subcontractors, and is operated by NPOL with support from the Max
Planck Institute for Extraterrestrial Physics (MPE). The development
and construction of the eROSITA X-ray instrument was led by MPE, with
contributions from the Dr. Karl Remeis Observatory Bamberg \& ECAP
(FAU Erlangen-Nuernberg), the University of Hamburg Observatory, the
Leibniz Institute for Astrophysics Potsdam (AIP), and the Institute
for Astronomy and Astrophysics of the University of T\"ubingen, with
the support of DLR and the Max Planck Society. The Argelander
Institute for Astronomy of the University of Bonn and the Ludwig
Maximilians Universit\"at Munich also participated in the science
preparation for eROSITA.

\end{acknowledgements}

% WARNING
%-------------------------------------------------------------------
% Please note that we have included the references to the file aa.dem in
% order to compile it, but we ask you to:
%
% - use BibTeX with the regular commands:
%   \bibliographystyle{aa} % style aa.bst
%   \bibliography{Yourfile} % your references Yourfile.bib
%
% - join the .bib files when you upload your source files
%-------------------------------------------------------------------

\bibliographystyle{aa} 
\bibliography{bibliography.bib} 

\begin{thebibliography}{66}
\expandafter\ifx\csname natexlab\endcsname\relax\def\natexlab#1{#1}\fi

\bibitem[{{Abdurro'uf} {et~al.}(2022){Abdurro'uf}, {Accetta}, {Aerts}, {Silva Aguirre}, {Ahumada}, {Ajgaonkar}, {Filiz Ak}, {Alam}, {Allende Prieto}, {Almeida}, {Anders}, {Anderson}, {Andrews}, {Anguiano}, {Aquino-Ort{\'\i}z}, {Arag{\'o}n-Salamanca}, {Argudo-Fern{\'a}ndez}, {Ata}, {Aubert}, {Avila-Reese}, {Badenes}, {Barb{\'a}}, {Barger}, {Barrera-Ballesteros}, {Beaton}, {Beers}, {Belfiore}, {Bender}, {Bernardi}, {Bershady}, {Beutler}, {Bidin}, {Bird}, {Bizyaev}, {Blanc}, {Blanton}, {Boardman}, {Bolton}, {Boquien}, {Borissova}, {Bovy}, {Brandt}, {Brown}, {Brownstein}, {Brusa}, {Buchner}, {Bundy}, {Burchett}, {Bureau}, {Burgasser}, {Cabang}, {Campbell}, {Cappellari}, {Carlberg}, {Wanderley}, {Carrera}, {Cash}, {Chen}, {Chen}, {Cherinka}, {Chiappini}, {Choi}, {Chojnowski}, {Chung}, {Clerc}, {Cohen}, {Comerford}, {Comparat}, {da Costa}, {Covey}, {Crane}, {Cruz-Gonzalez}, {Culhane}, {Cunha}, {Dai}, {Damke}, {Darling}, {Davidson}, {Davies}, {Dawson}, {De Lee}, {Diamond-Stanic}, {Cano-D{\'\i}az}, {S{\'a}nchez},
  {Donor}, {Duckworth}, {Dwelly}, {Eisenstein}, {Elsworth}, {Emsellem}, {Eracleous}, {Escoffier}, {Fan}, {Farr}, {Feng}, {Fern{\'a}ndez-Trincado}, {Feuillet}, {Filipp}, {Fillingham}, {Frinchaboy}, {Fromenteau}, {Galbany}, {Garc{\'\i}a}, {Garc{\'\i}a-Hern{\'a}ndez}, {Ge}, {Geisler}, {Gelfand}, {G{\'e}ron}, {Gibson}, {Goddy}, {Godoy-Rivera}, {Grabowski}, {Green}, {Greener}, {Grier}, {Griffith}, {Guo}, {Guy}, {Hadjara}, {Harding}, {Hasselquist}, {Hayes}, {Hearty}, {Hern{\'a}ndez}, {Hill}, {Hogg}, {Holtzman}, {Horta}, {Hsieh}, {Hsu}, {Hsu}, {Huber}, {Huertas-Company}, {Hutchinson}, {Hwang}, {Ibarra-Medel}, {Chitham}, {Ilha}, {Imig}, {Jaekle}, {Jayasinghe}, {Ji}, {Johnson}, {Jones}, {J{\"o}nsson}, {Katkov}, {Khalatyan}, {Kinemuchi}, {Kisku}, {Knapen}, {Kneib}, {Kollmeier}, {Kong}, {Kounkel}, {Kreckel}, {Krishnarao}, {Lacerna}, {Lane}, {Langgin}, {Lavender}, {Law}, {Lazarz}, {Leung}, {Leung}, {Lewis}, {Li}, {Li}, {Lian}, {Liang}, {Lin}, {Lin}, {Lin}, {Lintott}, {Long}, {Longa-Pe{\~n}a}, {L{\'o}pez-Cob{\'a}}, {Lu},
  {Lundgren}, {Luo}, {Mackereth}, {de la Macorra}, {Mahadevan}, {Majewski}, {Manchado}, {Mandeville}, {Maraston}, {Margalef-Bentabol}, {Masseron}, {Masters}, {Mathur}, {McDermid}, {Mckay}, {Merloni}, {Merrifield}, {Meszaros}, {Miglio}, {Di Mille}, {Minniti}, {Minsley}, {Monachesi}, {Moon}, {Mosser}, {Mulchaey}, {Muna}, {Mu{\~n}oz}, {Myers}, {Myers}, {Nadathur}, {Nair}, {Nandra}, {Neumann}, {Newman}, {Nidever}, {Nikakhtar}, {Nitschelm}, {O'Connell}, {Garma-Oehmichen}, {Luan Souza de Oliveira}, {Olney}, {Oravetz}, {Ortigoza-Urdaneta}, {Osorio}, {Otter}, {Pace}, {Padilla}, {Pan}, {Pan}, {Parikh}, {Parker}, {Peirani}, {Pe{\~n}a Ram{\'\i}rez}, {Penny}, {Percival}, {Perez-Fournon}, {Pinsonneault}, {Poidevin}, {Poovelil}, {Price-Whelan}, {B{\'a}rbara de Andrade Queiroz}, {Raddick}, {Ray}, {Rembold}, {Riddle}, {Riffel}, {Riffel}, {Rix}, {Robin}, {Rodr{\'\i}guez-Puebla}, {Roman-Lopes}, {Rom{\'a}n-Z{\'u}{\~n}iga}, {Rose}, {Ross}, {Rossi}, {Rubin}, {Salvato}, {S{\'a}nchez}, {S{\'a}nchez-Gallego}, {Sanderson}, {Santana
  Rojas}, {Sarceno}, {Sarmiento}, {Sayres}, {Sazonova}, {Schaefer}, {Schiavon}, {Schlegel}, {Schneider}, {Schultheis}, {Schwope}, {Serenelli}, {Serna}, {Shao}, {Shapiro}, {Sharma}, {Shen}, {Shetrone}, {Shu}, {Simon}, {Skrutskie}, {Smethurst}, {Smith}, {Sobeck}, {Spoo}, {Sprague}, {Stark}, {Stassun}, {Steinmetz}, {Stello}, {Stone-Martinez}, {Storchi-Bergmann}, {Stringfellow}, {Stutz}, {Su}, {Taghizadeh-Popp}, {Talbot}, {Tayar}, {Telles}, {Teske}, {Thakar}, {Theissen}, {Tkachenko}, {Thomas}, {Tojeiro}, {Hernandez Toledo}, {Troup}, {Trump}, {Trussler}, {Turner}, {Tuttle}, {Unda-Sanzana}, {V{\'a}zquez-Mata}, {Valentini}, {Valenzuela}, {Vargas-Gonz{\'a}lez}, {Vargas-Maga{\~n}a}, {Alfaro}, {Villanova}, {Vincenzo}, {Wake}, {Warfield}, {Washington}, {Weaver}, {Weijmans}, {Weinberg}, {Weiss}, {Westfall}, {Wild}, {Wilde}, {Wilson}, {Wilson}, {Wilson}, {Wolf}, {Wood-Vasey}, {Yan}, {Zamora}, {Zasowski}, {Zhang}, {Zhao}, {Zheng}, {Zheng}, \& {Zhu}}]{2022ApJS..259...35AsdssDR17}
{Abdurro'uf}, {Accetta}, K., {Aerts}, C., {et~al.} 2022, \apjs, 259, 35

\bibitem[{{Alexander} \& {Hickox}(2012)}]{2012NewAR..56...93A}
{Alexander}, D.~M. \& {Hickox}, R.~C. 2012, \nar, 56, 93

\bibitem[{{Arcodia} {et~al.}(2024){Arcodia}, {Merloni}, {Comparat}, {Dwelly}, {Seppi}, {Zhang}, {Buchner}, {Georgakakis}, {Haberl}, {Igo}, {Kyritsis}, {Liu}, {Nandra}, {Ni}, {Ponti}, {Salvato}, {Ward}, {Wolf}, \& {Zezas}}]{Arcodia2024}
{Arcodia}, R., {Merloni}, A., {Comparat}, J., {et~al.} 2024, \aap, 681, A97

\bibitem[{{Baldassare} {et~al.}(2018){Baldassare}, {Geha}, \& {Greene}}]{Baldassare18}
{Baldassare}, V.~F., {Geha}, M., \& {Greene}, J. 2018, \apj, 868, 152

\bibitem[{{Baldassare} {et~al.}(2020){Baldassare}, {Geha}, \& {Greene}}]{Baldassare2020}
{Baldassare}, V.~F., {Geha}, M., \& {Greene}, J. 2020, \apj, 896, 10

\bibitem[{Baldwin {et~al.}(1981)Baldwin, Phillips, \& Terlevich}]{baldwin1981classification}
Baldwin, J.~A., Phillips, M.~M., \& Terlevich, R. 1981, Publications of the Astronomical Society of the Pacific, 93, 5

\bibitem[{{Baldwin} {et~al.}(1981){Baldwin}, {Phillips}, \& {Terlevich}}]{BPT_Baldwin81}
{Baldwin}, J.~A., {Phillips}, M.~M., \& {Terlevich}, R. 1981, \pasp, 93, 5

\bibitem[{{Bauer} {et~al.}(2023){Bauer}, {Lira}, {Anguita}, {Arevalo}, {Assef}, {Barrientos}, {Berg}, {Bernal}, {Bian}, {Boquien}, {Buat}, {Chilingarian}, {Coppi}, {De Cicco}, {Diaz}, {Grishin}, {Hernandez-Garcia}, {Kakkad}, {Katkov}, {Krogager}, {L{\'o}pez-Navas}, {Mart{\'\i}nez-Ram{\'\i}rez}, {Mazzucchelli}, {Motta}, {Ricci}, {Ricci}, {Rojas}, {Rouse}, {S{\'a}nchez-S{\'a}ez}, {Toptun}, {Treister}, \& {Vito}}]{Bauer2023}
{Bauer}, F.~E., {Lira}, P., {Anguita}, T., {et~al.} 2023, The Messenger, 190, 34

\bibitem[{{Bellm} {et~al.}(2019){Bellm}, {Kulkarni}, {Graham}, {Dekany}, {Smith}, {Riddle}, {Masci}, {Helou}, {Prince}, {Adams}, {Barbarino}, {Barlow}, {Bauer}, {Beck}, {Belicki}, {Biswas}, {Blagorodnova}, {Bodewits}, {Bolin}, {Brinnel}, {Brooke}, {Bue}, {Bulla}, {Burruss}, {Cenko}, {Chang}, {Connolly}, {Coughlin}, {Cromer}, {Cunningham}, {De}, {Delacroix}, {Desai}, {Duev}, {Eadie}, {Farnham}, {Feeney}, {Feindt}, {Flynn}, {Franckowiak}, {Frederick}, {Fremling}, {Gal-Yam}, {Gezari}, {Giomi}, {Goldstein}, {Golkhou}, {Goobar}, {Groom}, {Hacopians}, {Hale}, {Henning}, {Ho}, {Hover}, {Howell}, {Hung}, {Huppenkothen}, {Imel}, {Ip}, {Ivezi{\'c}}, {Jackson}, {Jones}, {Juric}, {Kasliwal}, {Kaspi}, {Kaye}, {Kelley}, {Kowalski}, {Kramer}, {Kupfer}, {Landry}, {Laher}, {Lee}, {Lin}, {Lin}, {Lunnan}, {Giomi}, {Mahabal}, {Mao}, {Miller}, {Monkewitz}, {Murphy}, {Ngeow}, {Nordin}, {Nugent}, {Ofek}, {Patterson}, {Penprase}, {Porter}, {Rauch}, {Rebbapragada}, {Reiley}, {Rigault}, {Rodriguez}, {van Roestel}, {Rusholme}, {van
  Santen}, {Schulze}, {Shupe}, {Singer}, {Soumagnac}, {Stein}, {Surace}, {Sollerman}, {Szkody}, {Taddia}, {Terek}, {Van Sistine}, {van Velzen}, {Vestrand}, {Walters}, {Ward}, {Ye}, {Yu}, {Yan}, \& {Zolkower}}]{Bellm19}
{Bellm}, E.~C., {Kulkarni}, S.~R., {Graham}, M.~J., {et~al.} 2019, \pasp, 131, 018002

\bibitem[{{Birchall} {et~al.}(2020){Birchall}, {Watson}, \& {Aird}}]{2020MNRAS.492.2268Birchall}
{Birchall}, K.~L., {Watson}, M.~G., \& {Aird}, J. 2020, \mnras, 492, 2268

\bibitem[{{Bolton} {et~al.}(2012){Bolton}, {Schlegel}, {Aubourg}, {Bailey}, {Bhardwaj}, {Brownstein}, {Burles}, {Chen}, {Dawson}, {Eisenstein}, {Gunn}, {Knapp}, {Loomis}, {Lupton}, {Maraston}, {Muna}, {Myers}, {Olmstead}, {Padmanabhan}, {P{\^a}ris}, {Percival}, {Petitjean}, {Rockosi}, {Ross}, {Schneider}, {Shu}, {Strauss}, {Thomas}, {Tremonti}, {Wake}, {Weaver}, \& {Wood-Vasey}}]{2012AJ....144..144B}
{Bolton}, A.~S., {Schlegel}, D.~J., {Aubourg}, {\'E}., {et~al.} 2012, \aj, 144, 144

\bibitem[{{Brinchmann} {et~al.}(2004){Brinchmann}, {Charlot}, {White}, {Tremonti}, {Kauffmann}, {Heckman}, \& {Brinkmann}}]{2004MNRAS.351.1151B}
{Brinchmann}, J., {Charlot}, S., {White}, S.~D.~M., {et~al.} 2004, \mnras, 351, 1151

\bibitem[{{Buchner} {et~al.}(2024){Buchner}, {Starck}, {Salvato}, {Netzer}, {Igo}, {Laloux}, {Georgakakis}, {Gauger}, {Olechowska}, {Lopez}, {Shankar}, {Li}, {Nandra}, \& {Merloni}}]{Buchner2024}
{Buchner}, J., {Starck}, H., {Salvato}, M., {et~al.} 2024, arXiv e-prints, arXiv:2405.19297

\bibitem[{{Burke} {et~al.}(2022){Burke}, {Liu}, {Shen}, {Phadke}, {Yang}, {Hartley}, {Harrison}, {Palmese}, {Guo}, {Zhang}, {Kron}, {Turner}, {Giles}, {Lidman}, {Chen}, {Gruendl}, {Choi}, {Amon}, {Sheldon}, {Aguena}, {Allam}, {Andrade-Oliveira}, {Bacon}, {Bertin}, {Brooks}, {Rosell}, {Kind}, {Carretero}, {Conselice}, {Costanzi}, {da Costa}, {Pereira}, {Davis}, {De Vicente}, {Desai}, {Diehl}, {Everett}, {Ferrero}, {Flaugher}, {Garc{\'\i}a-Bellido}, {Gaztanaga}, {Gruen}, {Gschwend}, {Gutierrez}, {Hinton}, {Hollowood}, {Honscheid}, {Hoyle}, {James}, {Kuehn}, {Maia}, {Marshall}, {Menanteau}, {Miquel}, {Morgan}, {Paz-Chinch{\'o}n}, {Pieres}, {Malag{\'o}n}, {Reil}, {Romer}, {Sanchez}, {Schubnell}, {Serrano}, {Sevilla-Noarbe}, {Smith}, {Suchyta}, {Tarle}, {Thomas}, {To}, {Varga}, {Wilkinson}, \& {DES Collaboration}}]{Burke2022}
{Burke}, C.~J., {Liu}, X., {Shen}, Y., {et~al.} 2022, \mnras, 516, 2736

\bibitem[{{Cann} {et~al.}(2019){Cann}, {Satyapal}, {Abel}, {Blecha}, {Mushotzky}, {Reynolds}, \& {Secrest}}]{2019ApJ...870L...2Cann}
{Cann}, J.~M., {Satyapal}, S., {Abel}, N.~P., {et~al.} 2019, \apjl, 870, L2

\bibitem[{{Cappellari}(2017)}]{pPXF2017MNRAS.466..798C}
{Cappellari}, M. 2017, \mnras, 466, 798

\bibitem[{{Cappellari} \& {Emsellem}(2004)}]{pPXF2004PASP..116..138C}
{Cappellari}, M. \& {Emsellem}, E. 2004, \pasp, 116, 138

\bibitem[{{Cid Fernandes} {et~al.}(2011){Cid Fernandes}, {Stasi{\'n}ska}, {Mateus}, \& {Vale Asari}}]{Cid2011}
{Cid Fernandes}, R., {Stasi{\'n}ska}, G., {Mateus}, A., \& {Vale Asari}, N. 2011, \mnras, 413, 1687

\bibitem[{{de Jong} {et~al.}(2022){de Jong}, {Bellido-Tirado}, {Brynnel}, {Ezzati Amini}, {Frey}, {F{\"u}{\ss}lein}, {G{\"a}bler}, {Giannone}, {Johl}, {Kuba}, {Lemke}, {Micheva}, {Saviauk}, {Steinmetz}, {Walcher}, {Winkler}, {Lind}, {Loveday}, {Feltzing}, {McMahon}, {Mainieri}, {Pirard}, {Bensby}, {Bergemann}, {Chiappini}, {Christlieb}, {Cioni}, {Comparat}, {Driver}, {Hook}, {Irwin}, {Kneib}, {Liske}, {Merloni}, {Minchev}, {Richard}, {Starkenburg}, {Sullivan}, {Worley}, {Gaessler}, {Laurent}, {Pragt}, {Remillieux}, {Rothmaier}, {Smedley}, {Stilz}, {Walton}, {Alexander}, {Church}, {Croom}, {Davies}, {Heneka}, {Kacharov}, {Knoche}, {Kordopatis}, {Krumpe}, {Martell}, {Norberg}, {Pelisoli}, {Sharma}, {Storm}, \& {Tempel}}]{2022SPIE12184E..14D}
{de Jong}, R.~S., {Bellido-Tirado}, O., {Brynnel}, J.~G., {et~al.} 2022, in Society of Photo-Optical Instrumentation Engineers (SPIE) Conference Series, Vol. 12184, Ground-based and Airborne Instrumentation for Astronomy IX, ed. C.~J. {Evans}, J.~J. {Bryant}, \& K.~{Motohara}, 1218414

\bibitem[{{Edri} {et~al.}(2012){Edri}, {Rafter}, {Chelouche}, {Kaspi}, \& {Behar}}]{2012ApJ...756...73EmassNGC}
{Edri}, H., {Rafter}, S.~E., {Chelouche}, D., {Kaspi}, S., \& {Behar}, E. 2012, \apj, 756, 73

\bibitem[{{F{\"o}rster} {et~al.}(2021){F{\"o}rster}, {Cabrera-Vives}, {Castillo-Navarrete}, {Est{\'e}vez}, {S{\'a}nchez-S{\'a}ez}, {Arredondo}, {Bauer}, {Carrasco-Davis}, {Catelan}, {Elorrieta}, {Eyheramendy}, {Huijse}, {Pignata}, {Reyes}, {Reyes}, {Rodr{\'\i}guez-Mancini}, {Ruz-Mieres}, {Valenzuela}, {{\'A}lvarez-Maldonado}, {Astorga}, {Borissova}, {Clocchiatti}, {De Cicco}, {Donoso-Oliva}, {Hern{\'a}ndez-Garc{\'\i}a}, {Graham}, {Jord{\'a}n}, {Kurtev}, {Mahabal}, {Maureira}, {Mu{\~n}oz-Arancibia}, {Molina-Ferreiro}, {Moya}, {Palma}, {P{\'e}rez-Carrasco}, {Protopapas}, {Romero}, {Sabatini-Gacitua}, {S{\'a}nchez}, {San Mart{\'\i}n}, {Sep{\'u}lveda-Cobo}, {Vera}, \& {Vergara}}]{Forster21}
{F{\"o}rster}, F., {Cabrera-Vives}, G., {Castillo-Navarrete}, E., {et~al.} 2021, \aj, 161, 242

\bibitem[{{Gilmozzi} \& {Spyromilio}(2007)}]{2007Msngr.127...11GELT}
{Gilmozzi}, R. \& {Spyromilio}, J. 2007, The Messenger, 127, 11

\bibitem[{{Greene}(2012)}]{2012NatCo...3.1304G}
{Greene}, J.~E. 2012, Nature Communications, 3, 1304

\bibitem[{{Ho} {et~al.}(2001){Ho}, {Feigelson}, {Townsley}, {Sambruna}, {Garmire}, {Brandt}, {Filippenko}, {Griffiths}, {Ptak}, \& {Sargent}}]{2001ApJ...549L..51Ho_Xray}
{Ho}, L.~C., {Feigelson}, E.~D., {Townsley}, L.~K., {et~al.} 2001, \apjl, 549, L51

\bibitem[{{Hviding} {et~al.}(2024){Hviding}, {Hainline}, {Goulding}, \& {Greene}}]{2024AJ....167..169H}
{Hviding}, R.~E., {Hainline}, K.~N., {Goulding}, A.~D., \& {Greene}, J.~E. 2024, \aj, 167, 169

\bibitem[{{Inayoshi} {et~al.}(2020){Inayoshi}, {Visbal}, \& {Haiman}}]{2020ARA&A..58...27IBHseeds}
{Inayoshi}, K., {Visbal}, E., \& {Haiman}, Z. 2020, \araa, 58, 27

\bibitem[{{Ivezi{\'c}} {et~al.}(2019){Ivezi{\'c}}, {Kahn}, {Tyson}, {Abel}, {Acosta}, {Allsman}, {Alonso}, {AlSayyad}, {Anderson}, {Andrew}, {Angel}, {Angeli}, {Ansari}, {Antilogus}, {Araujo}, {Armstrong}, {Arndt}, {Astier}, {Aubourg}, {Auza}, {Axelrod}, {Bard}, {Barr}, {Barrau}, {Bartlett}, {Bauer}, {Bauman}, {Baumont}, {Bechtol}, {Bechtol}, {Becker}, {Becla}, {Beldica}, {Bellavia}, {Bianco}, {Biswas}, {Blanc}, {Blazek}, {Blandford}, {Bloom}, {Bogart}, {Bond}, {Booth}, {Borgland}, {Borne}, {Bosch}, {Boutigny}, {Brackett}, {Bradshaw}, {Brandt}, {Brown}, {Bullock}, {Burchat}, {Burke}, {Cagnoli}, {Calabrese}, {Callahan}, {Callen}, {Carlin}, {Carlson}, {Chandrasekharan}, {Charles-Emerson}, {Chesley}, {Cheu}, {Chiang}, {Chiang}, {Chirino}, {Chow}, {Ciardi}, {Claver}, {Cohen-Tanugi}, {Cockrum}, {Coles}, {Connolly}, {Cook}, {Cooray}, {Covey}, {Cribbs}, {Cui}, {Cutri}, {Daly}, {Daniel}, {Daruich}, {Daubard}, {Daues}, {Dawson}, {Delgado}, {Dellapenna}, {de Peyster}, {de Val-Borro}, {Digel}, {Doherty}, {Dubois},
  {Dubois-Felsmann}, {Durech}, {Economou}, {Eifler}, {Eracleous}, {Emmons}, {Fausti Neto}, {Ferguson}, {Figueroa}, {Fisher-Levine}, {Focke}, {Foss}, {Frank}, {Freemon}, {Gangler}, {Gawiser}, {Geary}, {Gee}, {Geha}, {Gessner}, {Gibson}, {Gilmore}, {Glanzman}, {Glick}, {Goldina}, {Goldstein}, {Goodenow}, {Graham}, {Gressler}, {Gris}, {Guy}, {Guyonnet}, {Haller}, {Harris}, {Hascall}, {Haupt}, {Hernandez}, {Herrmann}, {Hileman}, {Hoblitt}, {Hodgson}, {Hogan}, {Howard}, {Huang}, {Huffer}, {Ingraham}, {Innes}, {Jacoby}, {Jain}, {Jammes}, {Jee}, {Jenness}, {Jernigan}, {Jevremovi{\'c}}, {Johns}, {Johnson}, {Johnson}, {Jones}, {Juramy-Gilles}, {Juri{\'c}}, {Kalirai}, {Kallivayalil}, {Kalmbach}, {Kantor}, {Karst}, {Kasliwal}, {Kelly}, {Kessler}, {Kinnison}, {Kirkby}, {Knox}, {Kotov}, {Krabbendam}, {Krughoff}, {Kub{\'a}nek}, {Kuczewski}, {Kulkarni}, {Ku}, {Kurita}, {Lage}, {Lambert}, {Lange}, {Langton}, {Le Guillou}, {Levine}, {Liang}, {Lim}, {Lintott}, {Long}, {Lopez}, {Lotz}, {Lupton}, {Lust}, {MacArthur}, {Mahabal},
  {Mandelbaum}, {Markiewicz}, {Marsh}, {Marshall}, {Marshall}, {May}, {McKercher}, {McQueen}, {Meyers}, {Migliore}, {Miller}, {Mills}, {Miraval}, {Moeyens}, {Moolekamp}, {Monet}, {Moniez}, {Monkewitz}, {Montgomery}, {Morrison}, {Mueller}, {Muller}, {Mu{\~n}oz Arancibia}, {Neill}, {Newbry}, {Nief}, {Nomerotski}, {Nordby}, {O'Connor}, {Oliver}, {Olivier}, {Olsen}, {O'Mullane}, {Ortiz}, {Osier}, {Owen}, {Pain}, {Palecek}, {Parejko}, {Parsons}, {Pease}, {Peterson}, {Peterson}, {Petravick}, {Libby Petrick}, {Petry}, {Pierfederici}, {Pietrowicz}, {Pike}, {Pinto}, {Plante}, {Plate}, {Plutchak}, {Price}, {Prouza}, {Radeka}, {Rajagopal}, {Rasmussen}, {Regnault}, {Reil}, {Reiss}, {Reuter}, {Ridgway}, {Riot}, {Ritz}, {Robinson}, {Roby}, {Roodman}, {Rosing}, {Roucelle}, {Rumore}, {Russo}, {Saha}, {Sassolas}, {Schalk}, {Schellart}, {Schindler}, {Schmidt}, {Schneider}, {Schneider}, {Schoening}, {Schumacher}, {Schwamb}, {Sebag}, {Selvy}, {Sembroski}, {Seppala}, {Serio}, {Serrano}, {Shaw}, {Shipsey}, {Sick}, {Silvestri},
  {Slater}, {Smith}, {Smith}, {Sobhani}, {Soldahl}, {Storrie-Lombardi}, {Stover}, {Strauss}, {Street}, {Stubbs}, {Sullivan}, {Sweeney}, {Swinbank}, {Szalay}, {Takacs}, {Tether}, {Thaler}, {Thayer}, {Thomas}, {Thornton}, {Thukral}, {Tice}, {Trilling}, {Turri}, {Van Berg}, {Vanden Berk}, {Vetter}, {Virieux}, {Vucina}, {Wahl}, {Walkowicz}, {Walsh}, {Walter}, {Wang}, {Wang}, {Warner}, {Wiecha}, {Willman}, {Winters}, {Wittman}, {Wolff}, {Wood-Vasey}, {Wu}, {Xin}, {Yoachim}, \& {Zhan}}]{Ivezic2019}
{Ivezi{\'c}}, {\v{Z}}., {Kahn}, S.~M., {Tyson}, J.~A., {et~al.} 2019, \apj, 873, 111

\bibitem[{{Kauffmann} {et~al.}(2003{\natexlab{a}}){Kauffmann}, {Heckman}, {Tremonti}, {Brinchmann}, {Charlot}, {White}, {Ridgway}, {Brinkmann}, {Fukugita}, {Hall}, {Ivezi{\'c}}, {Richards}, \& {Schneider}}]{Kauffmann2003MNRAS.346.1055K}
{Kauffmann}, G., {Heckman}, T.~M., {Tremonti}, C., {et~al.} 2003{\natexlab{a}}, \mnras, 346, 1055

\bibitem[{{Kauffmann} {et~al.}(2003{\natexlab{b}}){Kauffmann}, {Heckman}, {White}, {Charlot}, {Tremonti}, {Brinchmann}, {Bruzual}, {Peng}, {Seibert}, {Bernardi}, {Blanton}, {Brinkmann}, {Castander}, {Cs{\'a}bai}, {Fukugita}, {Ivezic}, {Munn}, {Nichol}, {Padmanabhan}, {Thakar}, {Weinberg}, \& {York}}]{2003MNRAS.341...33K}
{Kauffmann}, G., {Heckman}, T.~M., {White}, S. D.~M., {et~al.} 2003{\natexlab{b}}, \mnras, 341, 33

\bibitem[{{Kewley} {et~al.}(2001){Kewley}, {Dopita}, {Sutherland}, {Heisler}, \& {Trevena}}]{Kewley2001ApJ...556..121K}
{Kewley}, L.~J., {Dopita}, M.~A., {Sutherland}, R.~S., {Heisler}, C.~A., \& {Trevena}, J. 2001, \apj, 556, 121

\bibitem[{{Kewley} {et~al.}(2006){Kewley}, {Groves}, {Kauffmann}, \& {Heckman}}]{Kewley2006MNRAS.372..961K}
{Kewley}, L.~J., {Groves}, B., {Kauffmann}, G., \& {Heckman}, T. 2006, \mnras, 372, 961

\bibitem[{{Kimura} {et~al.}(2020){Kimura}, {Yamada}, {Kokubo}, {Yasuda}, {Morokuma}, {Nagao}, \& {Matsuoka}}]{Kimura2020}
{Kimura}, Y., {Yamada}, T., {Kokubo}, M., {et~al.} 2020, \apj, 894, 24

\bibitem[{{Kormendy} \& {Ho}(2013)}]{2013ARA&A..51..511K}
{Kormendy}, J. \& {Ho}, L.~C. 2013, \araa, 51, 511

\bibitem[{{Latif} \& {Ferrara}(2016)}]{BHseeds2016PASA...33...51L}
{Latif}, M.~A. \& {Ferrara}, A. 2016, \pasa, 33, e051

\bibitem[{{L{\'o}pez-Navas} {et~al.}(2022){L{\'o}pez-Navas}, {Mart{\'\i}nez-Aldama}, {Bernal}, {S{\'a}nchez-S{\'a}ez}, {Ar{\'e}valo}, {Graham}, {Hern{\'a}ndez-Garc{\'\i}a}, {Lira}, \& {Rojas Lobos}}]{Lopez22}
{L{\'o}pez-Navas}, E., {Mart{\'\i}nez-Aldama}, M.~L., {Bernal}, S., {et~al.} 2022, \mnras, 513, L57

\bibitem[{{L{\'o}pez-Navas} {et~al.}(2023){L{\'o}pez-Navas}, {S{\'a}nchez-S{\'a}ez}, {Ar{\'e}valo}, {Bernal}, {Graham}, {Hern{\'a}ndez-Garc{\'\i}a}, {Homan}, {Krumpe}, {Lamer}, {Lira}, {Mart{\'\i}nez-Aldama}, {Merloni}, {R{\'\i}os}, {Salvato}, {Stern}, \& {Tub{\'\i}n-Arenas}}]{Lopez2023CLAGN}
{L{\'o}pez-Navas}, E., {S{\'a}nchez-S{\'a}ez}, P., {Ar{\'e}valo}, P., {et~al.} 2023, \mnras, 524, 188

\bibitem[{{Lyke} {et~al.}(2020){Lyke}, {Higley}, {McLane}, {Schurhammer}, {Myers}, {Ross}, {Dawson}, {Chabanier}, {Martini}, {Busca}, {Mas des Bourboux}, {Salvato}, {Streblyanska}, {Zarrouk}, {Burtin}, {Anderson}, {Bautista}, {Bizyaev}, {Brandt}, {Brinkmann}, {Brownstein}, {Comparat}, {Green}, {de la Macorra}, {Mu{\~n}oz Guti{\'e}rrez}, {Hou}, {Newman}, {Palanque-Delabrouille}, {P{\^a}ris}, {Percival}, {Petitjean}, {Rich}, {Rossi}, {Schneider}, {Smith}, {Vivek}, \& {Weaver}}]{Lyke2020}
{Lyke}, B.~W., {Higley}, A.~N., {McLane}, J.~N., {et~al.} 2020, \apjs, 250, 8

\bibitem[{{MacLeod} {et~al.}(2010){MacLeod}, {Ivezi{\'c}}, {Kochanek}, {Koz{\l}owski}, {Kelly}, {Bullock}, {Kimball}, {Sesar}, {Westman}, {Brooks}, {Gibson}, {Becker}, \& {de Vries}}]{2010ApJ...721.1014MacLeod_photoVar}
{MacLeod}, C.~L., {Ivezi{\'c}}, {\v{Z}}., {Kochanek}, C.~S., {et~al.} 2010, \apj, 721, 1014

\bibitem[{{Mart{\'\i}nez-Palomera} {et~al.}(2020){Mart{\'\i}nez-Palomera}, {Lira}, {Bhalla-Ladd}, {F{\"o}rster}, \& {Plotkin}}]{2020ApJ...889..113MP_IMBH_photo_var}
{Mart{\'\i}nez-Palomera}, J., {Lira}, P., {Bhalla-Ladd}, I., {F{\"o}rster}, F., \& {Plotkin}, R.~M. 2020, \apj, 889, 113

\bibitem[{{Masci} {et~al.}(2023){Masci}, {Laher}, {Rusholme}, {Shupe}, {Paladini}, {Groom}, {Wold}, {Miller}, \& {Drake}}]{Masci2023}
{Masci}, F.~J., {Laher}, R.~R., {Rusholme}, B., {et~al.} 2023, arXiv e-prints, arXiv:2305.16279

\bibitem[{{Masci} {et~al.}(2019){Masci}, {Laher}, {Rusholme}, {Shupe}, {Groom}, {Surace}, {Jackson}, {Monkewitz}, {Beck}, {Flynn}, {Terek}, {Landry}, {Hacopians}, {Desai}, {Howell}, {Brooke}, {Imel}, {Wachter}, {Ye}, {Lin}, {Cenko}, {Cunningham}, {Rebbapragada}, {Bue}, {Miller}, {Mahabal}, {Bellm}, {Patterson}, {Juri{\'c}}, {Golkhou}, {Ofek}, {Walters}, {Graham}, {Kasliwal}, {Dekany}, {Kupfer}, {Burdge}, {Cannella}, {Barlow}, {Van Sistine}, {Giomi}, {Fremling}, {Blagorodnova}, {Levitan}, {Riddle}, {Smith}, {Helou}, {Prince}, \& {Kulkarni}}]{masci18}
{Masci}, F.~J., {Laher}, R.~R., {Rusholme}, B., {et~al.} 2019, \pasp, 131, 018003

\bibitem[{{Mej{\'{\i}}a-Restrepo} {et~al.}(2016){Mej{\'{\i}}a-Restrepo}, {Trakhtenbrot}, {Lira}, {Netzer}, \& {Capellupo}}]{MejiaRestrepo16}
{Mej{\'{\i}}a-Restrepo}, J.~E., {Trakhtenbrot}, B., {Lira}, P., {Netzer}, H., \& {Capellupo}, D.~M. 2016, \mnras, 460, 187

\bibitem[{{Merloni} {et~al.}(2024){Merloni}, {Lamer}, {Liu}, {Ramos-Ceja}, {Brunner}, {Bulbul}, {Dennerl}, {Doroshenko}, {Freyberg}, {Friedrich}, {Gatuzz}, {Georgakakis}, {Haberl}, {Igo}, {Kreykenbohm}, {Liu}, {Maitra}, {Malyali}, {Mayer}, {Nandra}, {Predehl}, {Robrade}, {Salvato}, {Sanders}, {Stewart}, {Tub{\'\i}n-Arenas}, {Weber}, {Wilms}, {Arcodia}, {Artis}, {Aschersleben}, {Avakyan}, {Aydar}, {Bahar}, {Balzer}, {Becker}, {Berger}, {Boller}, {Bornemann}, {Br{\"u}ggen}, {Brusa}, {Buchner}, {Burwitz}, {Camilloni}, {Clerc}, {Comparat}, {Coutinho}, {Czesla}, {Dannhauer}, {Dauner}, {Dauser}, {Dietl}, {Dolag}, {Dwelly}, {Egg}, {Ehl}, {Freund}, {Friedrich}, {Gaida}, {Garrel}, {Ghirardini}, {Gokus}, {Gr{\"u}nwald}, {Grandis}, {Grotova}, {Gruen}, {Gueguen}, {H{\"a}mmerich}, {Hamaus}, {Hasinger}, {Haubner}, {Homan}, {Ider Chitham}, {Joseph}, {Joyce}, {K{\"o}nig}, {Kaltenbrunner}, {Khokhriakova}, {Kink}, {Kirsch}, {Kluge}, {Knies}, {Krippendorf}, {Krumpe}, {Kurpas}, {Li}, {Liu}, {Locatelli}, {Lorenz}, {M{\"u}ller},
  {Magaudda}, {Mannes}, {McCall}, {Meidinger}, {Michailidis}, {Migkas}, {Mu{\~n}oz-Giraldo}, {Musiimenta}, {Nguyen-Dang}, {Ni}, {Olechowska}, {Ota}, {Pacaud}, {Pasini}, {Perinati}, {Pires}, {Pommranz}, {Ponti}, {Poppenhaeger}, {P{\"u}hlhofer}, {Rau}, {Reh}, {Reiprich}, {Roster}, {Saeedi}, {Santangelo}, {Sasaki}, {Schmitt}, {Schneider}, {Schrabback}, {Schuster}, {Schwope}, {Seppi}, {Serim}, {Shreeram}, {Sokolova-Lapa}, {Starck}, {Stelzer}, {Stierhof}, {Suleimanov}, {Tenzer}, {Traulsen}, {Tr{\"u}mper}, {Tsuge}, {Urrutia}, {Veronica}, {Waddell}, {Willer}, {Wolf}, {Yeung}, {Zainab}, {Zangrandi}, {Zhang}, {Zhang}, \& {Zheng}}]{Merloni2024}
{Merloni}, A., {Lamer}, G., {Liu}, T., {et~al.} 2024, \aap, 682, A34

\bibitem[{{Mezcua}(2017)}]{Mezcua17}
{Mezcua}, M. 2017, International Journal of Modern Physics D, 26, 1730021

\bibitem[{{Mezcua} \& {Dom{\'\i}nguez S{\'a}nchez}(2020)}]{2020ApJ...898L..30M}
{Mezcua}, M. \& {Dom{\'\i}nguez S{\'a}nchez}, H. 2020, \apjl, 898, L30

\bibitem[{{Mezcua} \& {Dom{\'\i}nguez S{\'a}nchez}(2024)}]{2024MNRAS.528.5252Mezcua}
{Mezcua}, M. \& {Dom{\'\i}nguez S{\'a}nchez}, H. 2024, \mnras, 528, 5252

\bibitem[{{Moran} {et~al.}(2014){Moran}, {Shahinyan}, {Sugarman}, {V{\'e}lez}, \& {Eracleous}}]{2014AJ....148..136Moran}
{Moran}, E.~C., {Shahinyan}, K., {Sugarman}, H.~R., {V{\'e}lez}, D.~O., \& {Eracleous}, M. 2014, \aj, 148, 136

\bibitem[{{Panessa} {et~al.}(2006){Panessa}, {Bassani}, {Cappi}, {Dadina}, {Barcons}, {Carrera}, {Ho}, \& {Iwasawa}}]{2006A&A...455..173Panessa_X-ray}
{Panessa}, F., {Bassani}, L., {Cappi}, M., {et~al.} 2006, \aap, 455, 173

\bibitem[{{Pounds} {et~al.}(1994){Pounds}, {Nandra}, {Fink}, \& {Makino}}]{1994MNRAS.267..193P}
{Pounds}, K.~A., {Nandra}, K., {Fink}, H.~H., \& {Makino}, F. 1994, \mnras, 267, 193

\bibitem[{{Predehl} {et~al.}(2021){Predehl}, {Andritschke}, {Arefiev}, {Babyshkin}, {Batanov}, {Becker}, {B{\"o}hringer}, {Bogomolov}, {Boller}, {Borm}, {Bornemann}, {Br{\"a}uninger}, {Br{\"u}ggen}, {Brunner}, {Brusa}, {Bulbul}, {Buntov}, {Burwitz}, {Burkert}, {Clerc}, {Churazov}, {Coutinho}, {Dauser}, {Dennerl}, {Doroshenko}, {Eder}, {Emberger}, {Eraerds}, {Finoguenov}, {Freyberg}, {Friedrich}, {Friedrich}, {F{\"u}rmetz}, {Georgakakis}, {Gilfanov}, {Granato}, {Grossberger}, {Gueguen}, {Gureev}, {Haberl}, {H{\"a}lker}, {Hartner}, {Hasinger}, {Huber}, {Ji}, {Kienlin}, {Kink}, {Korotkov}, {Kreykenbohm}, {Lamer}, {Lomakin}, {Lapshov}, {Liu}, {Maitra}, {Meidinger}, {Menz}, {Merloni}, {Mernik}, {Mican}, {Mohr}, {M{\"u}ller}, {Nandra}, {Nazarov}, {Pacaud}, {Pavlinsky}, {Perinati}, {Pfeffermann}, {Pietschner}, {Ramos-Ceja}, {Rau}, {Reiffers}, {Reiprich}, {Robrade}, {Salvato}, {Sanders}, {Santangelo}, {Sasaki}, {Scheuerle}, {Schmid}, {Schmitt}, {Schwope}, {Shirshakov}, {Steinmetz}, {Stewart}, {Str{\"u}der},
  {Sunyaev}, {Tenzer}, {Tiedemann}, {Tr{\"u}mper}, {Voron}, {Weber}, {Wilms}, \& {Yaroshenko}}]{2021A&A...647A...1PeROSITA}
{Predehl}, P., {Andritschke}, R., {Arefiev}, V., {et~al.} 2021, \aap, 647, A1

\bibitem[{{Reines} {et~al.}(2013){Reines}, {Greene}, \& {Geha}}]{reines2013}
{Reines}, A.~E., {Greene}, J.~E., \& {Geha}, M. 2013, \apj, 775, 116

\bibitem[{{Reines} \& {Volonteri}(2015)}]{Reines2015}
{Reines}, A.~E. \& {Volonteri}, M. 2015, \apj, 813, 82

\bibitem[{{Rosa Gonz{\'a}lez} {et~al.}(2009){Rosa Gonz{\'a}lez}, {Terlevich}, {Jim{\'e}nez Bail{\'o}n}, {Terlevich}, {Ranalli}, {Comastri}, {Laird}, \& {Nandra}}]{2009MNRAS.399..487RG}
{Rosa Gonz{\'a}lez}, D., {Terlevich}, E., {Jim{\'e}nez Bail{\'o}n}, E., {et~al.} 2009, \mnras, 399, 487

\bibitem[{{S{\'a}nchez-S{\'a}ez} {et~al.}(2023){S{\'a}nchez-S{\'a}ez}, {Arredondo}, {Bayo}, {Ar{\'e}valo}, {Bauer}, {Cabrera-Vives}, {Catelan}, {Coppi}, {Est{\'e}vez}, {F{\"o}rster}, {Hern{\'a}ndez-Garc{\'\i}a}, {Huijse}, {Kurtev}, {Lira}, {Mu{\~n}oz Arancibia}, \& {Pignata}}]{Sanchez-Saez23}
{S{\'a}nchez-S{\'a}ez}, P., {Arredondo}, J., {Bayo}, A., {et~al.} 2023, \aap, 675, A195

\bibitem[{{S{\'a}nchez-S{\'a}ez} {et~al.}(2019){S{\'a}nchez-S{\'a}ez}, {Lira}, {Cartier}, {Mirand a}, {Ho}, {Ar{\'e}valo}, {Bauer}, {Coppi}, \& {Yovaniniz}}]{Sanchez-Saez19}
{S{\'a}nchez-S{\'a}ez}, P., {Lira}, P., {Cartier}, R., {et~al.} 2019, \apjs, 242, 10

\bibitem[{{S{\'a}nchez-S{\'a}ez} {et~al.}(2021{\natexlab{a}}){S{\'a}nchez-S{\'a}ez}, {Reyes}, {Valenzuela}, {F{\"o}rster}, {Eyheramendy}, {Elorrieta}, {Bauer}, {Cabrera-Vives}, {Est{\'e}vez}, {Catelan}, {Pignata}, {Huijse}, {De Cicco}, {Ar{\'e}valo}, {Carrasco-Davis}, {Abril}, {Kurtev}, {Borissova}, {Arredondo}, {Castillo-Navarrete}, {Rodriguez}, {Ruz-Mieres}, {Moya}, {Sabatini-Gacit{\'u}a}, {Sep{\'u}lveda-Cobo}, \& {Camacho-I{\~n}iguez}}]{Sanchez-Saez20LC}
{S{\'a}nchez-S{\'a}ez}, P., {Reyes}, I., {Valenzuela}, C., {et~al.} 2021{\natexlab{a}}, \aj, 161, 141

\bibitem[{{S{\'a}nchez-S{\'a}ez} {et~al.}(2021{\natexlab{b}}){S{\'a}nchez-S{\'a}ez}, {Reyes}, {Valenzuela}, {F{\"o}rster}, {Eyheramendy}, {Elorrieta}, {Bauer}, {Cabrera-Vives}, {Est{\'e}vez}, {Catelan}, {Pignata}, {Huijse}, {De Cicco}, {Ar{\'e}valo}, {Carrasco-Davis}, {Abril}, {Kurtev}, {Borissova}, {Arredondo}, {Castillo-Navarrete}, {Rodriguez}, {Ruz-Mieres}, {Moya}, {Sabatini-Gacit{\'u}a}, {Sep{\'u}lveda-Cobo}, \& {Camacho-I{\~n}iguez}}]{Sanchez-Saez21}
{S{\'a}nchez-S{\'a}ez}, P., {Reyes}, I., {Valenzuela}, C., {et~al.} 2021{\natexlab{b}}, \aj, 161, 141

\bibitem[{{Schawinski} {et~al.}(2007){Schawinski}, {Thomas}, {Sarzi}, {Maraston}, {Kaviraj}, {Joo}, {Yi}, \& {Silk}}]{Schawinski2007MNRAS.382.1415S}
{Schawinski}, K., {Thomas}, D., {Sarzi}, M., {et~al.} 2007, \mnras, 382, 1415

\bibitem[{{Shen} {et~al.}(2008){Shen}, {Greene}, {Strauss}, {Richards}, \& {Schneider}}]{2008ApJ...680..169S}
{Shen}, Y., {Greene}, J.~E., {Strauss}, M.~A., {Richards}, G.~T., \& {Schneider}, D.~P. 2008, \apj, 680, 169

\bibitem[{{Shi} {et~al.}(2010){Shi}, {Rieke}, {Smith}, {Rigby}, {Hines}, {Donley}, {Schmidt}, \& {Diamond-Stanic}}]{2010ApJ...714..115Shi_Xray}
{Shi}, Y., {Rieke}, G.~H., {Smith}, P., {et~al.} 2010, \apj, 714, 115

\bibitem[{{Tachibana} \& {Miller}(2018)}]{Tachibana18}
{Tachibana}, Y. \& {Miller}, A.~A. 2018, \pasp, 130, 128001

\bibitem[{{Thomas} {et~al.}(2013){Thomas}, {Steele}, {Maraston}, {Johansson}, {Beifiori}, {Pforr}, {Str{\"o}mb{\"a}ck}, {Tremonti}, {Wake}, {Bizyaev}, {Bolton}, {Brewington}, {Brownstein}, {Comparat}, {Kneib}, {Malanushenko}, {Malanushenko}, {Oravetz}, {Pan}, {Parejko}, {Schneider}, {Shelden}, {Simmons}, {Snedden}, {Tanaka}, {Weaver}, \& {Yan}}]{2013MNRAS.431.1383T}
{Thomas}, D., {Steele}, O., {Maraston}, C., {et~al.} 2013, \mnras, 431, 1383

\bibitem[{{Tremonti} {et~al.}(2004){Tremonti}, {Heckman}, {Kauffmann}, {Brinchmann}, {Charlot}, {White}, {Seibert}, {Peng}, {Schlegel}, {Uomoto}, {Fukugita}, \& {Brinkmann}}]{2004ApJ...613..898T}
{Tremonti}, C.~A., {Heckman}, T.~M., {Kauffmann}, G., {et~al.} 2004, \apj, 613, 898

\bibitem[{{Vazdekis} {et~al.}(2016){Vazdekis}, {Koleva}, {Ricciardelli}, {R{\"o}ck}, \& {Falc{\'o}n-Barroso}}]{EMILES2016MNRAS.463.3409V}
{Vazdekis}, A., {Koleva}, M., {Ricciardelli}, E., {R{\"o}ck}, B., \& {Falc{\'o}n-Barroso}, J. 2016, \mnras, 463, 3409

\bibitem[{{Volonteri}(2010)}]{2010A&ARv..18..279VFsmbh}
{Volonteri}, M. 2010, \aapr, 18, 279

\bibitem[{{Ward} {et~al.}(2022){Ward}, {Gezari}, {Nugent}, {Bellm}, {Dekany}, {Drake}, {Duev}, {Graham}, {Kasliwal}, {Kool}, {Masci}, \& {Riddle}}]{Ward2022}
{Ward}, C., {Gezari}, S., {Nugent}, P., {et~al.} 2022, \apj, 936, 104

\end{thebibliography}

\begin{appendix}
\section{Higher redshift AGN}\label{ap:redshifts}
AGN candidates at higher redshift that are included in the NSA low-stellar mass sample because their redshift was incorrect are listed in Table \ref{tab:wrong_z}, together with the spectroscopic redshifts and classification provided by the SDSS pipeline.  

Another group of six AGN candidates also have spectra consistent with higher redshift QSO but have been miss-classified by the SDSS pipeline and assigned lower redshifts. These are J032620.06-064221.7, J120948.84+224432.5, J023259.60+004801.7, J163844.20+242601.9, J144540.13+113521.0, all classified as GALAXY by the SDSS pipeline and J101336.37+561536.3, classified as low-redshift QSO. The corresponding spectra are plotted below. These targets appear in the SDSS DR16Q quasar catalog with visual classifications \citep{Lyke2020}, the redshifts and broad type classifications from this table are summarized in Table \ref{tab:wrong_z2}. The objects in these two tables are considered correct AGN identifications for purposes of establishing the success rate of the classifications but are not included in the study of black hole masses, since their host-galaxy stellar masses are under-predicted. 
\FloatBarrier
\begin{table}[h!]
    \centering
        \caption{Spectroscopic redshifts from SDSS}

    \begin{tabular}{l|l|l}
    IAU name     & SDSS $z_{\rm spec}$ \\
    \hline
    SDSSJ125201.36-022831.8&0.225 &QSO\\
    SDSSJ025105.27-070230.4&0.327&QSO\\
    SDSSJ160648.09+291047.9&0.328&QSO\\
    SDSSJ154226.99+294202.9&0.457&QSO\\
  SDSSJ142526.19-011826.0 & 0.920 &QSO  \\
  SDSSJ120548.50+005344.0 & 0.932 &QSO    \\
  SDSSJ123618.90+032456.0 & 1.008 &QSO    \\
  SDSSJ023615.09-012132.8 & 1.093 &QSO   \\ 
  SDSSJ081459.59+280936.0 & 1.130 &QSO      \\
  SDSSJ081421.29+294020.0 & 1.260 &QSO    \\
  SDSSJ011517.09-012704.8 & 1.371 &QSO    \\
  SDSSJ030641.70+000107.9 & 1.400 &QSO    \\
  SDSSJ022943.49+002214.9 & 1.715 &QSO   \\
  SDSSJ121928.22+053028.3 & 1.818 &QSO    \\
  SDSSJ113900.49-020139.9 & 1.908 &QSO   \\
  SDSSJ110148.93-014539.9 & 2.427 &QSO   \\
  \hline
  SDSSJ092331.29+412526.9 & 0.521 &GALAXY             
    \end{tabular}
    \label{tab:wrong_z}
\end{table}
%\FloatBarrier

\begin{table}[h!]
    \centering
        \caption{Spectroscopic redshifts from SDSS DR16 Quasar catalog with visual classifications of \citet{Lyke2020}.}
    \begin{tabular}{l|l|l}
    IAU name     & z &type\\
    \hline
SDSSJ032620.06-064221.7& 1.912& A\\ SDSSJ120948.84+224432.5& 1.737& AR\\ SDSSJ023259.60+004801.7& 2.250& QX\\ SDSSJ163844.20+242601.9& 1.965&A\\ 
SDSSJ144540.13+113521.0& 2.052& A\\ SDSSJ101336.37+561536.3&3.650&QR \\
SDSSJ105816.18+544310.1&0.47942&QR 
  \end{tabular}
    \label{tab:wrong_z2}
    \begin{tablenotes}
        \item {\bf Note.} Type A: AGN, type-I Seyferts/host-dominated, type Q: QSO, type-I broad-line core-dominated, type R: radio association, type X: X-ray association.
    \end{tablenotes}
\end{table}
\FloatBarrier

\begin{figure}

    \centering
    \caption{AGN candidates with low-redshift GALAXY or QSO classification by the SDSS pipeline, that we identify as higher redshift AGN.}
    \includegraphics[width=0.4\textwidth]{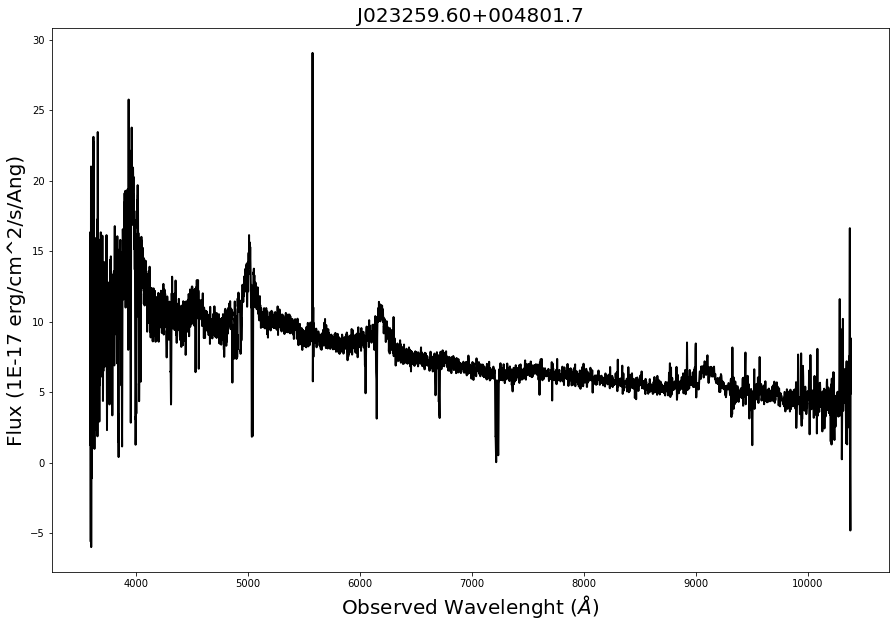}
    \includegraphics[width=0.4\textwidth]{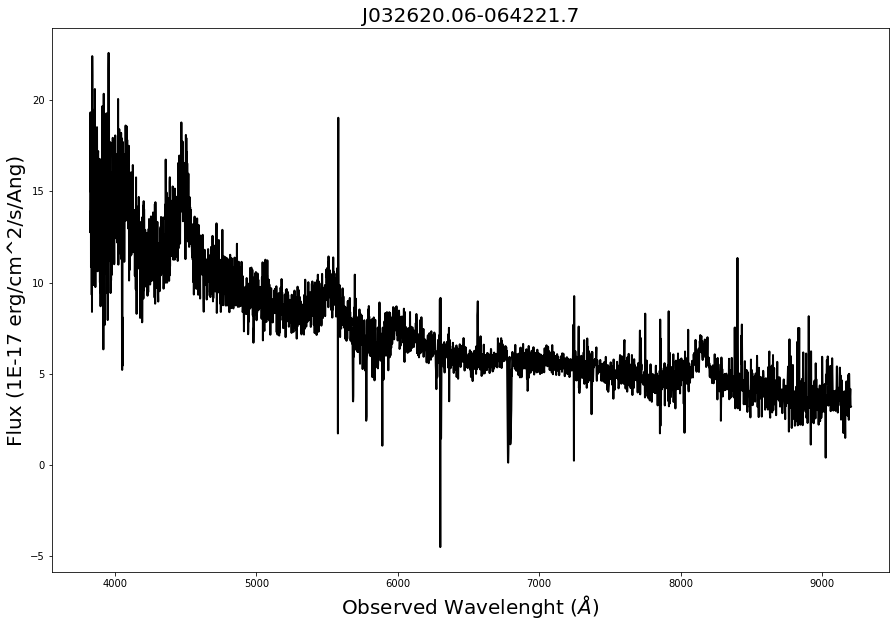}
    \includegraphics[width=0.4\textwidth]{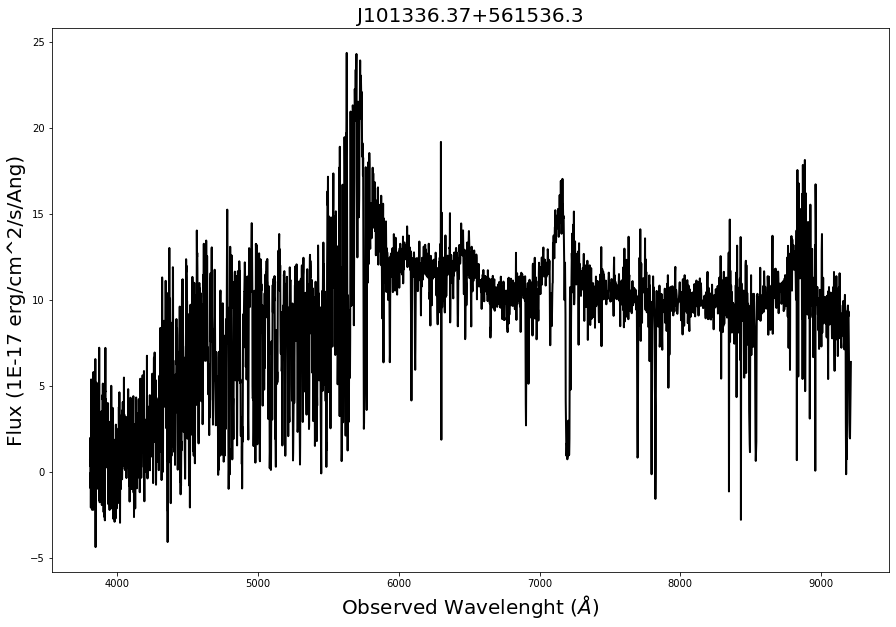}
    \includegraphics[width=0.4\textwidth]{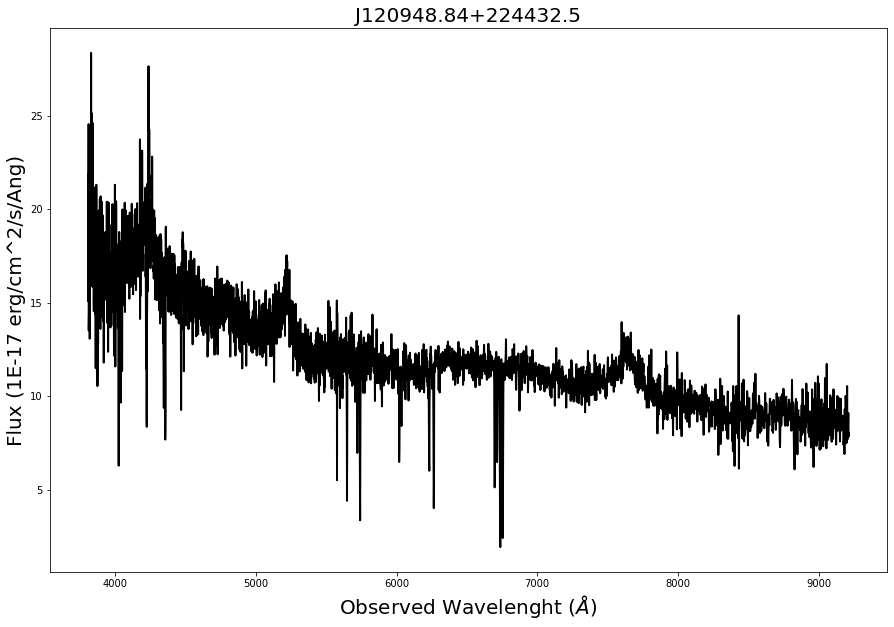}
\end{figure}
\addtocounter{figure}{-1} % Restablece el contador
\begin{figure}
    \centering
    \caption{continued }
    \includegraphics[width=0.4\textwidth]{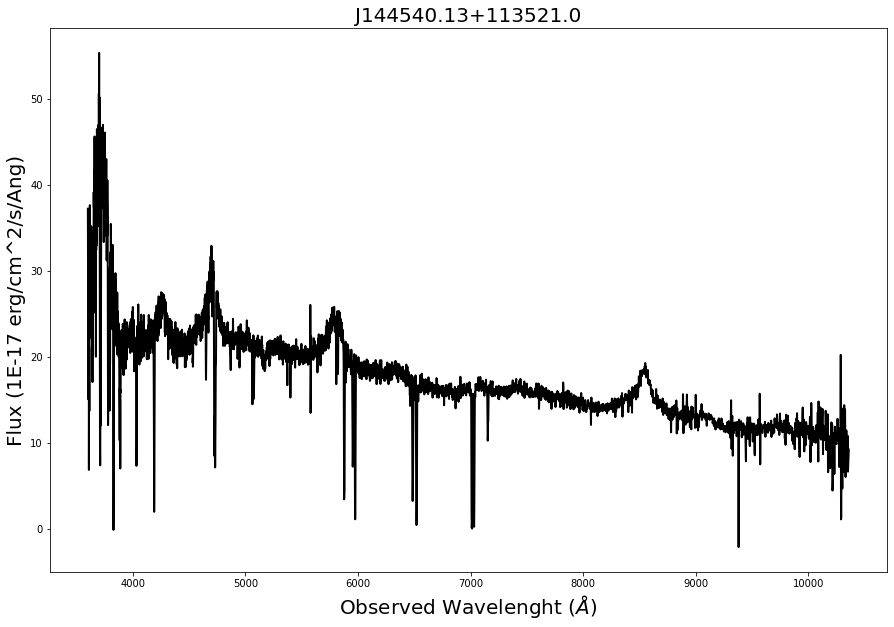}
    \includegraphics[width=0.4\textwidth]{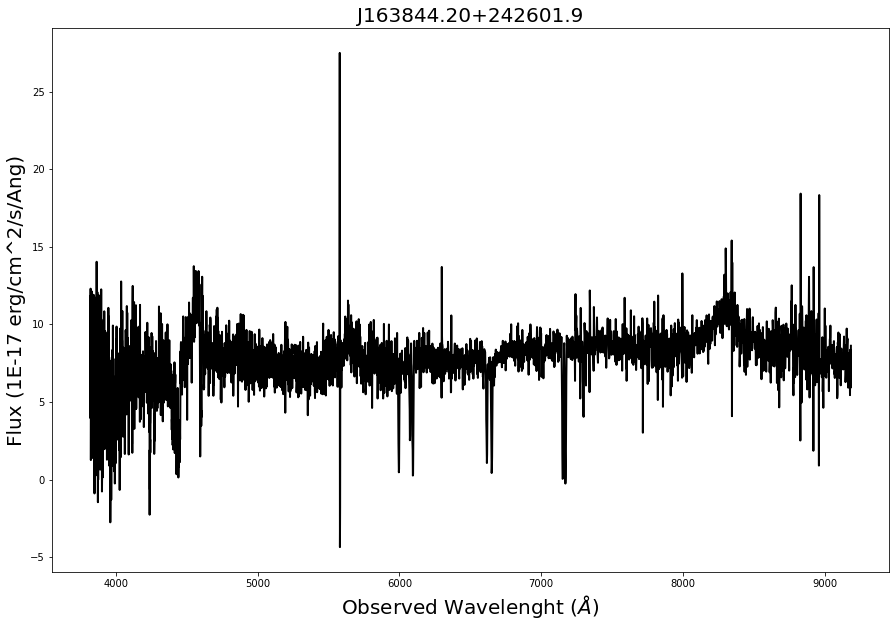}
    
    \label{fig:enter-label}
\end{figure}
\FloatBarrier

\section{Suspected AGN}\label{ap: suspected AGN}

Here we show the spectra of suspected AGN objects. The list of objects and the corresponding set are presented in table \ref{tab:suspected}. The visual inspection of light curves and spectra reveals variability and profiles that can potentially be attributed to an AGN. The spectra profiles are presented in Fig.~\ref{fig:suspected-agn}. 
\FloatBarrier
\begin{table}[h!]
\small
    \centering
        \caption{List of suspected AGN}
    \begin{tabular}{l|cccc}
    IAU name     & Alerts & Forced & ZTFDR11 & ZTFDR11\\
    & & photometry & $g$-band & $r$-band \\
    \hline
J151825.66+431514.8  & \checkmark & X & X & X\\ %BAL QSO 1
J115518.61+264559.8 & \checkmark & X & X & X\\ %BAL QSO 1
J123341.34-014423.7 & \checkmark  & \checkmark & \checkmark & X\\ %Blazar 1
J141927.49+044513.7 & \checkmark & \checkmark & \checkmark & \checkmark\\  %Blazar 1
J103916.55+040536.9  & X & \checkmark & \checkmark & X %609_000469_zg_c06_q1
  \end{tabular}
    \label{tab:suspected}
    \begin{tablenotes}
        \item {\bf Note.} Columns with sets names indicates in which set or sets the objects where selected.
    \end{tablenotes}
\end{table}
\FloatBarrier

\begin{figure}
    \centering
    \caption{Objects classified as suspected AGN. The spectral profiles show BELs and/or blue continuum.}
    \includegraphics[width=0.4\textwidth]{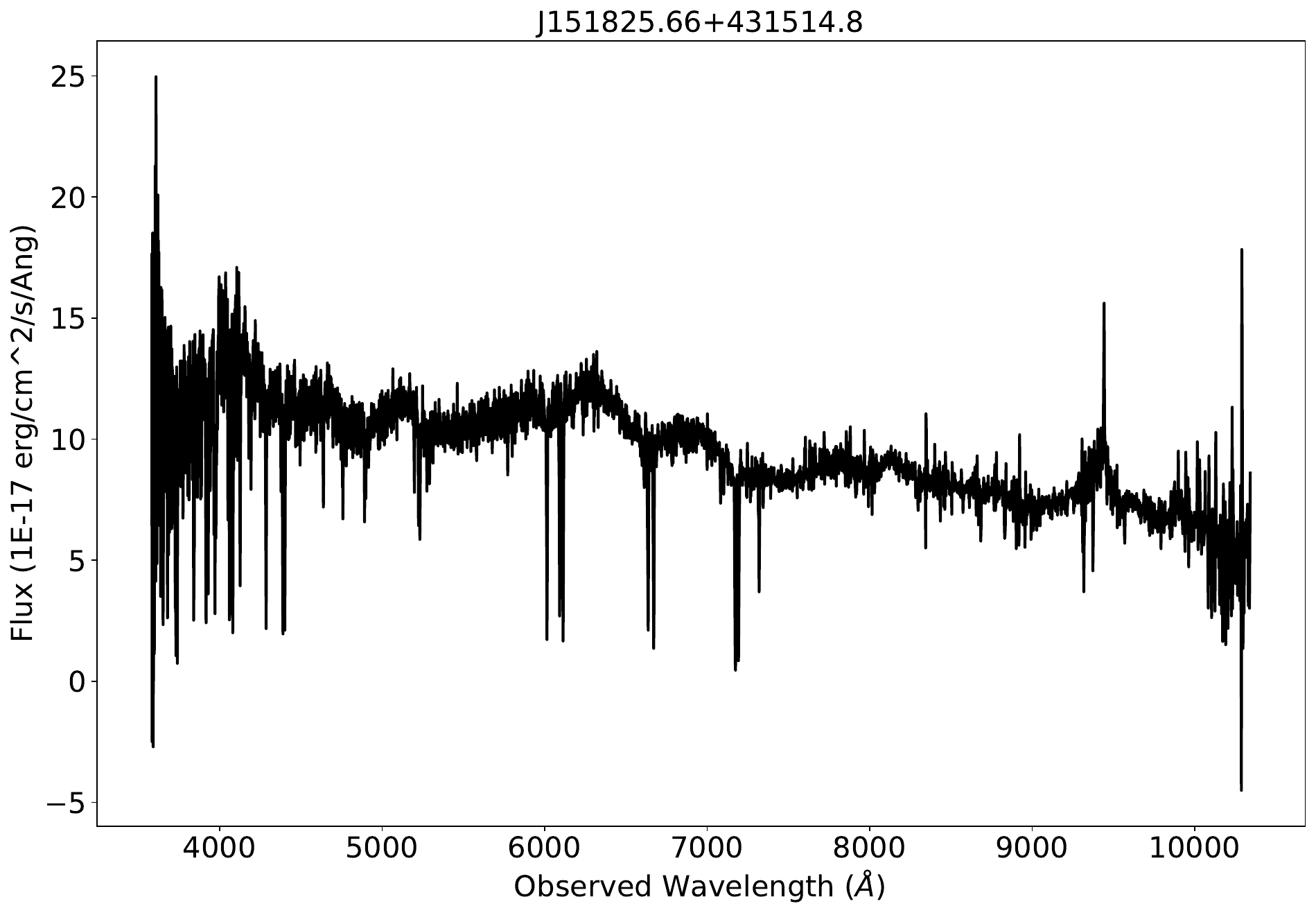}
    \includegraphics[width=0.4\textwidth]{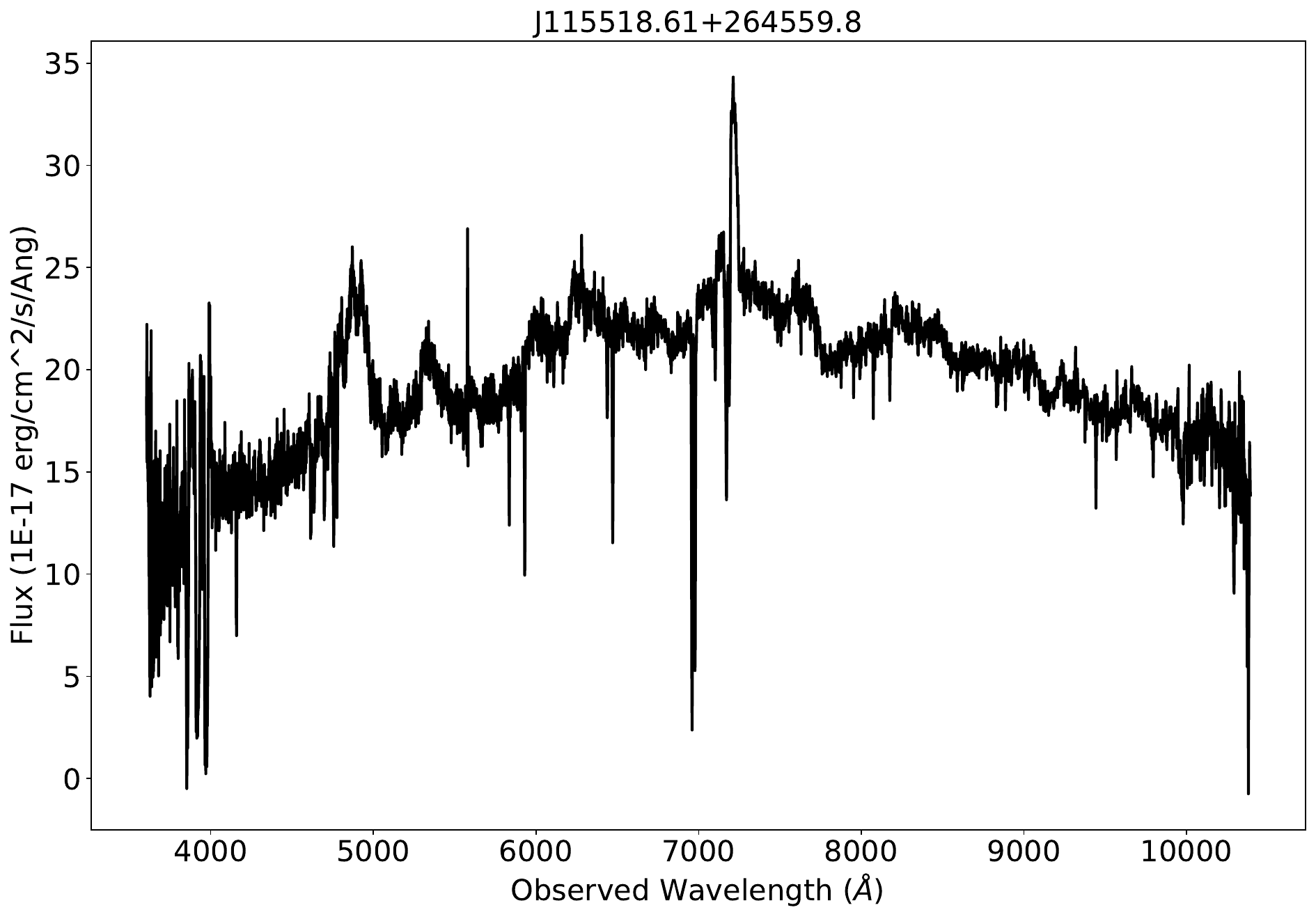}
    \includegraphics[width=0.4\textwidth]{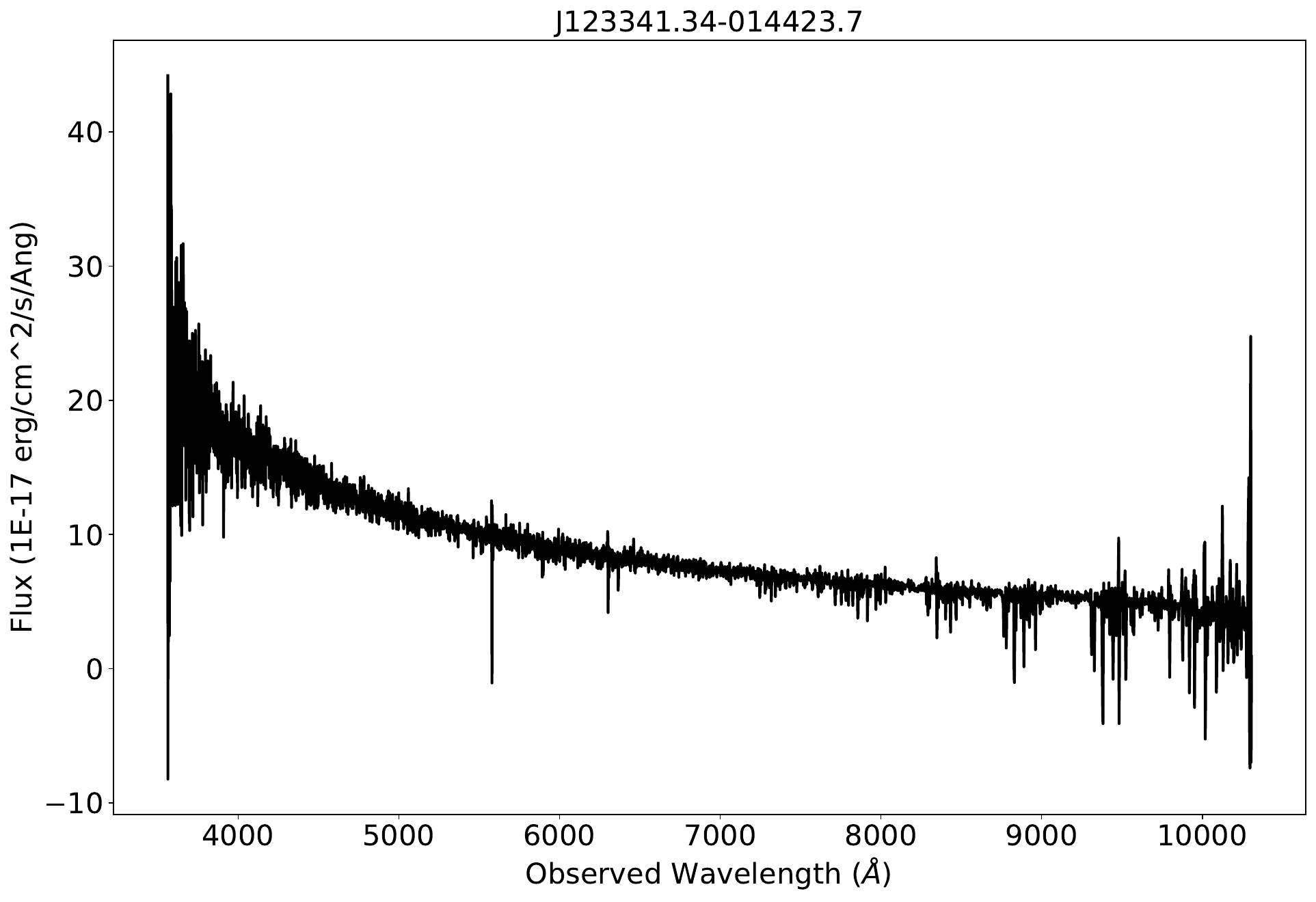}
    \includegraphics[width=0.4\textwidth]{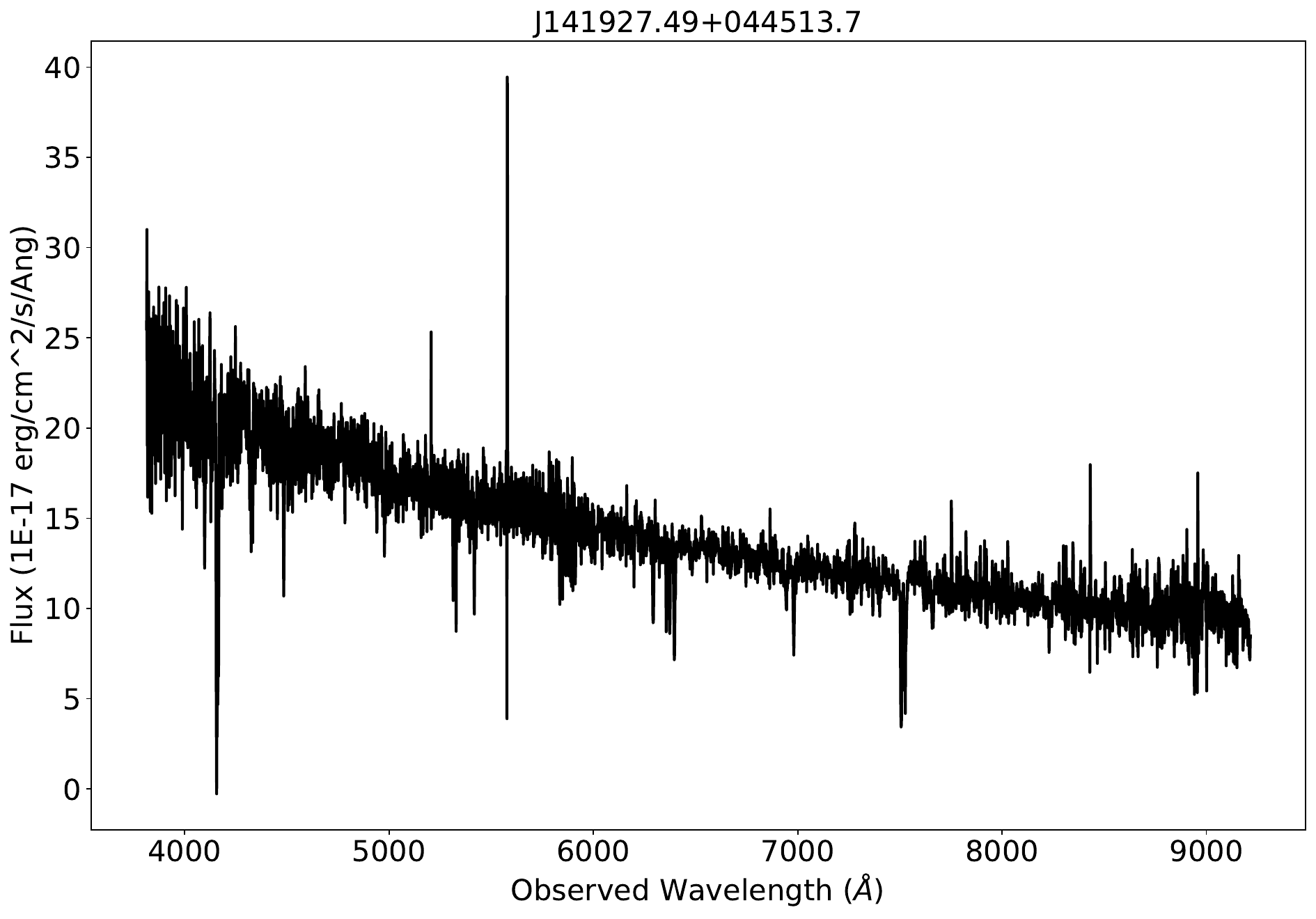}
    \end{figure}
\addtocounter{figure}{-1} % Restablece el contador
\begin{figure}
\centering
    \caption{continued }
    \includegraphics[width=0.4\textwidth]{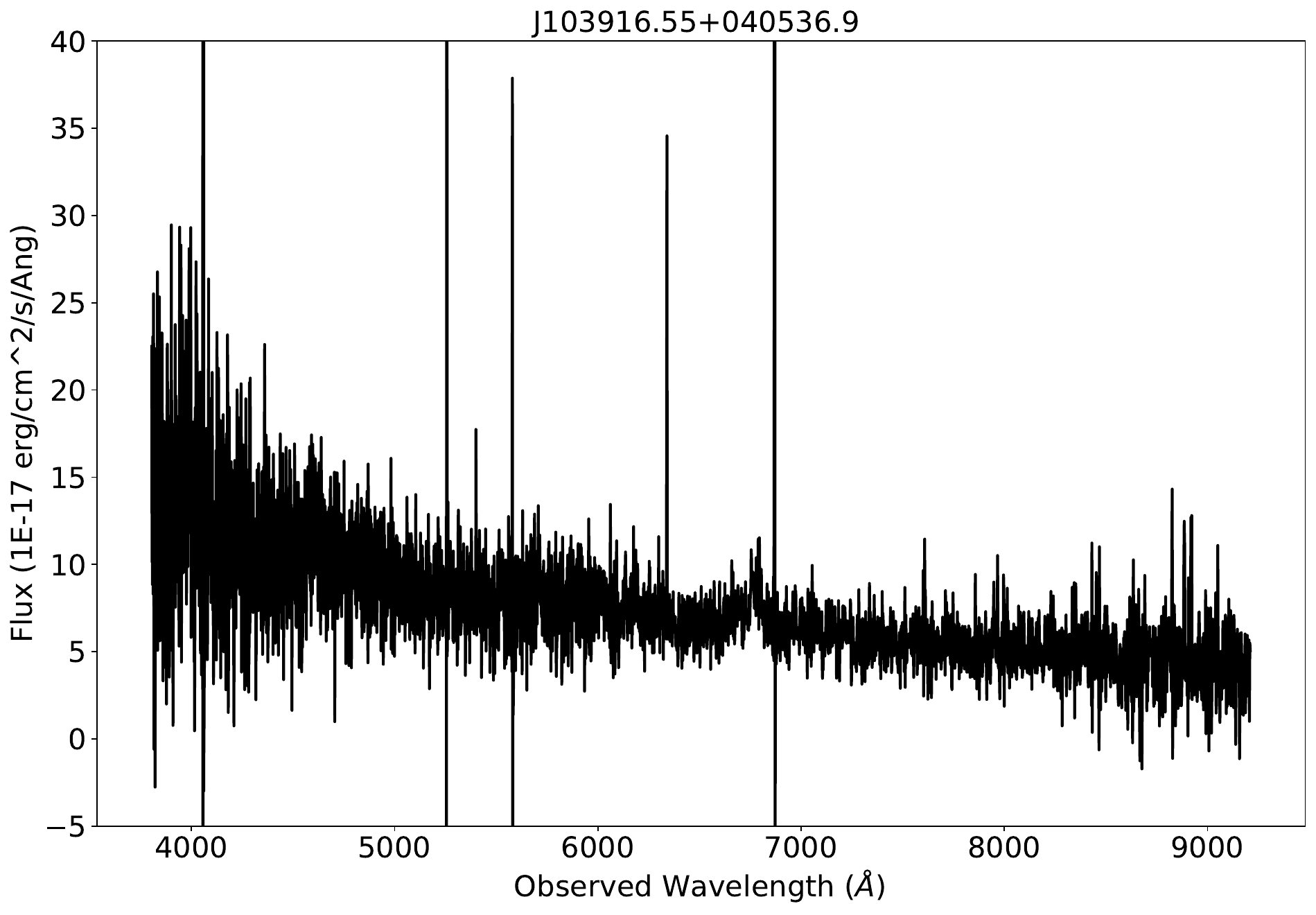}

    \label{fig:suspected-agn}
\end{figure}
\FloatBarrier

\section{Comparison with the BPT classification results of the MPA-JHU group}\label{ap: BPT comparison} 

In table \ref{tab:BPT_class} we compare the BPT classification obtained in this study with the Value added catalogs of SDSS. We found that the major difference is for 117 objects that we classify as Seyfert and that are classified as SF in the results of the MPA-JHU group. To explore this difference we compare the flux measured for the different emission-lines used to calculate the ratios, meaning, H$_\alpha$, H$_\beta$, [O{\sc iii}]$\lambda 5007$, [N{\sc ii}]$\lambda 6584$. In Fig.~\ref{fig: Ap_fluxes_for_BPT} it is evident that flux measurements are different for emission-lines where a BEL is present. In the comparison of [O{\sc iii}]$\lambda 5007$, the agreement is clear with the exception of one object (J000926.40+001932.1). For this and the objects that show more difference in the fluxes, we display the quality of our fits in Fig.~\ref{fig:ap_best_fit_flux_comp}.
This comparison and the quality of fits leads us to interpret the difference in classification as a product of the difference in the method used to measure the flux of the H$_\alpha$ and H$_\beta$. In particular, the method used by the MPA-JHU group does not distinguish between NELs and BELs, resulting in an increase in the flux of the narrow H$_\alpha$ and H$_\beta$ emission lines.
%\FloatBarrier
\begin{figure}[h!]
\centering
    \caption{Comparison of fluxes, between this work and the MPA-JHU group, for different emission-lines. The fluxes obtained by this study are in the x-axes and the ones obtained by the MPA-JHU group are in the y-axes. We show the comparison with only the narrow component of our model (Flux narrow), and for the cases of H$_\beta$, H$_\alpha$ and H$_\alpha$ complex we also compare with the total narrow+broad components (Flux narrow+broad). Flux units in all cases are $10^{-17} erg\ cm^{-2}\ s^{-1}$. The grey lines indicate the one-to-one relation.}
    \includegraphics[width=0.48\textwidth]{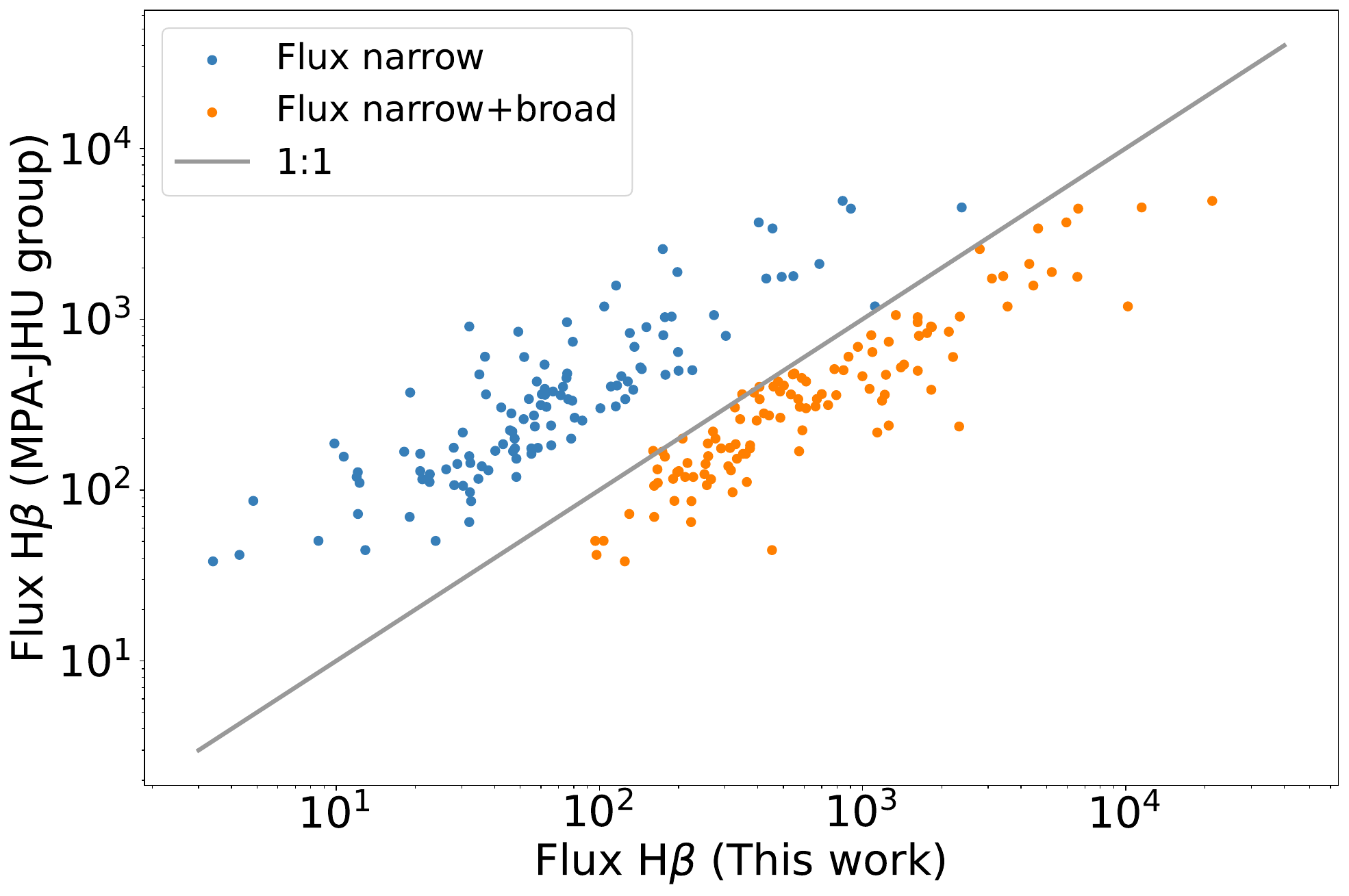}
    \label{fig: Ap_fluxes_for_BPT}
\end{figure}
\FloatBarrier
%\addtocounter{figure}{-1} % Restablece el contador
\begin{figure}[h!]
\centering
    
    \includegraphics[width=0.48\textwidth]{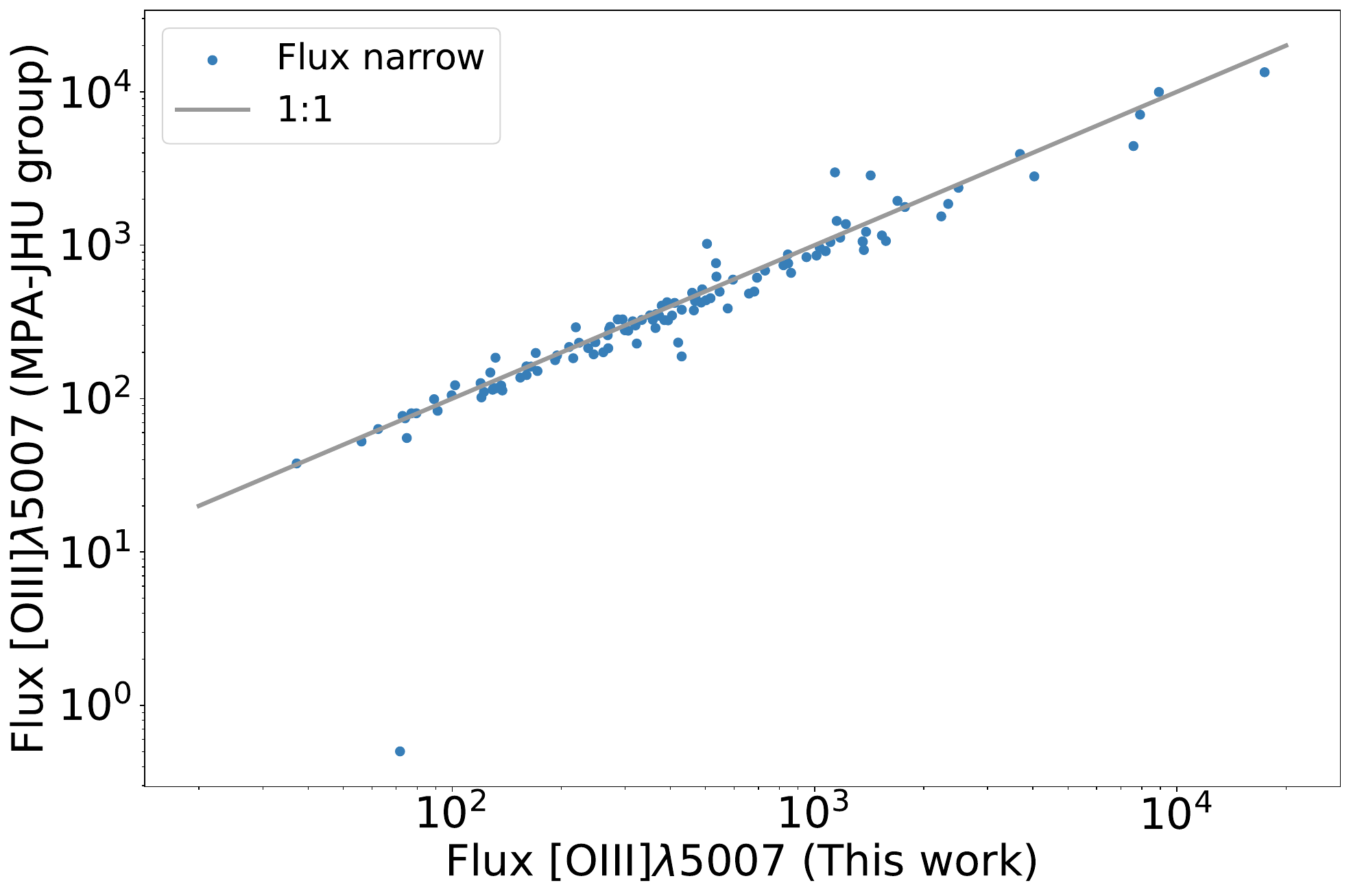}
    \includegraphics[width=0.48\textwidth]{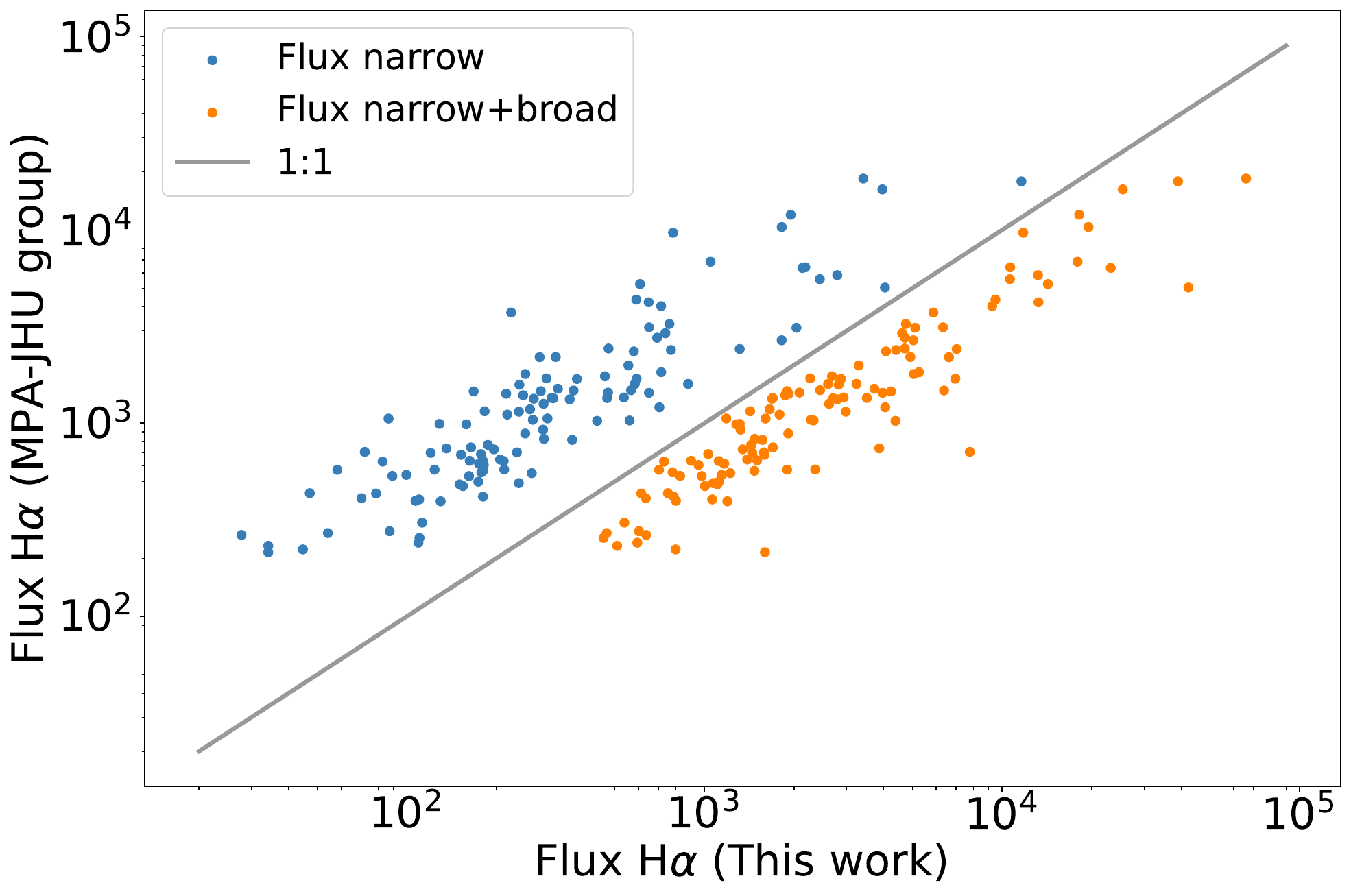}
    \includegraphics[width=0.48\textwidth]{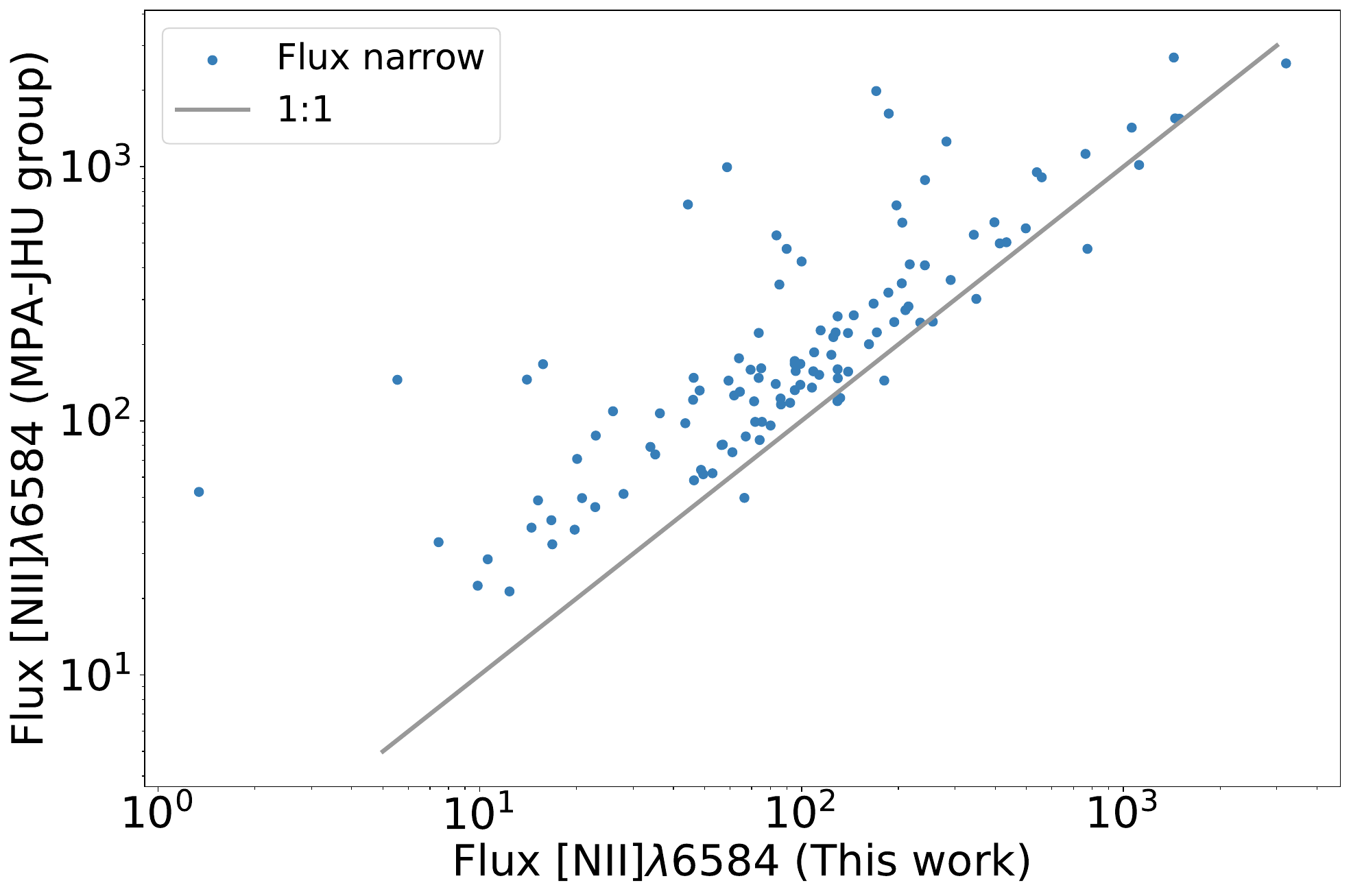}
    \includegraphics[width=0.48\textwidth]{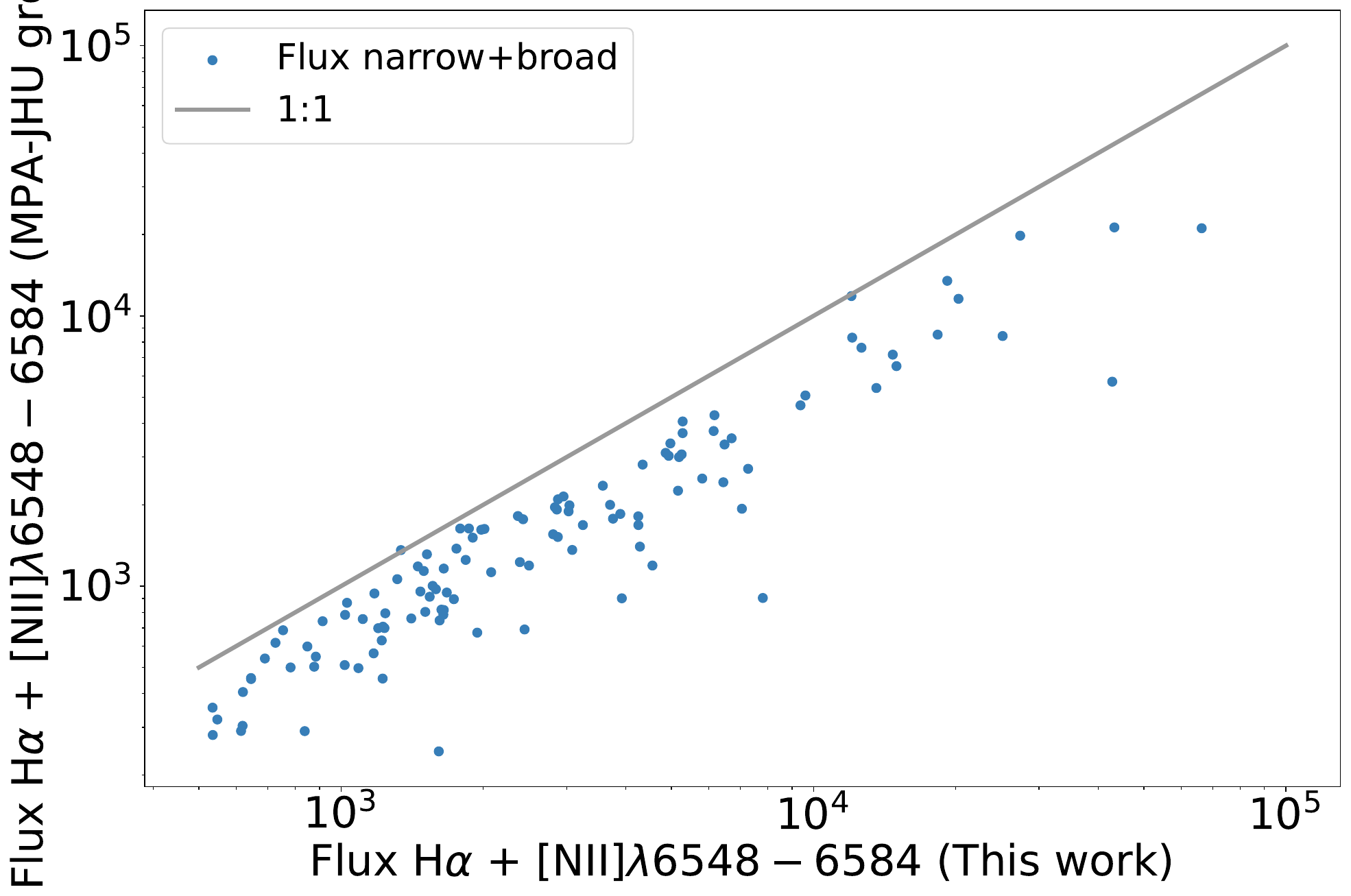}  
\end{figure}
\FloatBarrier

\begin{figure}

    \centering
    \caption{Best-fit models for objects that show more difference in the comparison of H$_\beta$ flux between our fits and the MPA-JHU group results.}
    \includegraphics[width=0.48\textwidth]{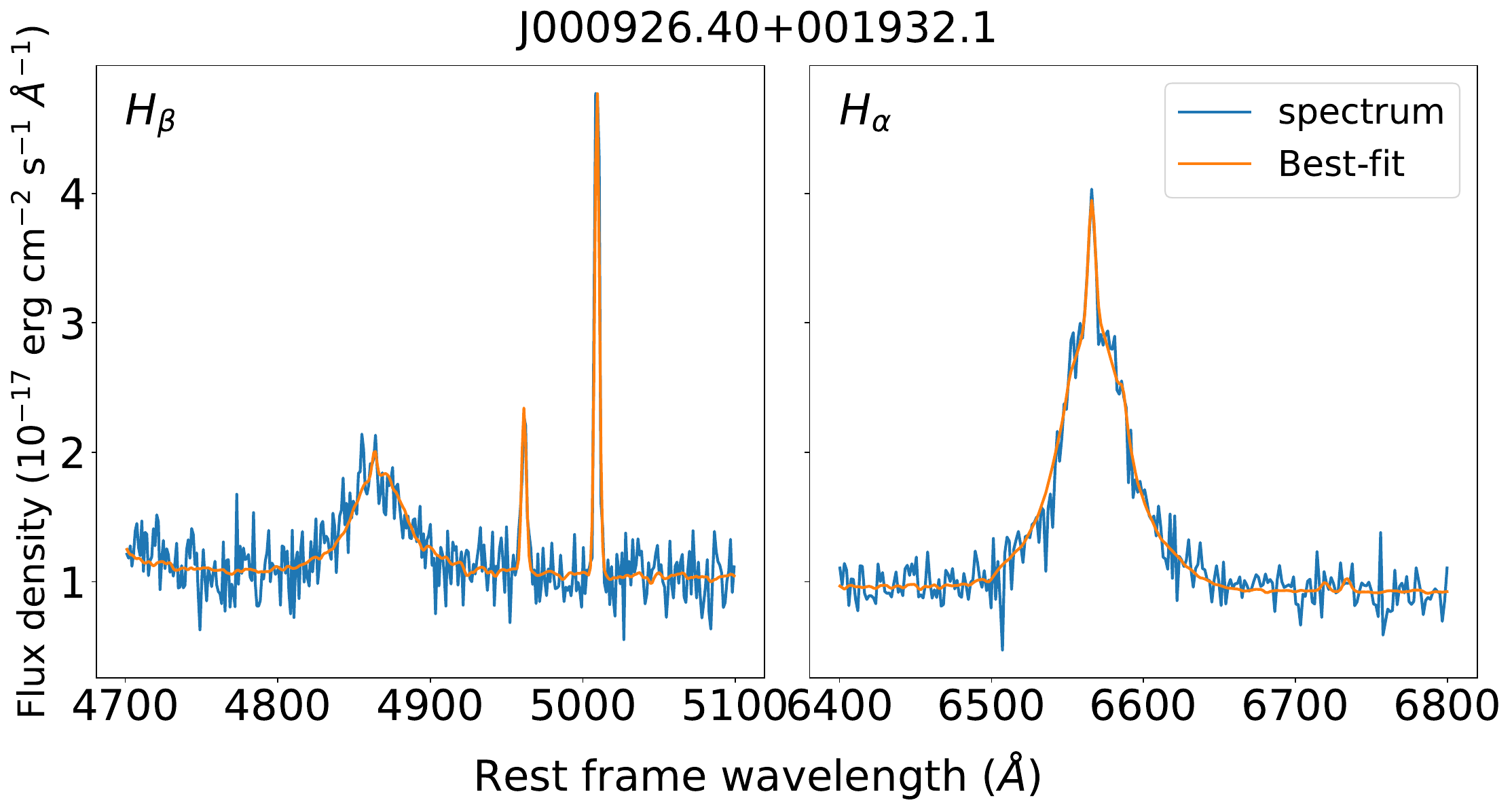}
    \includegraphics[width=0.48\textwidth]{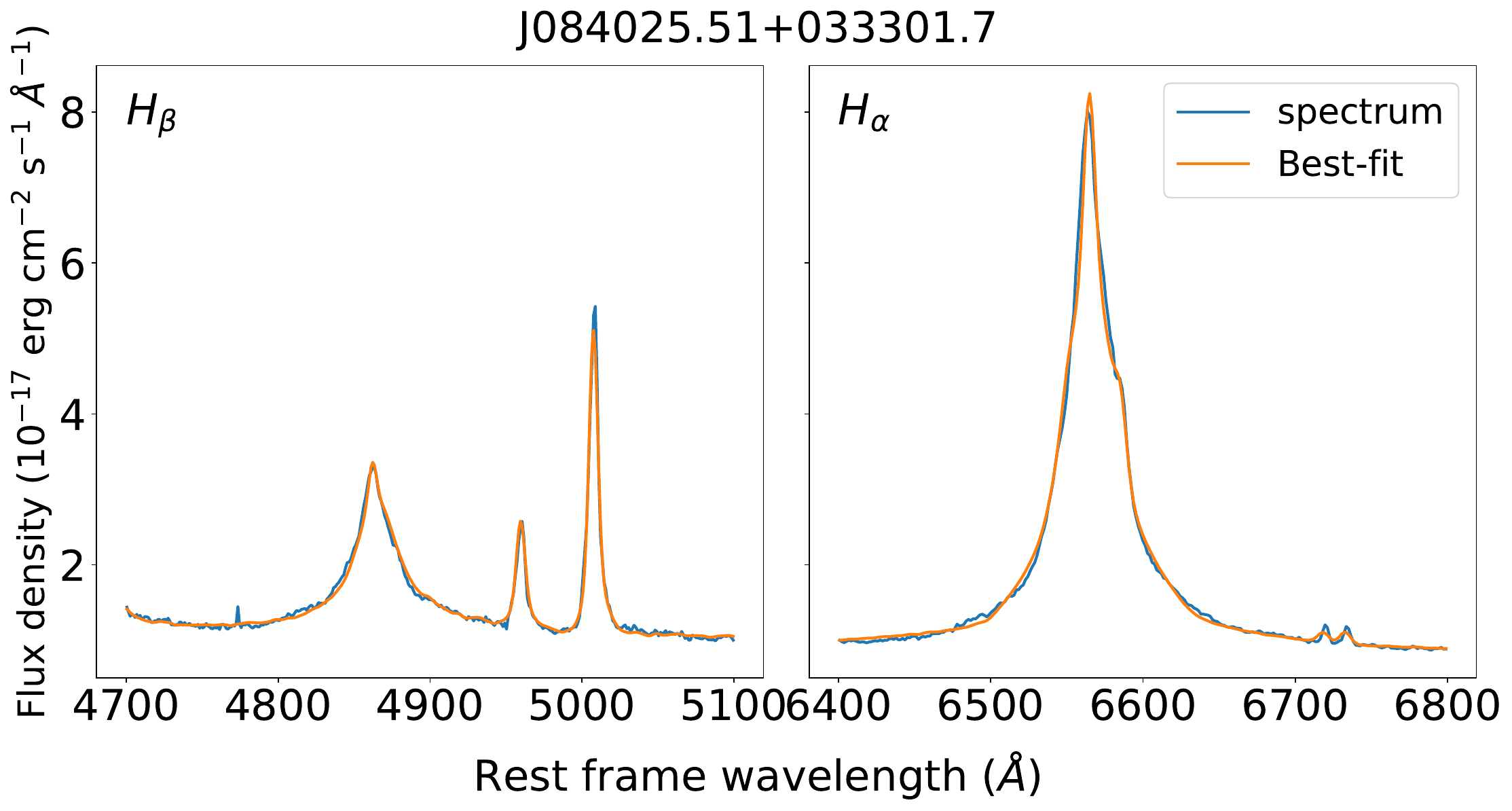}
    \includegraphics[width=0.48\textwidth]{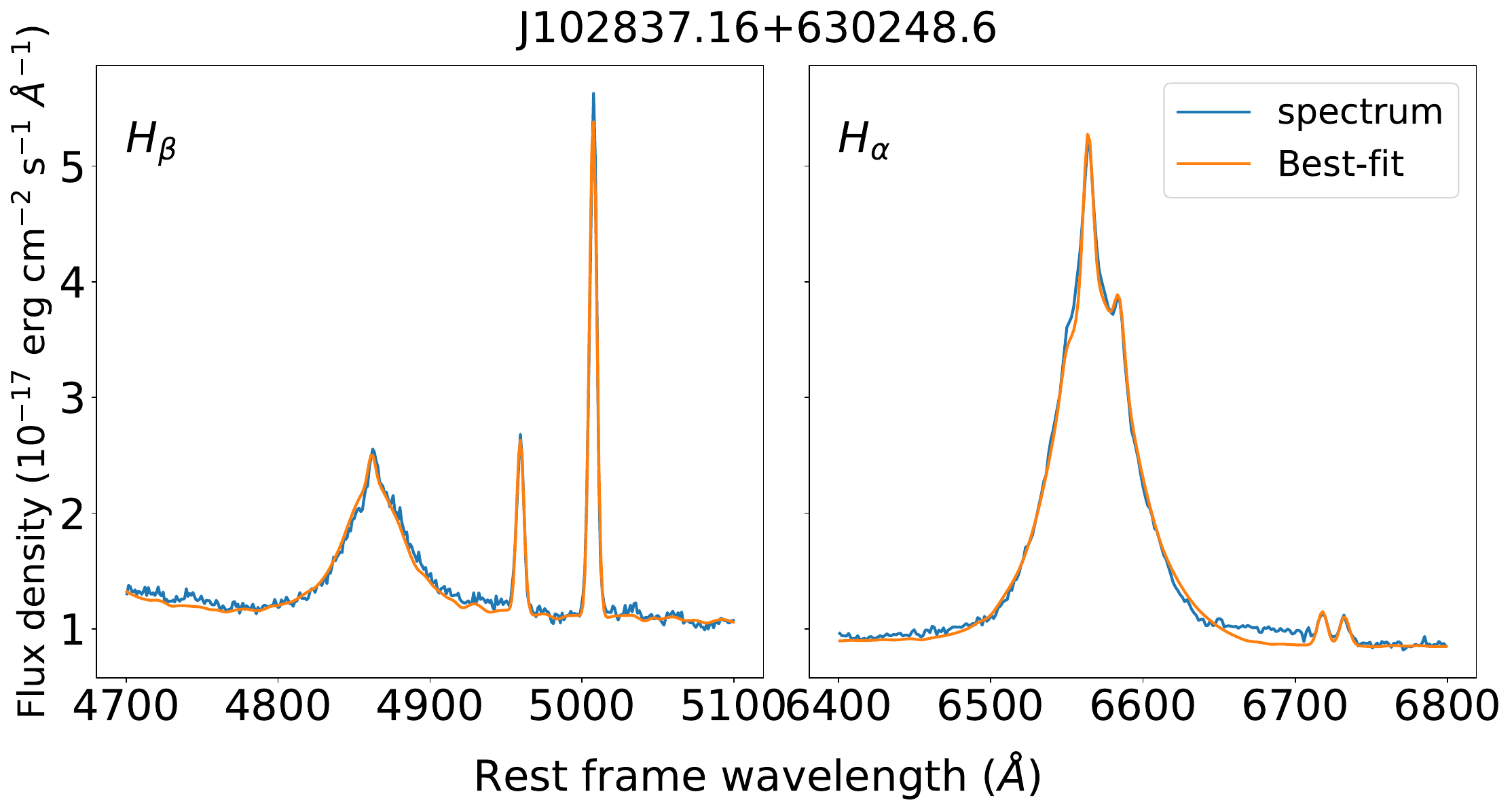}
    \includegraphics[width=0.48\textwidth]{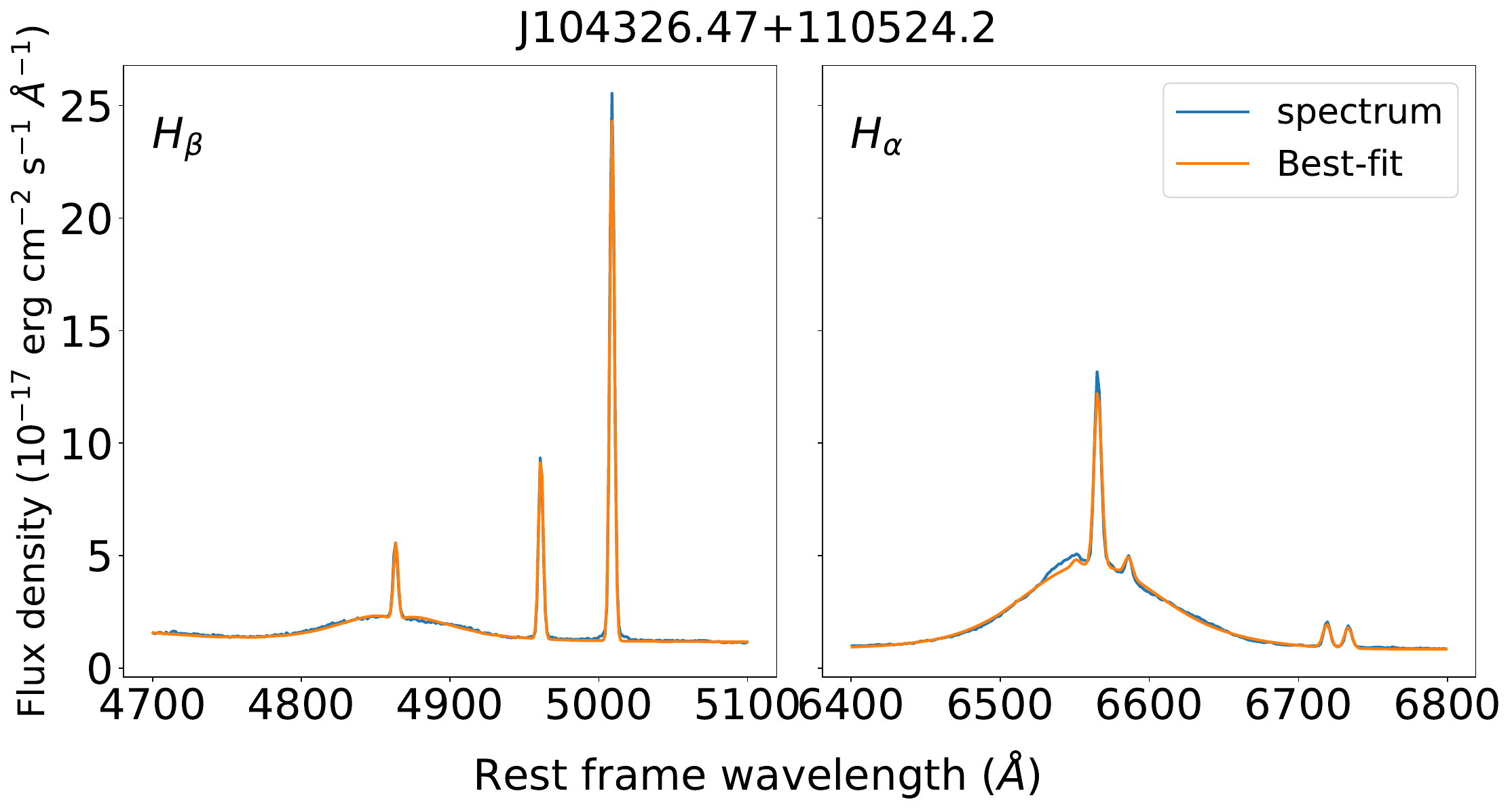}
    \includegraphics[width=0.48\textwidth]{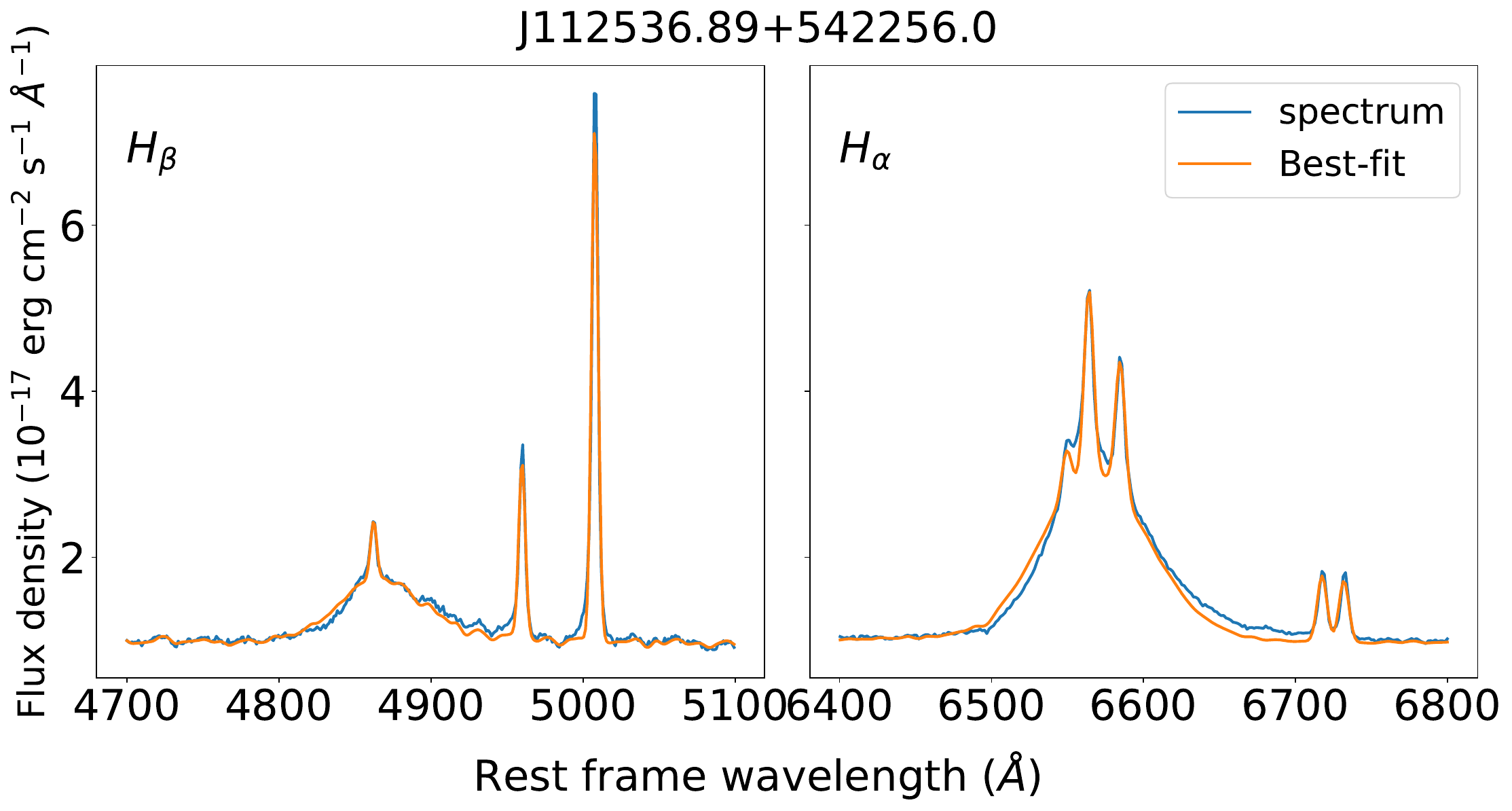}
    \label{fig:ap_best_fit_flux_comp}
    \end{figure}
\begin{figure}
    \centering

    \includegraphics[width=0.48\textwidth]{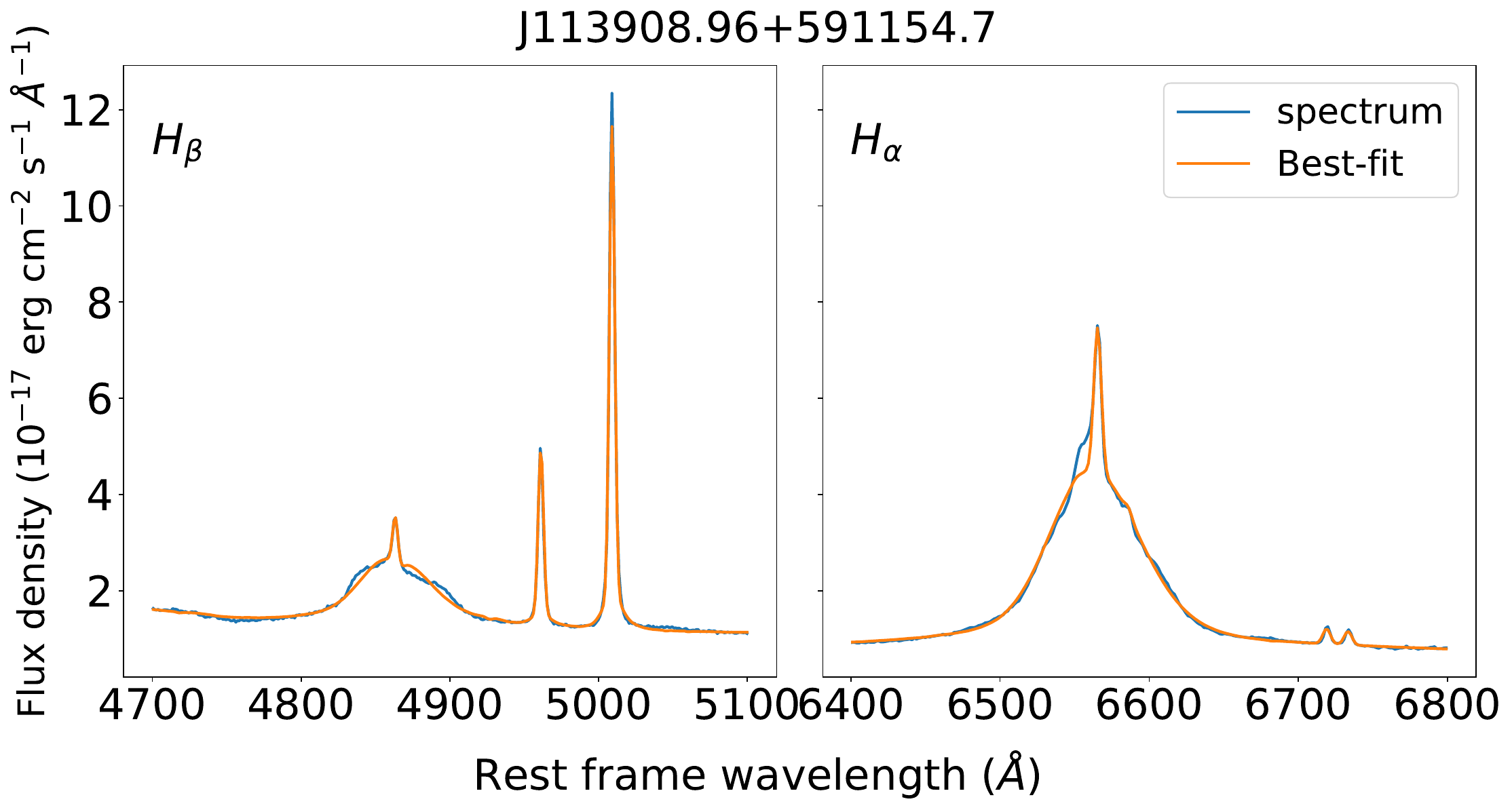}
    \includegraphics[width=0.48\textwidth]{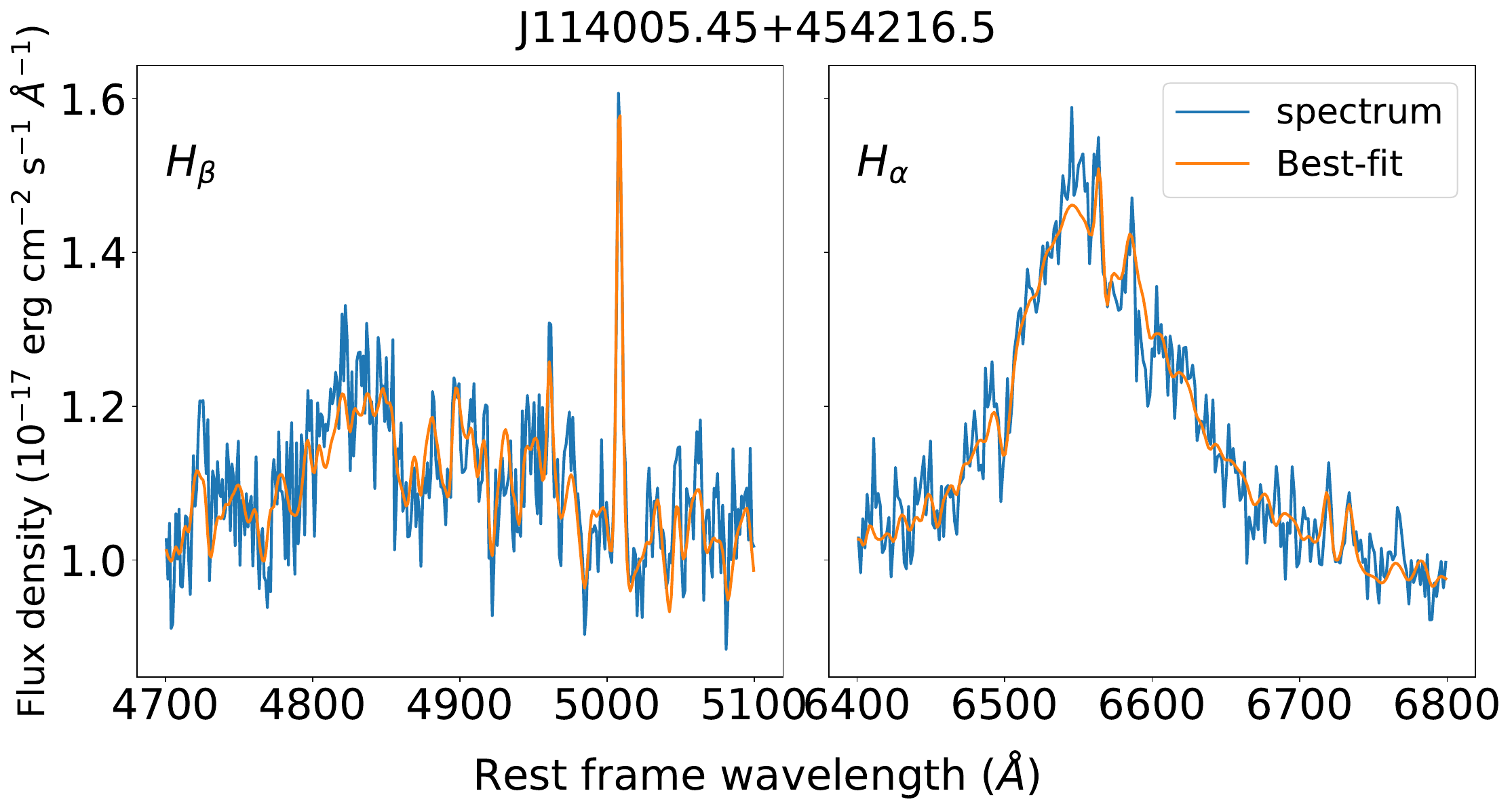}
    \includegraphics[width=0.48\textwidth]{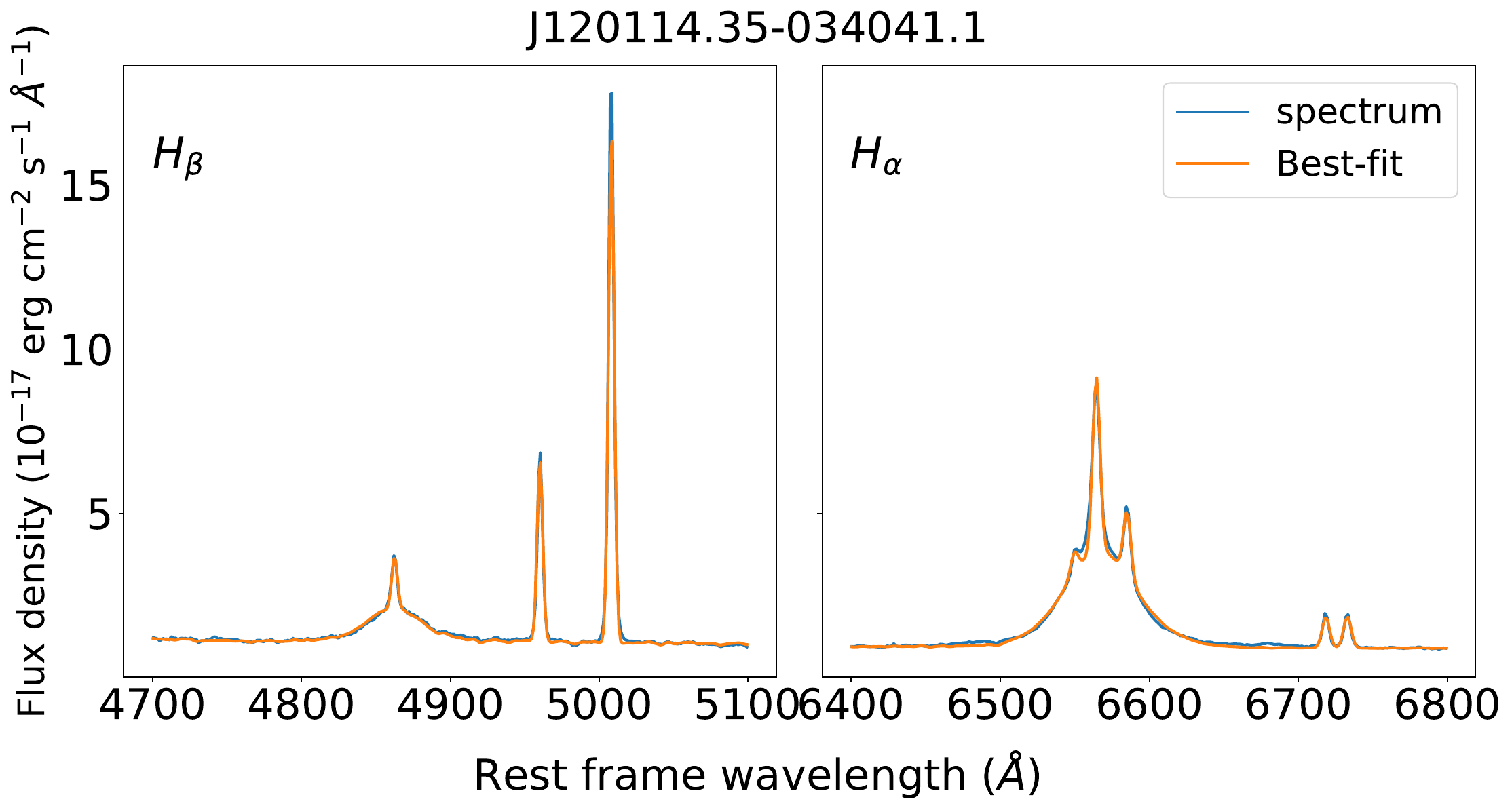}
    \includegraphics[width=0.48\textwidth]{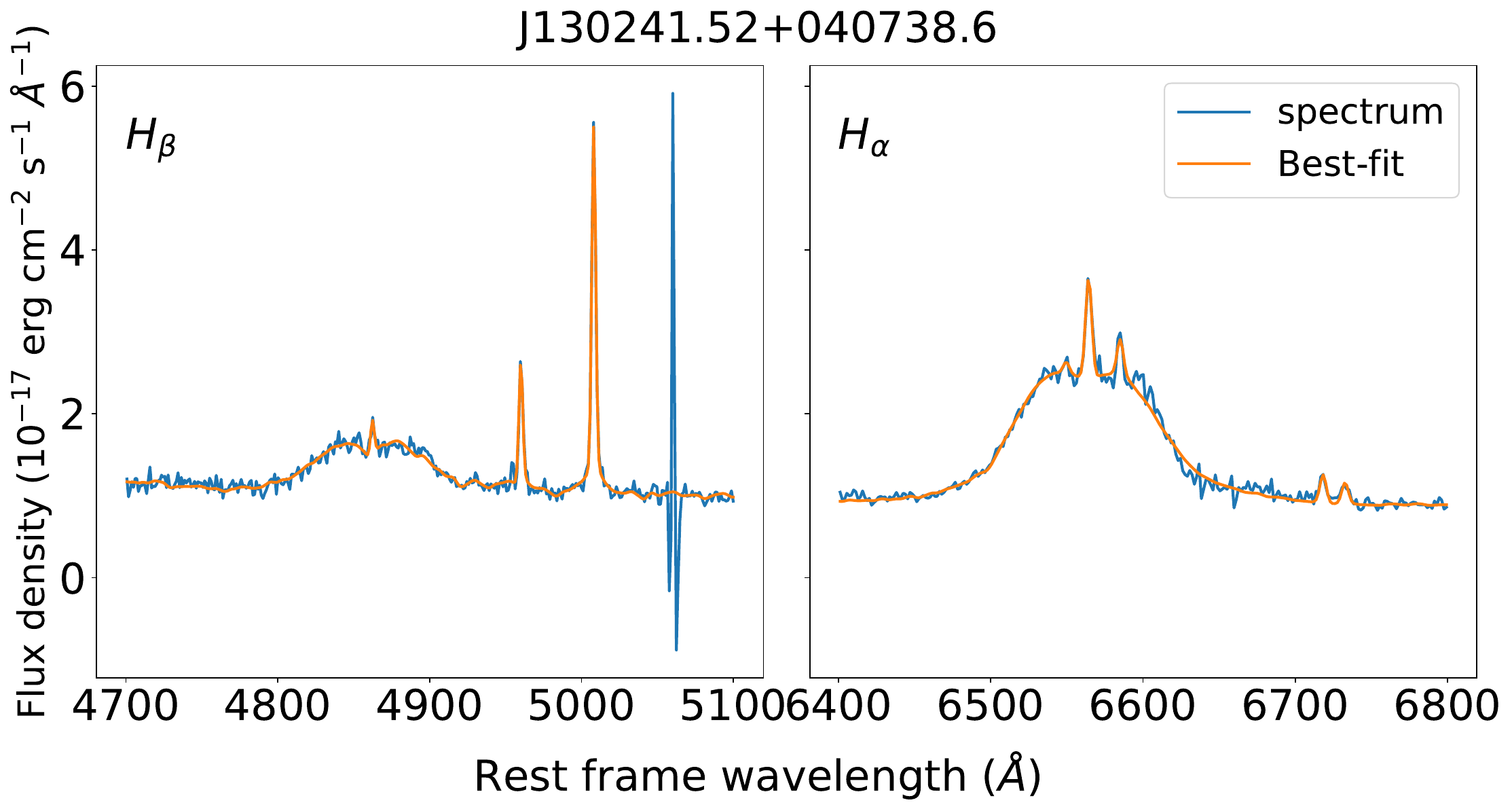}
    \includegraphics[width=0.48\textwidth]{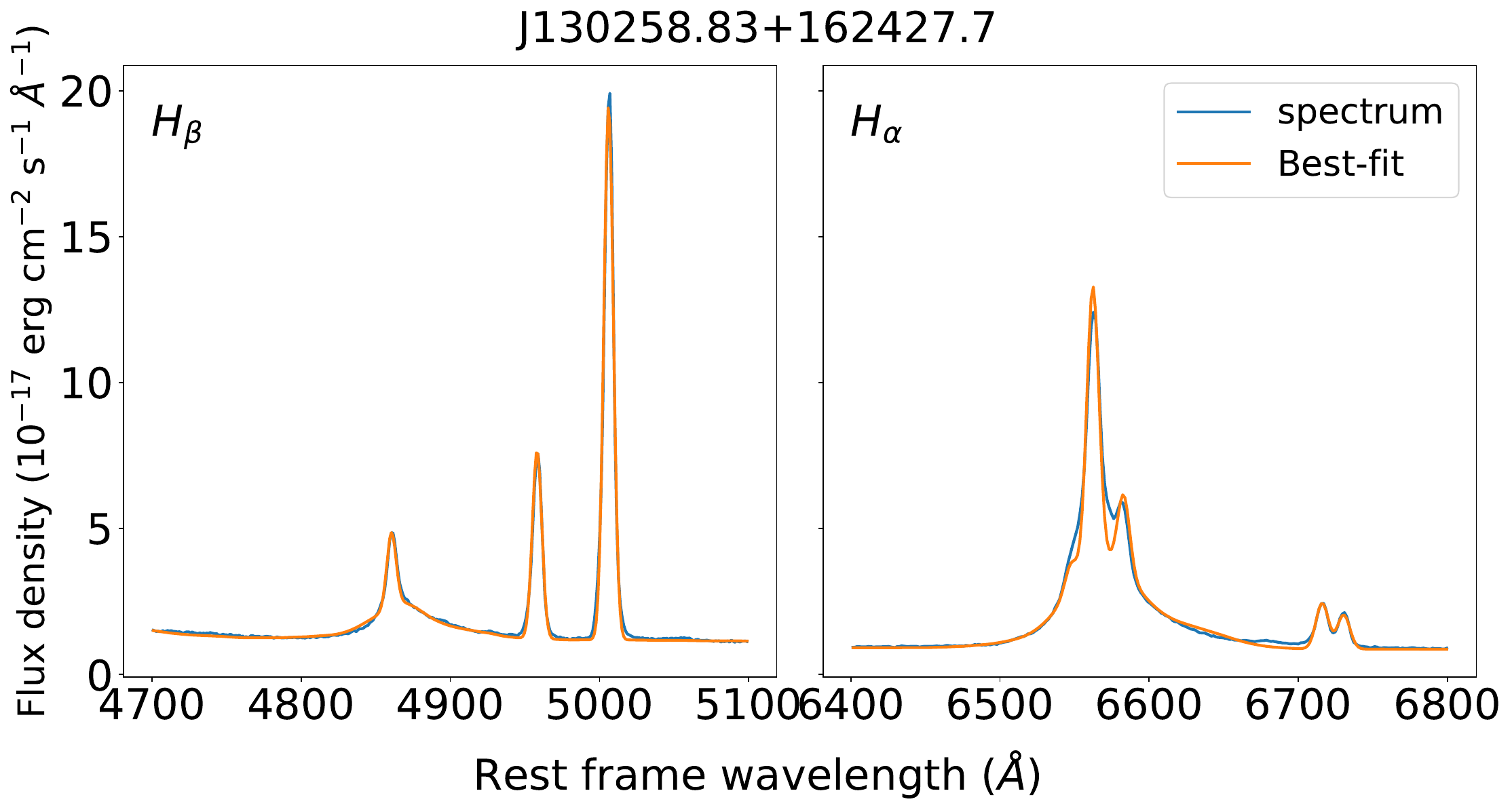}
    \end{figure}
\addtocounter{figure}{-1} % Restablece el contador
\begin{figure}
    \centering
    \caption{continued}
    \includegraphics[width=0.48\textwidth]{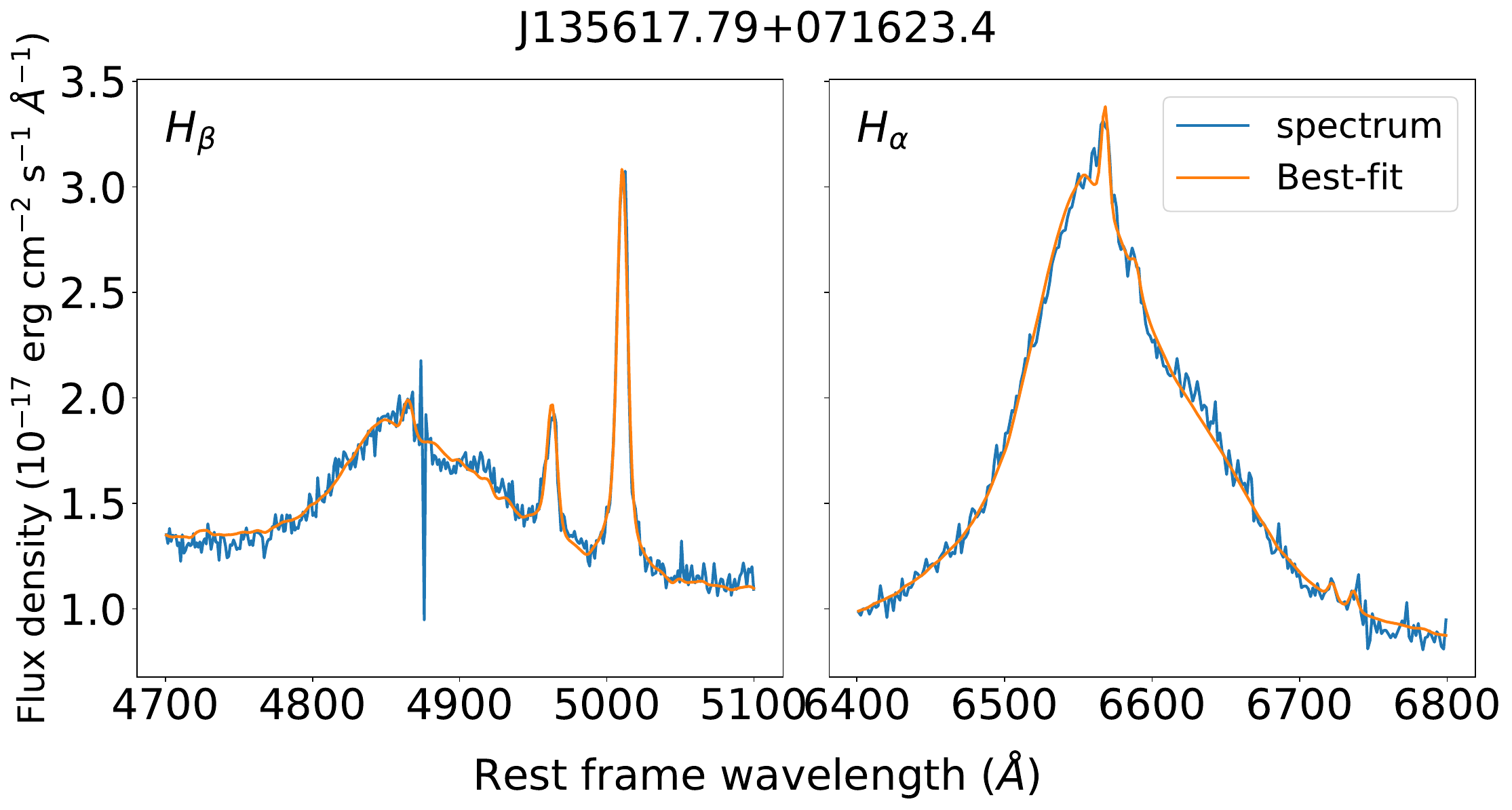}
\end{figure}
\FloatBarrier

\section{Flat spectra with relative high AGN contribution}\label{ap:AGN contribution}

In section \ref{sec: AGN contribution} we compare the relative contribution of the AGN continuum. From the analysis, we found objects that have a flat profile and where the best-fit model results in a power law slope equal to zero ($\alpha=0$). For these objects, we show the best-fit model in Fig.~\ref{fig:ap_flat_spectra}. We notice that the stellar absorption lines are well-fitted, this indicates that the decomposition of the continuum is highly accurate.
\FloatBarrier
\begin{figure}[h!]
    \centering
    \caption{Best-fit models for objects that show flat spectra and high AGN relative contribution. The dashed grey line correspond to zero Flux.}
    \includegraphics[width=0.42\textwidth]{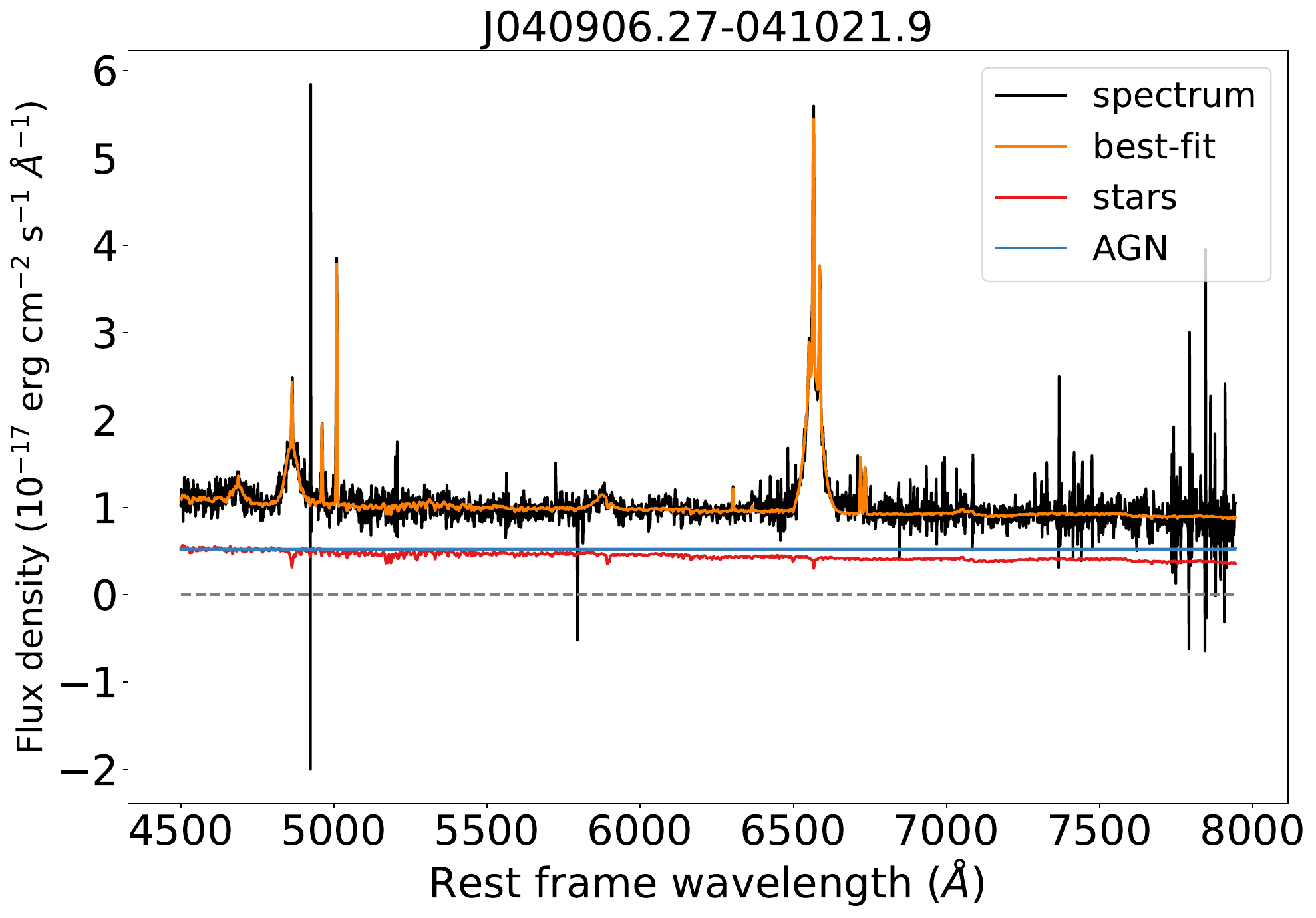}
    \includegraphics[width=0.42\textwidth]{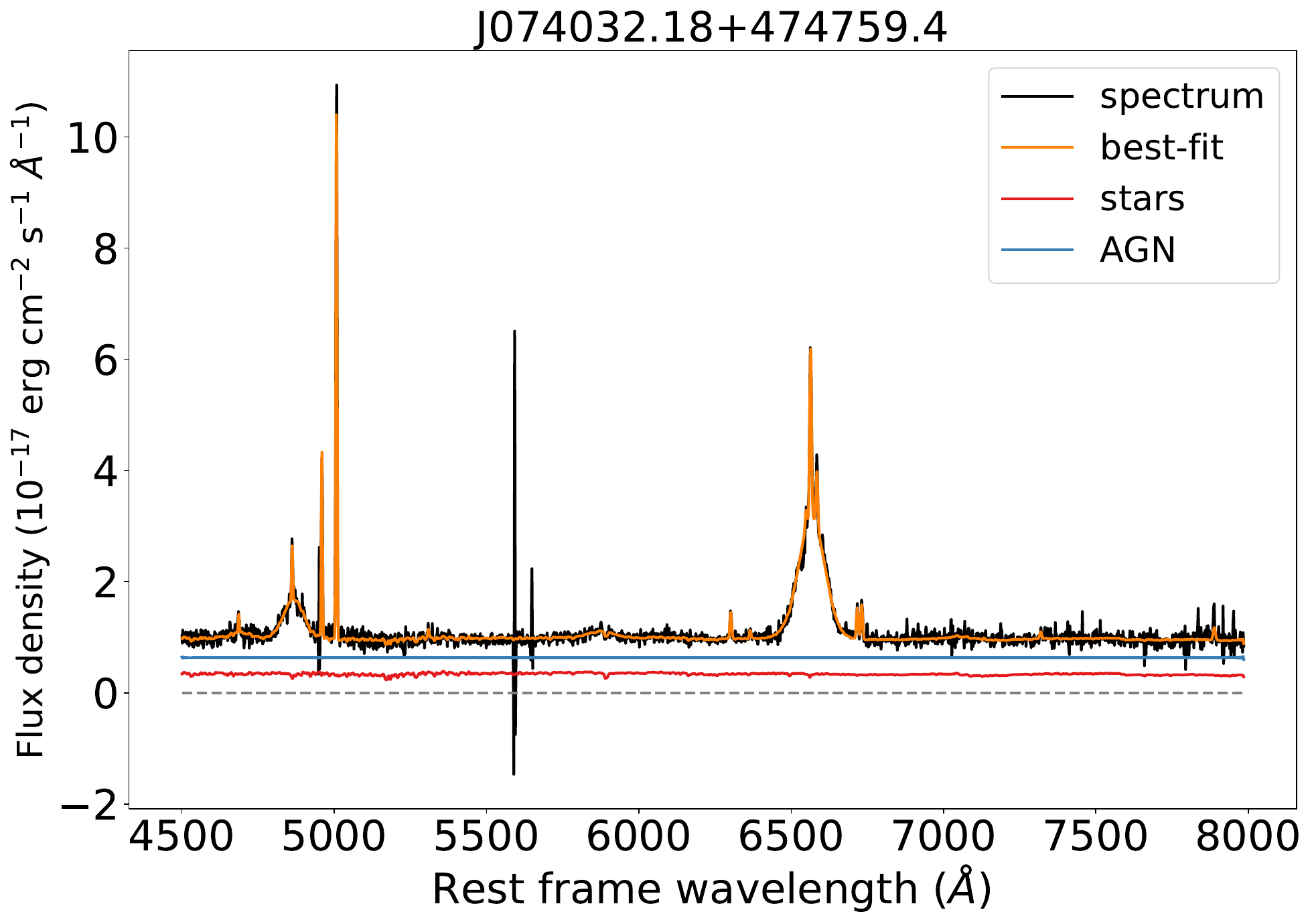}
    \label{fig:ap_flat_spectra}
    \end{figure}
\begin{figure}
    \centering
    \includegraphics[width=0.42\textwidth]{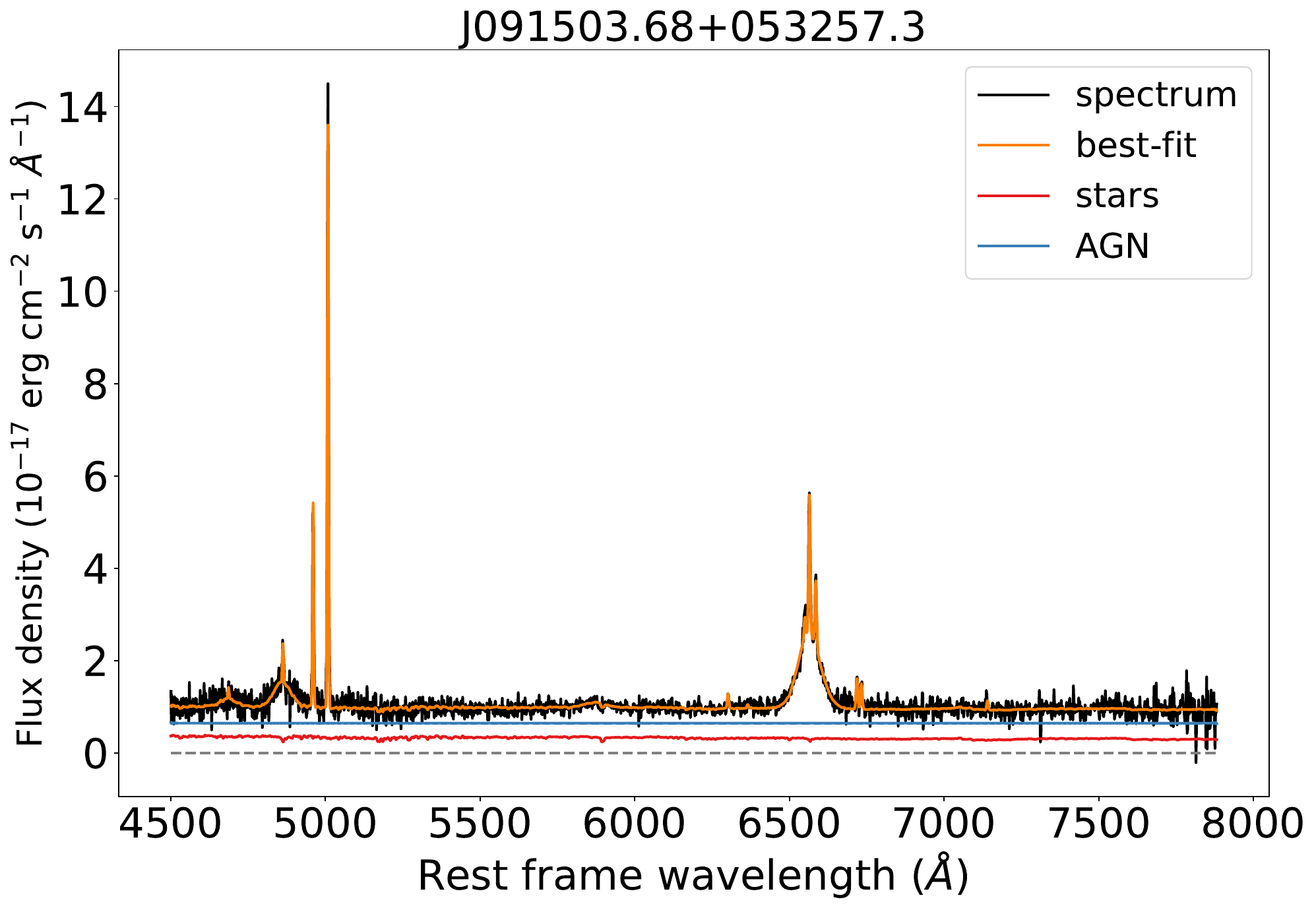}
    \includegraphics[width=0.42\textwidth]{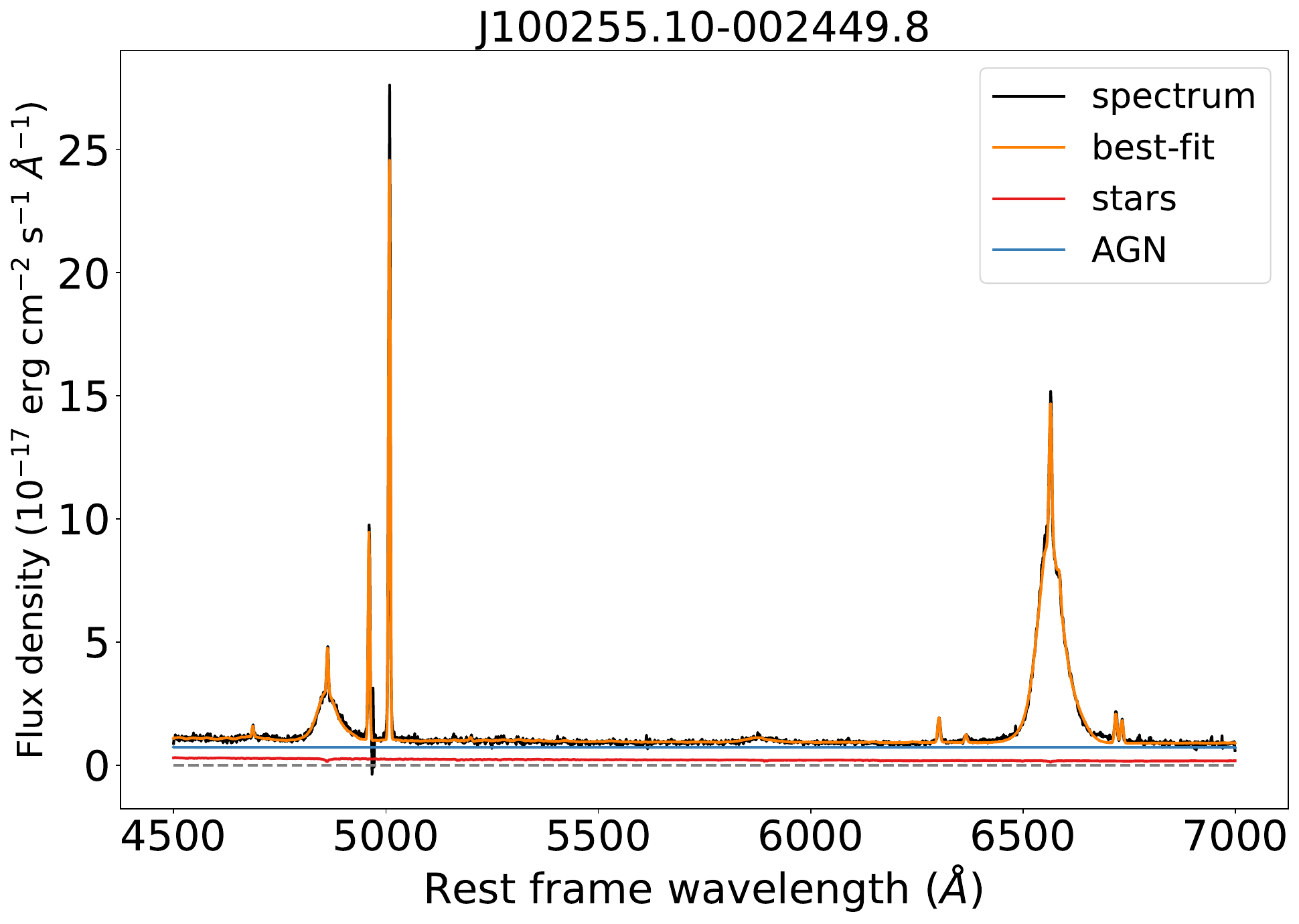}
    \includegraphics[width=0.42\textwidth]{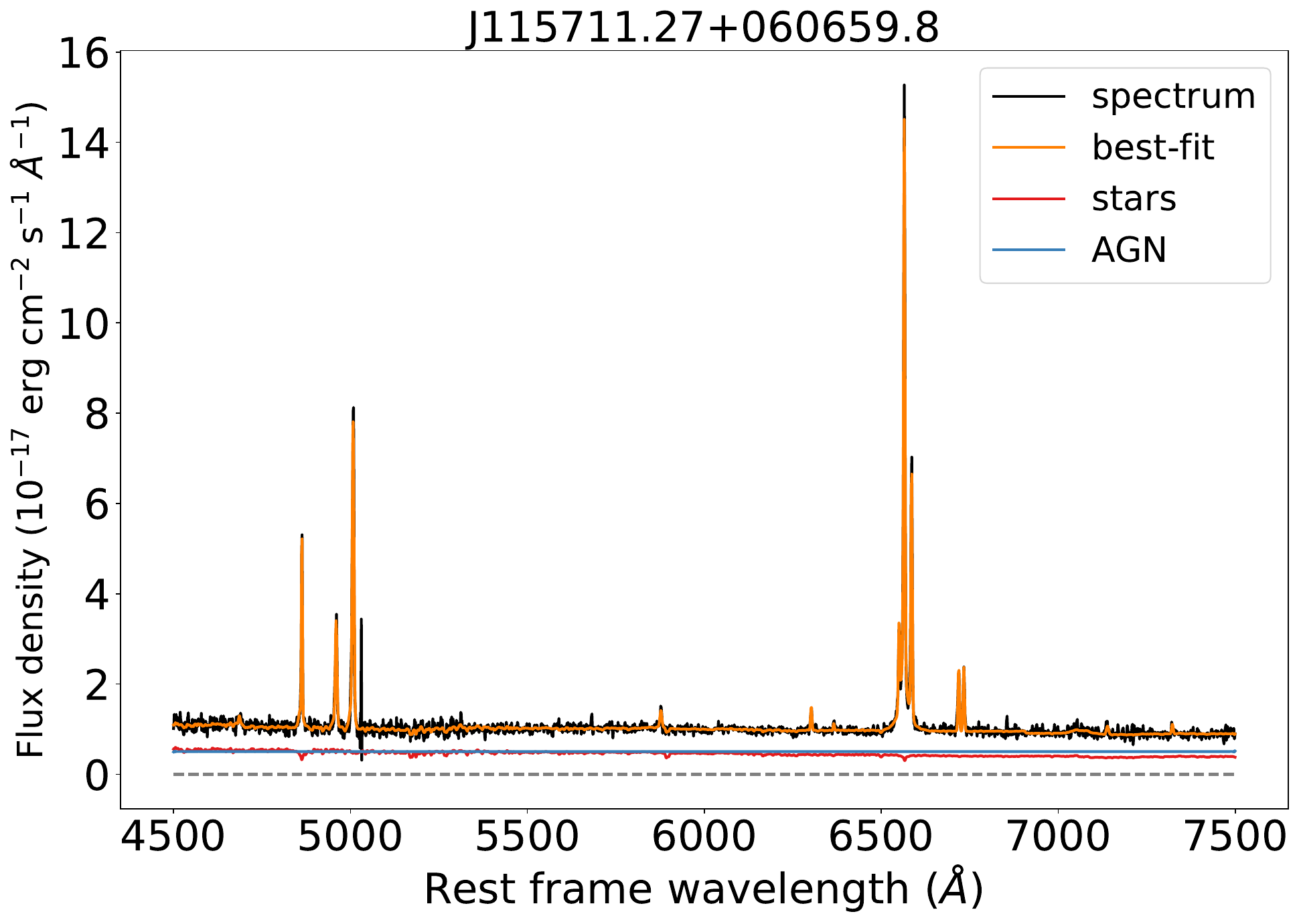}
    \includegraphics[width=0.42\textwidth]{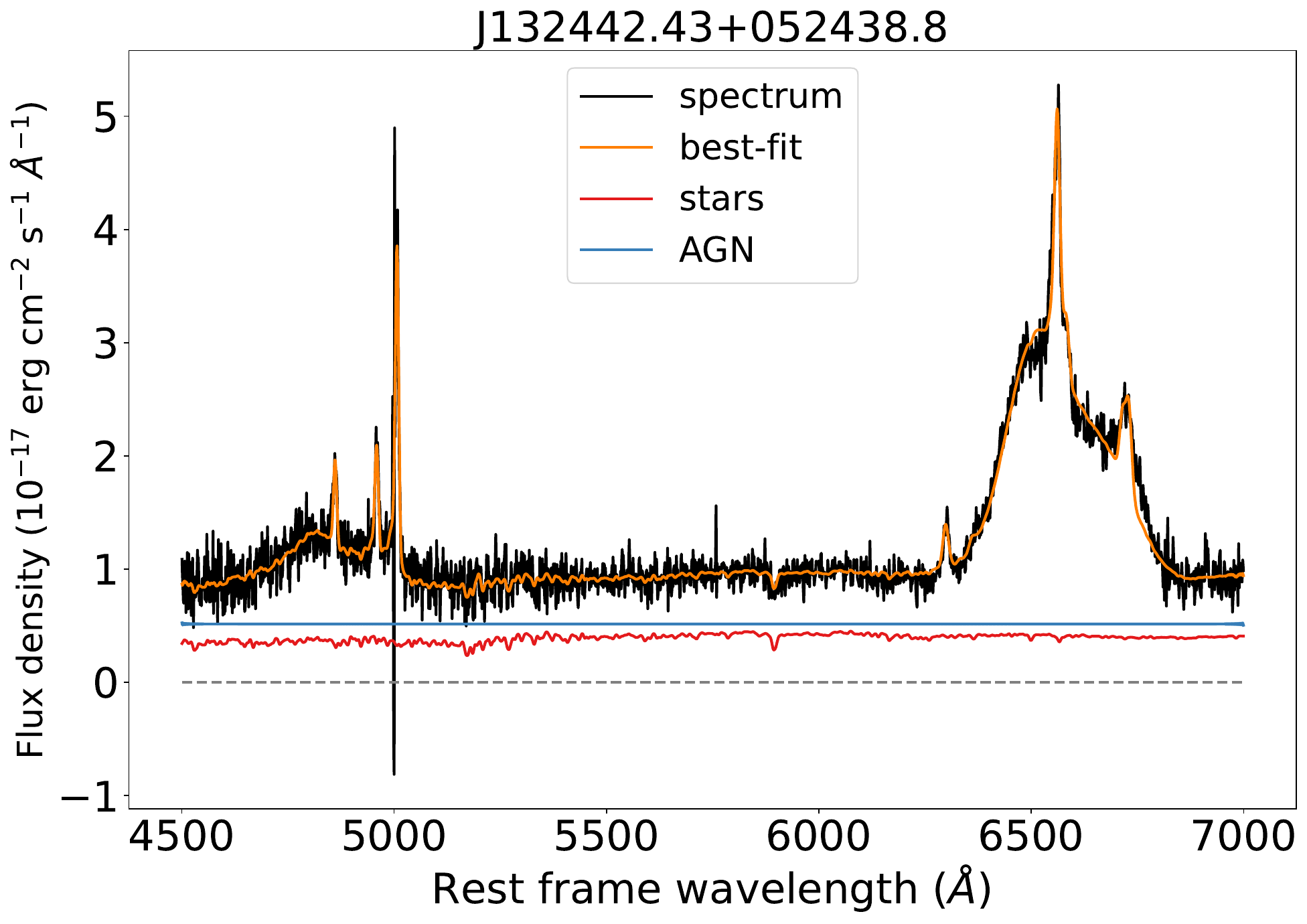}
    \end{figure}
    \addtocounter{figure}{-1} % Restablece el contador
\begin{figure}
    \centering
    \caption{continued}
    \includegraphics[width=0.42\textwidth]{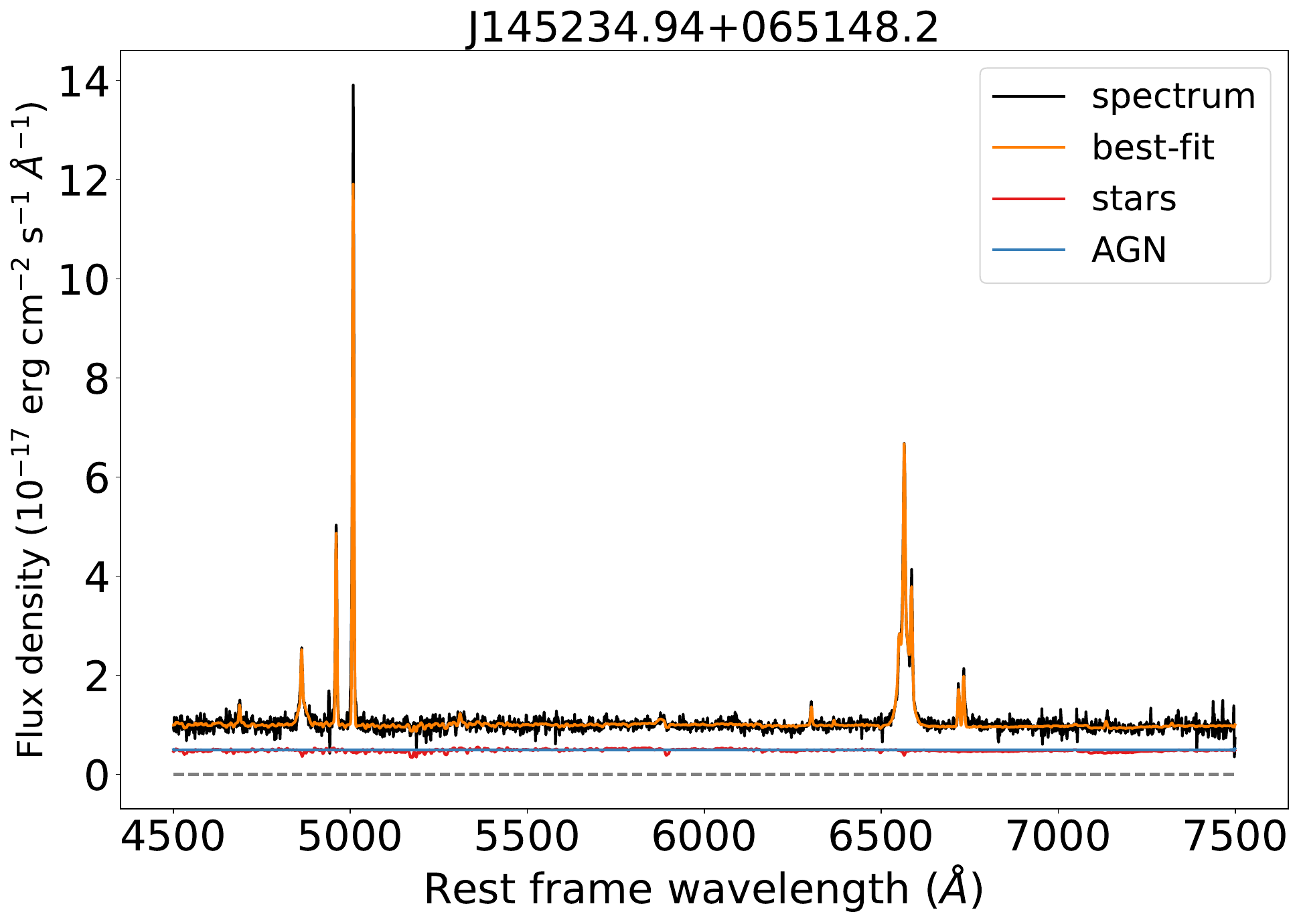}
    \includegraphics[width=0.42\textwidth]{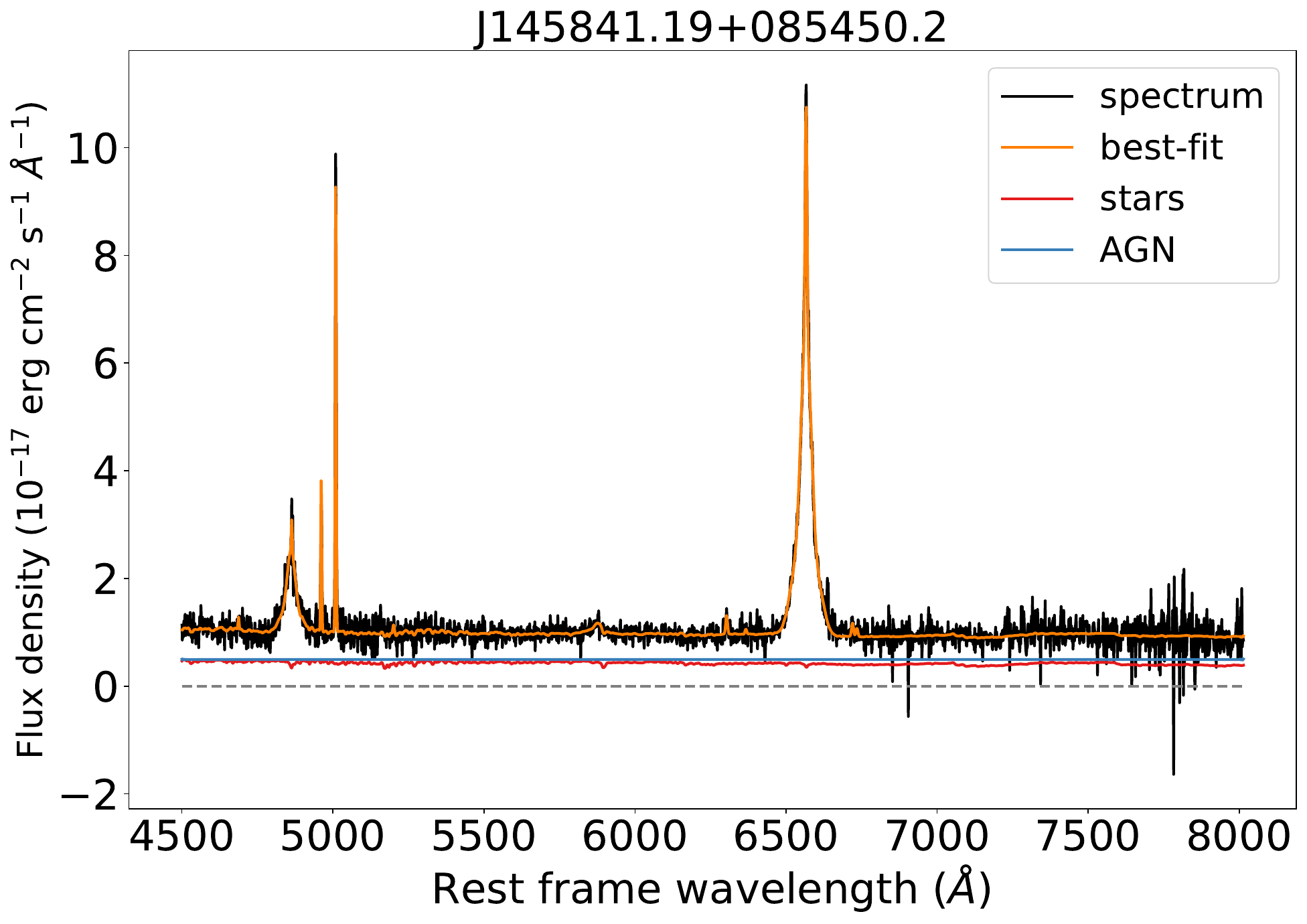}
    \includegraphics[width=0.42\textwidth]{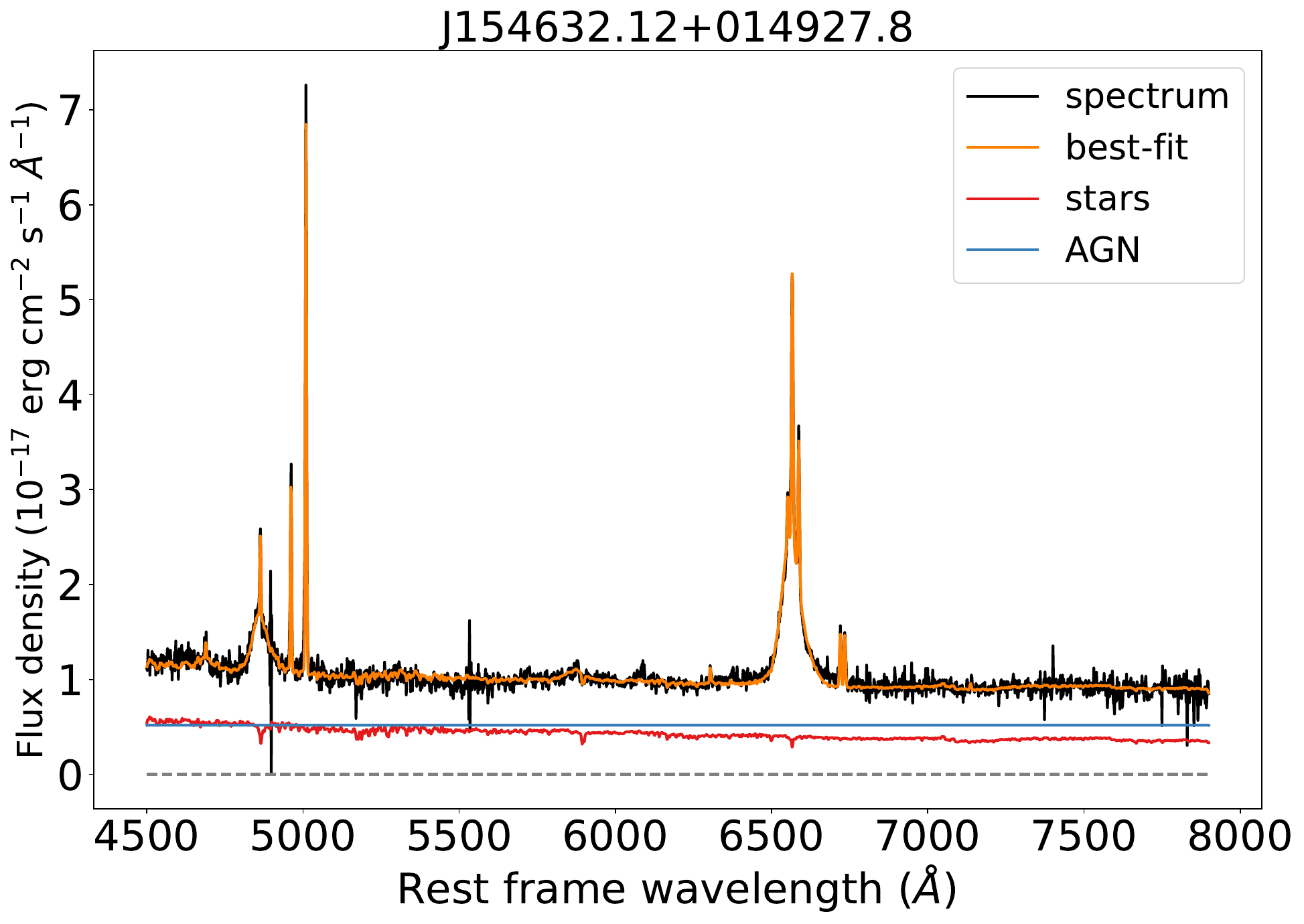}
    \includegraphics[width=0.42\textwidth]{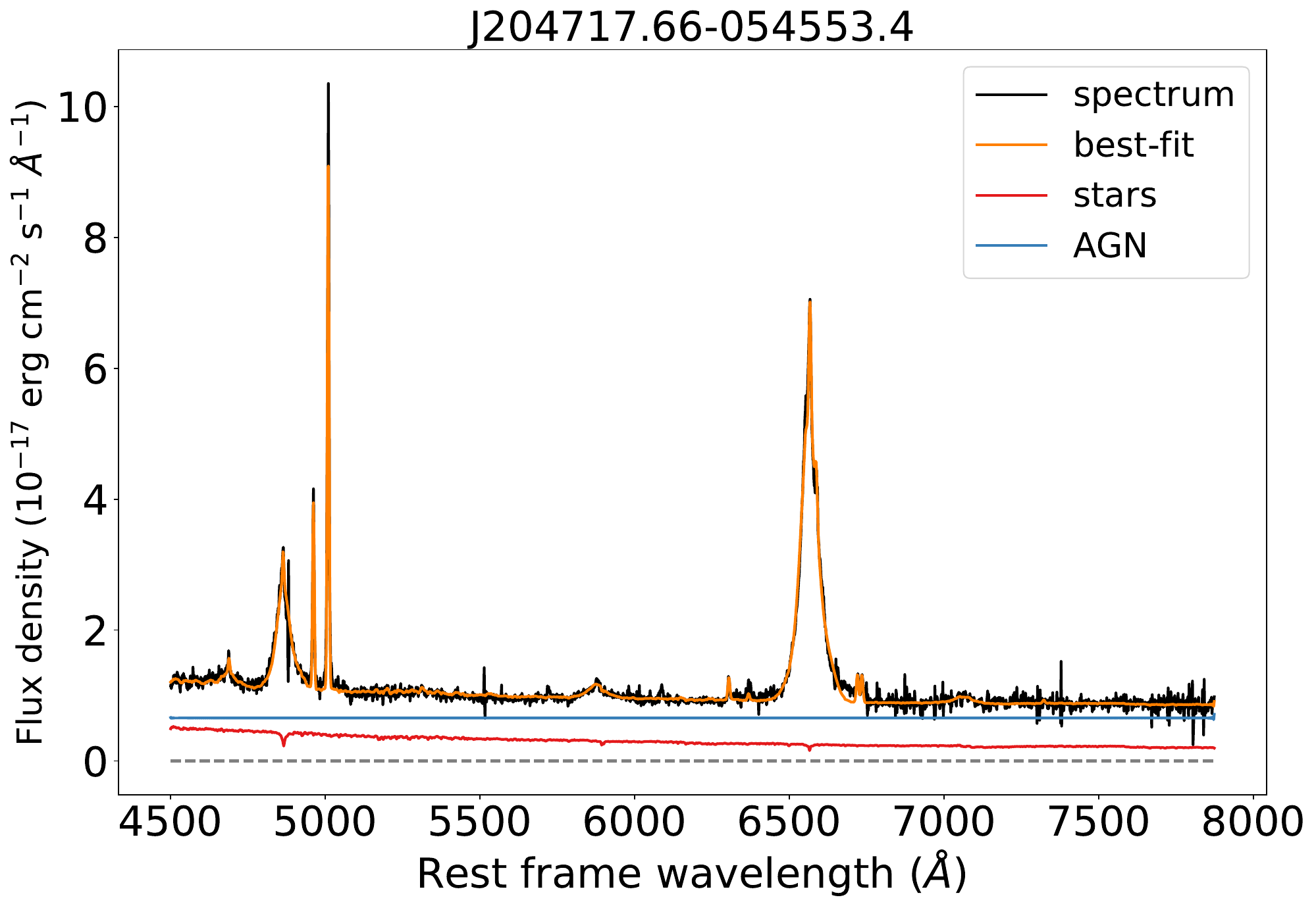}
    
\end{figure}
\FloatBarrier

\end{appendix}
\end{document}